%% file: main.tex
\documentclass[prl,superscriptaddress,reprint,preprintnumbers,longbibliography]{revtex4-2}

\usepackage{textgreek}
\usepackage{braket}
\usepackage{physics}
\input{preamble.tex}

\begin{document}
\title{Simulating Heisenberg Interactions in the Ising Model with Strong Drive Fields}
 \author{Anthony N. Ciavarella \,\orcidlink{0000-0003-3918-4110}}
 \email{aciavare@uw.edu}
 \affiliation{InQubator for Quantum Simulation (IQuS), Department of Physics, University of Washington, Seattle, Washington 98195-1550, USA}
 \author{Stephan Caspar\,\orcidlink{0000-0002-3658-9158}}
 \email{caspar@uw.edu}
 \affiliation{InQubator for Quantum Simulation (IQuS), Department of Physics, University of Washington, Seattle, Washington 98195-1550, USA}
 \author{Hersh Singh\,\orcidlink{0000-0002-2002-6959}}
 \email{hershsg@uw.edu}
 \affiliation{InQubator for Quantum Simulation (IQuS), Department of Physics, University of Washington, Seattle, Washington 98195-1550, USA}
 \author{Martin J. Savage\,\orcidlink{0000-0001-6502-7106}}
 \email{mjs5@uw.edu. On leave from the Institute for Nuclear Theory.}
 \affiliation{InQubator for Quantum Simulation (IQuS), Department of Physics, University of Washington, Seattle, Washington 98195-1550, USA}
 \author{Pavel Lougovski\,\orcidlink{0000-0002-9229-1444}}
 \affiliation{AWS Center for Quantum Networking, Seattle, Washington 98121, USA}
 \preprint{IQuS@UW-21-028, INT-PUB-22-020}

\begin{abstract}
The time-evolution of an Ising model with large driving fields
over discrete time intervals is shown to be reproduced by an effective XXZ-Heisenberg model at leading order in the inverse field strength. For specific orientations of the drive field, the dynamics of the XXX-Heisenberg model is reproduced.
These approximate equivalences, valid above a critical driving field strength
set by dynamical phase transitions in the Ising model,
are expected to enable quantum devices that natively evolve qubits
according to the Ising model to simulate 
more complex systems.
\end{abstract}

\maketitle
%\glsresetall

%%%%%%%%%%%%
\section{Introduction}
\noindent
With the recent advances in digital and analog quantum computers and simulators, there is a growing effort toward mapping quantum field theories onto arrays of qubits and, more generally, qudits~\cite{jordan2019quantum,jordan2014quantum,preskill2018simulating,https://doi.org/10.48550/arxiv.2108.13357,Lamm_2020,Mazza_2012,Yeter_Aydeniz_2019,Kokail_2019,Kharzeev_2020,Zohar:2012ay,Zohar:2012xf,Banerjee:2012xg,Banerjee:2012pg,Martinez2016,Muschik:2016tws,Zohar:2016iic,Banuls:2017ena,Kaplan:2018vnj,Zache:2018jbt,Stryker:2018efp,Raychowdhury:2018tfj,Davoudi:2019bhy,Haase:2020kaj,Paulson_2021,Davoudi:2020yln,PhysRevA.103.042410,Buser:2020cvn,Raychowdhury:2018osk,Raychowdhury:2019iki,Kreshchuk:2020kcz,Tagliacozzo:2012df,Ji:2020kjk,Bender_2018,Klco:2018zqz,10.5555/3179430.3179434,PhysRevA.98.042312,PhysRevA.103.042410,Gharibyan_2021,Ciavarella_2021,klco2021standard,dejong2021quantum,icons2020,meurice2021theoretical,PhysRevD.63.085007,PhysRevD.60.094502,Alam_2022,PhysRevD.105.114508,zohar2021quantum,armon2021photonmediated,klco2021hierarchical,Hostetler_2021,Andrade_2022,PhysRevLett.105.190404,Byrnes_2006,kan2021lattice,stryker2021shearing,stetina2022simulating,https://doi.org/10.48550/arxiv.2207.01731,Paulson_2021,davoudi2021simulating,PhysRevD.100.054505,PhysRevLett.126.172001,https://doi.org/10.48550/arxiv.1911.12353,https://doi.org/10.48550/arxiv.2203.00059,https://doi.org/10.48550/arxiv.2203.15766,PhysRevD.103.114505,PhysRevD.99.074502,https://doi.org/10.48550/arxiv.2108.08248}.
While classical computers are used successfully in studies of the static properties of quantum systems, 
time dependent evolution and finite-density systems typically suffer from non-polynomial scaling computational resource requirements~\cite{PhysRevD.86.105012,PhysRevD.66.074507,DEFORCRAND2002290}. 
This places important quantities of interest in Standard Model physics, and in many other areas,
beyond the capabilities of classical computation. 
In contrast, quantum simulations  are expected to be able to address some of these quantities with polynomially scaling computational costs~\cite{Feynman1982,Benioff:1980},
and 
quantum advantages in scientific applications
are being sought.
A number of platforms for digital quantum computation are being developed, such as those utilizing superconducting qubits, optical qubits, quantum dots, trapped ions and Rydberg atoms~\cite{huang2020superconducting,2019Photon,barthelemy2013quantum,2019Trap,PRXQuantum.2.030322,Weimer_2011,PhysRevLett.115.137002,PhysRevLett.105.223601,https://doi.org/10.48550/arxiv.2206.15467,wallraff2004strong,chiorescu2004coherent}. 
In addition to performing universal digital quantum computations, these platforms can be used for analog simulations 
of systems that can be mapped onto their native Hamiltonians. 
This approach is more limited than universal digital quantum computation, but recent work suggests that the error rates on existing hardware may be 
low enough to enable a useful quantum advantage through analog simulation~\cite{https://doi.org/10.48550/arxiv.2204.13644}.
While the study of spin systems,
and 
systems comprised of finite-dimensional Hilbert spaces more generally,
are interesting by themselves, 
they are now central to 
advancing quantum simulations for scientific applications.

Despite the apparent simplicity
of a regular lattice of qubits or qudits, 
it has been widely appreciated since the 
discovery of quantum mechanics that 
spin systems exhibit a wide variety of non-trivial properties and dynamics.
One of the most studied spin systems is the Heisenberg model which is able to describe critical points and phase transitions in magnetic materials. 
In addition to condensed matter applications, the Heisenberg model also has an important role in high energy physics as, for example, 
it can be used to study the lattice $O(3)$ nonlinear $\sigma$ model~\cite{CHANDRASEKHARAN2002388,BROWER2004149,PhysRevD.99.074501,PhysRevD.100.054505,PhysRevLett.126.172001,https://doi.org/10.48550/arxiv.1911.12353,https://doi.org/10.48550/arxiv.2203.00059,https://doi.org/10.48550/arxiv.2203.15766}. This theory is one of the  ``sandboxes" used to better understand quantum chromodynamics (QCD) as it shares a number of qualitative aspects such as asymptotic freedom and $\theta$-vacua~\cite{https://doi.org/10.48550/arxiv.1911.12353,https://doi.org/10.48550/arxiv.2203.00059,https://doi.org/10.48550/arxiv.2203.15766}. In the electroweak sector, the dynamics of 
coherent collective flavor oscillations of neutrinos can be mapped to 
the Heisenberg model~\cite{Duan:2005cp,Balantekin_2006}. 
This is of particular importance in extreme astrophysical environments, where neutrino-neutrino interactions can be significant. After extensive studies using classical computers, 
including of entanglement,
for example Refs.~\cite{PhysRevD.99.123013,PhysRevD.104.123035,PhysRevC.101.065805,Martin:2019dof,PhysRevD.104.103016,Martin:2021xyl,https://doi.org/10.48550/arxiv.2203.02783,Illa:2022zgu,Martin:2023ljq},
it is expected that quantum simulations~\cite{PhysRevD.104.063009,PhysRevD.104.123023,PhysRevD.105.083020,Yeter-Aydeniz:2021olz,Amitrano:2022yyn,Illa:2022zgu} 
of the Heisenberg model will enable the study of dynamics beyond the reach of classical computers and provide new insights into these problems.
Due to this wide applicability, a number of methods for quantum simulation of the Heisenberg model have been developed, including digital approaches \cite{Childs_2018,Childs_2021}, hybrid digital-analog approaches \cite{PRXQuantum.2.020328}, and analog simulation on trapped ions \cite{PhysRevB.95.024431}, Rydberg atoms using dipole-dipole interactions \cite{PRXQuantum.3.020303} and nuclear spins \cite{MADI1997300,PhysRevLett.124.030601}. Dissipative versions of the Heisenberg model have also been studied on digital quantum computers~\cite{rost2021demonstrating,delre2022robust}.

In this work, the potential for using Ising systems for analog simulations of physical systems that can be mapped to the Heisenberg model is investigated. 
The Ising model is considered because a number of platforms available for quantum simulation, such as Rydberg atoms, trapped ions and superconducting qubits, can be natively described by this Hamiltonian~\cite{ebadi2021quantum,doi:10.1126/science.abi8794,bluvstein2022quantum,PhysRevLett.92.207901,grass2014trapped,RevModPhys.93.025001,johnson2011quantum,PhysRevB.82.024511,6802426}. Unlike previous approaches, this method of analog simulation of the Heisenberg model can be implemented with a time-independent Hamiltonian which could be beneficial for some platforms.
There is an extensive literature regarding 
the time evolution of Ising models in background fields~\cite{PhysRevLett.122.130603,kormos2017real,Richerme_2014,Jurcevic_2014,Wall_2017,Kiely_2018,Robinson_2021,PhysRevLett.110.135704,PhysRevB.87.195104,PhysRevLett.126.040602,RAKOVSZKY2016805,Heyl_2015,PhysRevResearch.4.013250,https://doi.org/10.48550/arxiv.2111.05586,https://doi.org/10.48550/arxiv.2203.06188,sieberer2019digital,Heyl_2019}, including recent work related to transitions to chaotic phases induced by finite-time steps in Trotterized time evolution in digital quantum simulations~\cite{sieberer2019digital,Heyl_2019}, identified through studying the system at periodic times. Previous work has shown that Ising interactions with a transverse field generate time evolution according to the XY model~\cite{Richerme_2014,Jurcevic_2014,Wall_2017,Kiely_2018,https://doi.org/10.48550/arxiv.2208.01869}. Studying the time evolution of the Ising model has also shown that the Ising model undergoes confinement analogous to QCD near its critical point with a field in the $\hat{x}$ and $\hat{z}$ directions~\cite{kormos2017real,PhysRevLett.122.130603,PhysRevB.99.195108}.
Known to be universal in a computational sense~\cite{Cubitt_2018},
developing analog simulations of the Heisenberg model is expected to also advance simulations of other physical systems.
We show how constant driving fields in the Ising model 
generate an effective Heisenberg model Hamiltonian to leading order in the inverse field strength at periodic times. The systematic errors associated with the periodic dynamics are quantified, revealing 
that the convergence of the expansion in the inverse field strength is limited by the location of 
dynamical quantum phase transitions (DQPTs) in the Ising model with an external field.

%%%%%%%%%%%%%%%%%%%%%%%%%
\section{Heisenberg from Ising with Strong Fields}
\begin{figure}[!th]
    \centering
    \includegraphics[width=8.6cm,trim=0cm 2cm 0cm 0cm, clip]{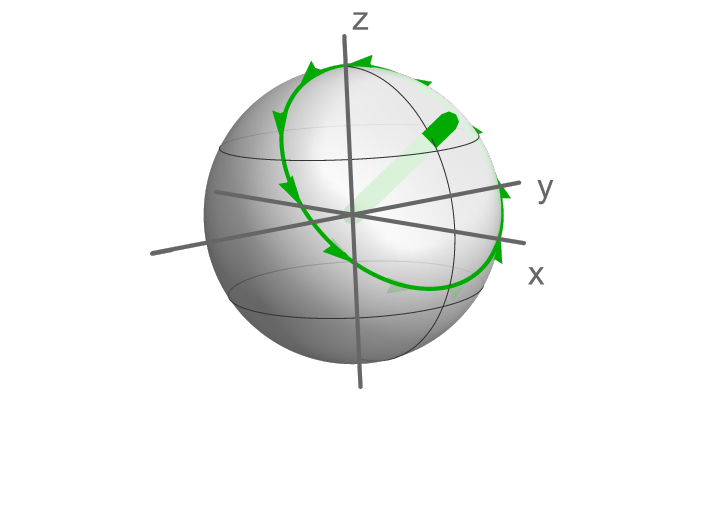}
    \caption{
    A representative path on the unit sphere taken by $\Vec{e}(t)$ that generates time evolution according to the XXX-Heisenberg model. The green line corresponds to the direction of the driving field.}
    \label{fig:Zpath}
\end{figure}

\noindent
The Ising Hamiltonian with constant global driving fields is given by
\begin{equation}
	\hat{H}^{\rm Ising} = \sum_{i > j} J_{i j} \hat{Z}_i \hat{Z}_j + \frac{1}{2} \sum_i \Omega_x \hat{X}_i + \Omega_y \hat{Y}_i + \Omega_z \hat{Z}_i \ \ \ .
\end{equation}
To analyze the evolution of this Hamiltonian, it is helpful to transform into the 
interaction picture, where the driving fields are taken to be the ``free'' term in the Hamiltonian. 
The interaction-picture Hamiltonian is
\begin{equation}
	\hat{H}_{I}^{\rm Ising}(t) = \sum_{i > j} J_{i j} \hat{Z}_{I,i}(t) \hat{Z}_{I,j}(t) \ \ \ ,
\end{equation}
where $ \hat{Z}_{I,i}(t) = \hat{U}^\dagger_0(t) \hat{Z}_i \hat{U}_0(t) = \Vec{e}(t) \cdot \Vec{S}$, $\hat{U}_0(t) = \prod_{j} e^{-i t \left(  \Omega_x \hat{X}_j + \Omega_y \hat{Y}_j + \Omega_z \hat{Z}_j \right)}$, $\Vec{S}$ is a vector of Pauli matrices and $\Vec{e}(t)$ is a unit vector. 
From this perspective, the driving fields can be viewed as rotating $\Vec{e}(t)$ from the north pole to other points on the unit sphere. 
By choosing periodic driving fields that generate closed paths on the sphere, it is possible to engineer evolution according to different Hamiltonians. 
The use of periodic dynamics to generate different Hamiltonians, known as Floquet engineering, has been used to simulate a range of interactions~\cite{PhysRevX.4.031027,PhysRev.138.B979,RevModPhys.76.1037,PhysRevX.5.021027,PhysRevLett.111.185301,doi:10.1126/science.aad4568,PhysRevLett.116.205301,schweizer2019floquet,RevModPhys.89.011004,wintersperger2020realization,PRXQuantum.3.020303,PhysRevB.95.024431,PhysRevX.10.031002},
including the Ising Hamiltonian from the Heisenberg interaction in quantum-dot systems~\cite{qiao2021floquet,https://doi.org/10.48550/arxiv.2204.13354}. Floquet engineering has also been previously applied to static Hamiltonians in the interaction picture to understand how some systems prethermalize to a Hamiltonian that is not the generator of their evolution~\cite{abanin2017rigorous,PhysRevX.7.011026,PhysRevB.100.020301,https://doi.org/10.48550/arxiv.2103.07485}. In particular, it has been used to show that the dynamics of the XYZ-Heisenberg model with a strong external field are approximated by the XXZ-Heisenberg model for times that are exponential in the driving field~\cite{abanin2017rigorous}. We will show that in the Ising model, evolution according to the XXZ-Heisenberg Hamiltonian can be approximated by taking $\Omega_x = \Omega \sin{\theta}$, $\Omega_y = 0$, and $\Omega_z = \Omega \cos{\theta}$. 
With these driving fields, the interaction Hamiltonian becomes periodic over time intervals $\frac{2\pi}{\Omega}$,
and the Schrodinger picture becomes equivalent to the interaction picture at these periods.
A representative path generated by such fields on the unit sphere is shown in Fig.~\ref{fig:Zpath}. 
The time-evolution of the system
after discrete time intervals, and the associated 
Magnus expansion is given by
\begin{eqnarray}
\hat{U}_F  & = & \mathcal{T} \exp{-i\int_{0}^{\frac{2\pi}{\Omega}}\!\! dt^\prime\ \hat{H}_{I}^{\rm Ising}(t^\prime)} 
\nonumber\\
& = & \hat{U}^\dagger_B
\exp{
-i \frac{2\pi}{\Omega} \left( \hat{H}_1 \ +\ O\left({1\over \Omega}\right)\ \right)} \hat{U}_B
\ ,
\label{eq:floq}
\end{eqnarray}
where
\begin{eqnarray}
\hat{H}_1 & = &
\sum_{i > j} J_{ij}
\left[\cos^2\!\theta\ \hat{Z}_i \hat{Z}_j  +\frac{\sin^2\!\theta}{2}\left(\hat{X}_i \hat{X}_j + \hat{Y}_i \hat{Y}_j\right) \right]
,
\label{eq:A1}
\end{eqnarray}
where $\hat{U}_B$ is a local change of basis given by
$\hat{U}_B = 
\prod_j e^{+i \theta\hat{Y}_j/2} $
(that aligns the driving field with the $z$ axis). 
Therefore, time evolution 
of the Ising model with this choice of driving fields 
approximates that of the XXZ-Heisenberg model between discrete intervals of 
$\Delta t = \frac{2\pi}{\Omega}$.
Note that while the formalism of Floquet engineering was used to derive this result, 
the Hamiltonian is time independent and the periodicity is only manifest in the 
interaction picture. Also, the Ising model with a fast oscillating drive field could be used to simulate an XXZ-Heisenberg model  because the dynamics of a transversely driven Ising model are equivalent to that of a time-independent Ising model with external fields in the $\hat{z}$ and $\hat{x}$ directions~\cite{Robinson_2021}.
The $O\left(\frac{1}{\Omega^2}\right)$ higher-order terms in the Magnus expansion
of the Floquet operator
in Eq.~(\ref{eq:floq})
have one-body and three-body operators.  The one-body operators can be 
eliminated by renormalization of the ``free'' Hamiltonian employed to transform to the interaction picture, but the three-body terms are a genuine deviation from the Heisenberg Hamiltonian. Such higher-order terms in the Magnus expansion can be removed through the use of time-dependent driving fields \cite{PhysRevLett.125.080602}.

This approach to simulating the XXZ-Heisenberg model is similar in spirit to recent proposals for simulating gauge theories by adding terms to the Hamiltonian that generate gauge symmetries~\cite{PRXQuantum.2.040311,halimeh2022gauge,https://doi.org/10.48550/arxiv.2108.02203,PhysRevLett.82.2417,https://doi.org/10.48550/arxiv.2012.08620}. 
In these proposals, an energy penalty for breaking gauge invariance decouples the gauge invariant sector from the rest of Hilbert space analogously to how dynamical decoupling can be used to decouple systems from their enviroment~\cite{PhysRevLett.82.2417}. In this work, the addition of driving fields to the Ising model can be interpreted as adding an energy penalty for violating the global $U(1)$ symmetry generated by the driving fields. This causes the different symmetry sectors to decouple, leading to time evolution that can be described by the XXZ-Heisenberg model.

%%%%%%%%%%%%%%%%%%%%%%%%%%%%%%%%%%
\section{Beyond Leading Order in the  Magnus Expansion and Dynamical Phase Transitions}
\noindent
The derivation of the approximate Heisenberg time evolution indicates that systematic errors from the Magnus expansion
are suppressed by $\Omega^{-1}$ compared to leading order.
However, the Magnus expansion is known to have a finite radius of convergence, and a priori it is not obvious what the minimum value of $\Omega$ is for the leading order term to accurately describe dynamics. 
As mentioned previously, numerical studies of digital quantum simulations have been used to show that Trotterized time evolution transitions into chaotic dynamics for sufficiently large time steps~\cite{sieberer2019digital,Heyl_2019}. In this context, the Floquet-period
$\Delta t = \frac{2\pi}{\Omega}$ is analogous to a Trotter time step, and we show that at small $\Omega$ the breakdown of the Magnus expansion is associated with a dynamical quantum phase transition in the Ising model.

\begin{figure}[!t]
    \centering
    \includegraphics[width=8.6cm]{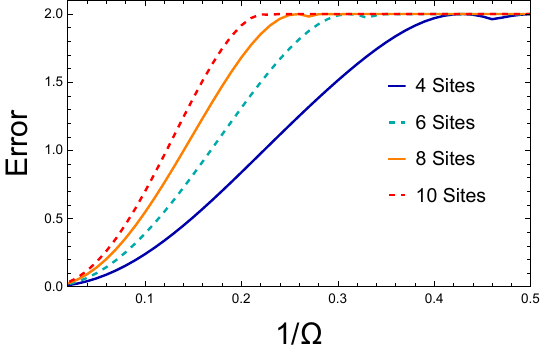}
    \caption{
    The spectral norm (magnitude of the largest eigenvalue) $||e^{-i \frac{2\pi}{\Omega} \hat{H}^\text{Heis.}_{\text{XXX}}} - \hat{U}_F||$ of the difference between the XXX-Heisenberg chain time-evolution operator 
    derived from Eq.~(\ref{eq:XXXHeis})
    and the Floquet engineered approximation in
    Eqs.~(\ref{eq:floq}) and (\ref{eq:A1})
    as a function of the driving field strength $\Omega$, 
    for a selection of chain lengths.
    }
    \label{fig:FloquetError}
\end{figure}

As an example, we focus on the special point $\theta= \tan^{-1} \sqrt{2}$, where the leading order Eq.~\eqref{eq:A1} becomes an XXX-Heisenberg Hamiltonian with enhanced non-abelian $O(3)$-symmetry
\begin{equation}
    \hat{H}^{\text{Heis.}}_{\text{XXX}} = {1\over 3} \sum_i \hat{X}_i \hat{X}_{i+1} + \hat{Y}_i \hat{Y}_{i+1} + \hat{Z}_i \hat{Z}_{i+1}.
    \label{eq:XXXHeis}
\end{equation}
%
% The spectral norm (magnitude of the largest eigenvalue) of the difference between the exact Heisenberg time-evolution operator and the Floquet engineered approximation given in Eqs.~(\ref{eq:floq}) and (\ref{eq:A1}) bounds the systematic errors in the time evolution (of any state).
The systematic errors in the time evolution (of any state) are bounded by the spectral norm (magnitude of the largest eigenvalue) of the difference between the exact Heisenberg time-evolution operator and the Floquet engineered approximation given in Eqs.~(\ref{eq:floq}) and (\ref{eq:A1}),
$||e^{-i \frac{2\pi}{\Omega} \hat{H}^\text{Heis.}_{\text{XXX}}} - \hat{U}_F||$.
This is shown for the one-dimensional XXX-Heisenberg model with $J=1/3$ in Fig. \ref{fig:FloquetError} as a function of $\Omega$ for varying chain lengths. 
At large values of $\Omega$, 
systematic deviations in the spectral norm decrease with increasing $\Omega$ as predicted by the Magnus expansion. 
At small values of $\Omega$, the spectral norm saturates below a critical value
$\Omega_c$. 
Unfortunately, the lattice sizes for which the spectral norm can be efficiently computed are not large enough to determine the scaling of $\Omega_c$  with chain length.
\begin{figure}[!th]
    \centering
    \includegraphics[width=8.6cm]{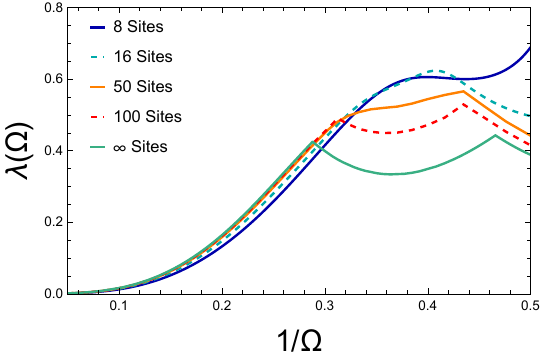}
    \caption{
    The rate function, defined in Eq. \ref{eq:rateFn}, for the ground state of XXX-Heisenberg chains of different lengths.
    }
    \label{fig:VacEcho}
\end{figure}
For longer chain lengths, 
Loschmidt echoes of the 
ground state of the XXX-Heisenberg model
in Eq.~(\ref{eq:XXXHeis}),
$\ket{\psi_G}$,
time evolved over $t=\frac{2\pi}{\Omega}$ with the driven Ising model,
\begin{equation}
    \hat{H}^{\rm Ising} = \sum_i \hat{Z}_i \hat{Z}_{i+1} + \frac{\Omega}{2\sqrt{3}} \left( \hat{Z}_i +\sqrt{2} \hat{X}_i \right)
    \ ,
\end{equation}
are computed.
If the time evolution of the XXX-Heisenberg model were perfectly reproduced by the driven Ising model, 
the Loschmidt echo, 
defined as the probability to return to the initial state, i.e.,
\begin{equation}
    \mathcal{L}(\Omega) = \abs{\bra{\psi_G} e^{-i \frac{2\pi}{\Omega} \hat{H}^{\rm Ising}}\ket{\psi_G}}^2 
    \ ,
\end{equation}
would equal unity,
and deviation from unity provide an estimate of contributions beyond leading order in the Magnus expansion.
As $\log \mathcal{L}(\Omega)$ is an extensive quantity, the rate function
\begin{equation}
    \lambda(\Omega) = -\log\left(\mathcal{L}(\Omega) \right)/L \ \ \ ,
    \label{eq:rateFn}
\end{equation}
is computed 
to compare chains of different lengths $L$, as shown in Fig.~\ref{fig:VacEcho}.
For chains of $L\le 16$, time evolution was computed using exact diagonalization. 
The ground states of the $L=50$ and $L=100$ chains were computed using DMRG  
and time evolution was performed using TDVP~\cite{itensor,PhysRevLett.69.2863,PhysRevB.48.10345,PhysRevLett.107.070601,PhysRevB.94.165116,PhysRevB.102.094315}. 
The ground state and time evolution of the infinite Heisenberg chain was computed using iTEBD~\cite{2003,2004Eff,2004MPS}. 
At large $\Omega$, $\lambda(\Omega)$ 
decreases with increasing $\Omega$, 
indicating that the leading order Magnus expansion is correctly describing the dynamics of the model.
This asymptotic behavior only occurs beyond  a ``kink'' in $\lambda(\Omega)$, 
indicating that at small values of $\Omega$ the Magnus expansion is failing to converge.
The presence of a kink (non-analytic behavior) in $\lambda(\Omega)$ is the defining characteristic of a dynamical quantum phase transition \cite{PhysRevLett.110.135704}. Note that other inequivalent definitions of dynamical phase transitions have been introduced in the literature~\cite{Heyl_2018}. 
DQPTs have previously been studied in spin systems and have been shown to be associated with unstable renormalization group fixed points~\cite{PhysRevLett.110.135704,PhysRevB.87.195104,PhysRevLett.126.040602,RAKOVSZKY2016805,Heyl_2015,PhysRevResearch.4.013250,alvarez2006environmentally}. 
Our results show that the Ising model with a constant driving field $\vec\Omega = \Omega \left(  \frac{1}{\sqrt{3}} \hat{z} + \sqrt{\frac{2}{3}}\hat{x} \right)$ 
undergoes a DQPT into a regime with an approximate $O(3)$ symmetry at discrete time intervals, 
as seen  in Fig. \ref{fig:VacEcho}. 

\begin{figure}[!th]
    \centering
    \includegraphics[width=8.6cm]{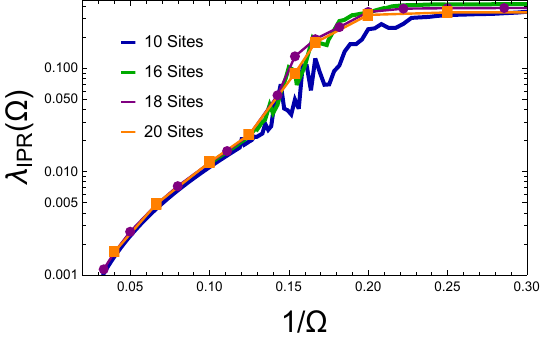}
    \caption{
    The log IPR for the ground state of XXX-Heisenberg chains of different lengths. The IPR for the chain of length 10 was computed with exact diagonalization and the IPR for larger chains was computed by averaging the Loschmidt echo over 1000 periods.
    }
    \label{fig:LongEcho}
\end{figure}
These calculations demonstrate that for short time scales Heisenberg evolution is being successfully simulated, provided a sufficiently large driving field strength is used. However, this does not guarantee that the dynamics are reproduced at long times. Generically, periodically driven systems are expected to heat at late times~\cite{PhysRevE.90.012110,Abanin_2015,PhysRevLett.116.120401}, however in the context of digital quantum simulation it has been shown that quantum localization prevents this in Trotterized time evolution~\cite{sieberer2019digital,Heyl_2019}. The Floquet engineering technique used in this work uses a static Ising Hamiltonian so one would expect that at large field strengths the eigenstates of the Ising model are perturbatively close to those of the Heisenberg Hamiltonian. This would guarantee that long time dynamics are correctly reproduced as in the case of Trotterized time evolution. This perturbative argument can be verified through the calculation of the inverse participation ratio (IPR). For a given state, $\ket{\psi}$, and eigenstates, $\ket{n}$ of some Hamiltonian, the IPR is defined by
\begin{equation}
    \text{IPR} = \sum_n \abs{\bra{n}\ket{\psi}}^4 \ \ \ .
    \label{eq:IPR}
\end{equation}
The IPR measures how localized $\ket{\psi}$ is relative to the eigenbasis $\ket{n}$. In practice, it can be evaluated by averaging the Loschmidt echo over long periods of time. To compare systems of different sizes, a normalized IPR, defined by $\lambda_{\text{IPR}}(L) = -\frac{1}{L}\log(\text{IPR})$ for a chain of length $L$, was computed for the ground state of the XXX-Heisenberg model in Fig. \ref{fig:LongEcho}. For the chain of length 10, the IPR was computed by explicitly evaluating Eq.~(\ref{eq:IPR}), while for the larger chains the IPR was computed by averaging Loschmidt echos. As this figure shows, for large $\Omega$ the log IPR is small which indicates the perturbative argument holds and the XXX-Heisenberg ground state is localized with respect to the Ising Hamiltonian. This indicates that the long time dynamics of the XXX-Heisenberg model is being successfully simulated with this technique.

\begin{figure}[tb]
    \centering
    \includegraphics[width=8.6cm]{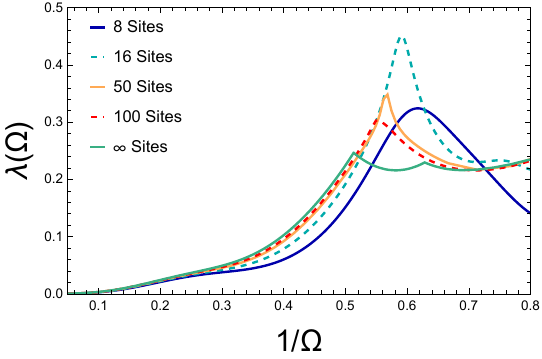}
    \caption{The rate function for the ground state of XY-Heisenberg chains of different lengths.}
    \label{fig:VacEchoYZ}
\end{figure}
When the constant driving field is taken to be in another direction, 
an approximate $O(2)$ symmetry emerges. 
The arguments above suggest that there should be a DQPT that occurs in this case as well. 
As an example, the traditional one-dimensional transverse field Ising model with the driving field purely in the $\hat{x}$ direction will generate evolution according to the XY-Heisenberg model (up to a change of basis), 
\begin{equation}
    \hat{H}^{\rm Heis.}_{XY}  = {1\over 2}\sum_i  \hat{Y}_i \hat{Y}_{i+1} + \hat{Z}_i \hat{Z}_{i+1} 
    \  .
\end{equation}
The rate function $\lambda(\Omega)$ for the ground state of $\hat{H}_{XY}$ evolved under the transverse field Ising Hamiltonian for one period is shown in Fig. \ref{fig:VacEchoYZ}. 
As is the case for the XXX-Heisenberg model, there is a series of kinks indicating a DQPT before the rate function begins to decrease. 
It is interesting to note that the final kink  is at $\Omega^* \approx 1.948$ which is close to, but not quite at the critical point of the transverse field Ising model at $\Omega = 2$. While this work shows that the Ising model can be used to simulate $\hat{H}_{XY}$ in the strong field limit, it has been shown in previous work that in $2+1$ dimensions, the weak field limit of the Ising model also reproduces the dynamics of the XY-Heisenberg model~\cite{https://doi.org/10.48550/arxiv.2111.05586,https://doi.org/10.48550/arxiv.2203.06188}.

These results explicitly show that in 1D, this technique can be used to simulate Heisenberg model physics for long times with a driving field that is not extensive with the system size. While there are classical computational tools that enable the study of large 1D systems such as tensor networks, simulating real time evolution in 1D systems still has computational costs that grow exponentially with time and analog quantum simulation may be of practical use. While these calculation were only performed for 1D, this technique can also be applied to simulate higher dimensional Heisenberg models and it is likely that the required driving field strength for simulating dynamics accurately is not extensive with the system size as well. Even if the required driving field strength is extensive with system size in higher dimensions, this technique may still be of practical importance as real time evolution for even modest sized 2D systems is difficult for classical computers. In addition to enabling analog quantum simulation of the Heisenberg model on platforms with Ising interactions such as Rydberg atoms and superconducting qubits, this technique could be combined with the results of Ref.~\cite{Cubitt_2018} to potentially perform analog quantum simulation of an arbitrary Hamiltonian. This could potentially enable analog simulations of any quantum field theory of physical interest on these platforms.

%%%%%%%%%%%%%%%%%%%%%%%%
\section{Discussion}
\noindent
In this work, a new method for analog simulation of the Heisenberg model has been proposed that can be implemented on platforms whose natural evolution is described by the Ising model with constant external fields. For a specific driving field, the time-evolution operator of the Ising model is approximately that of the Heisenberg model over periodic time intervals.
Interestingly, the leading-order effective Heisenberg operator has enhanced symmetry over the intrinsic Hamiltonian.

The systematic errors associated with this method at small external-field strength are limited by non-analytic behavior 
in the Ising model,
associated with dynamical quantum phase transitions, which indicates the Magnus expansion is failing to converge.   
Beyond a critical value of the driving field, the effective Hamiltonian describing the time-evolution can be determined from a Magnus expansion, with each increasing order in $1/\Omega$ introducing operators involving an increasing number of spins.

The technique presented in this work could be implementable on a range of quantum devices, including systems of Rydberg atoms or even superconducting qubits, in any dimension. While Rydberg systems are very promising as analog quantum simulators, one of the main challenges is that natively they offer a very narrow class of interactions, which severely limits their applicability. Therefore, being able to engineer new interactions makes an important step forward in expanding the systems that can be studied on these platforms.
Furthermore, unlike previous proposals for quantum simulations of Heisenberg models, this method can be implemented using time-independent fields, which is extremely important for experimental platforms with a limited slew rate for the external fields. 
In the near term, analog quantum simulations will be the only method of probing the long time dynamics of large systems. The dynamics of the Heisenberg model are of particular interest, not only for condensed matter applications, but also for high-energy physics, such as in coherent neutrino oscillations and as a lattice regularization of the O(3) nonlinear $\sigma$ model which will be important for developing quantum simulations of QCD. 
Importantly, this technique easily scales to higher dimensions.  By enabling analog simulation of the Heisenberg model on Rydberg atoms and superconducting qubits, systems beyond the reach of classical computers can be simulated and one may be able to achieve a scientifically useful quantum advantage.

%%%%%%%%%%%%%%%%
\section*{Acknowledgments}
We would like to thank Marc Illa, Roland Farrell, Henry Froland, and Nikita Zemlevskiy for useful discussions. We would also like
to acknowledge Peter Komar and Cedric Lin from the Amazon Braket team for contributions to developing this technique. 
The
views expressed are those of the authors and do not reflect the official policy or position of AWS. 
The material presented here was funded in part by the DOE QuantISED program through the theory  consortium ``Intersections of QIS and Theoretical Particle Physics'' at Fermilab with Fermilab Subcontract No. 666484,
in part by Institute for Nuclear Theory with US Department of Energy Grant DE-FG02-00ER41132,
and in part by U.S. Department of Energy, Office of Science, Office of Nuclear Physics, Inqubator for Quantum Simulation (IQuS) under Award Number DOE (NP) Award DE-SC0020970.
This work was enabled, in part, by
the use of advanced computational, storage and networking infrastructure provided by the Hyak supercomputer system at the University of Washington 
\footnote{\url{https://itconnect.uw.edu/research/hpc}}.
This work is also supported, in part, through the Department of Physics 
\footnote{\url{https://phys.washington.edu}}
and the College of Arts and Sciences 
\footnote{\url{https://www.artsci.washington.edu}}
at the University of Washington.

\bibliography{refs}
\end{document}

%% file: preamble.tex
\usepackage{multirow}
\usepackage{siunitx}

\usepackage{tikz}
\usetikzlibrary{decorations.markings, arrows, positioning, calc}

\usepackage{graphicx,times}
\usepackage{latexsym}
\usepackage{mathtools}

\usepackage{amsmath,amssymb,amsbsy,amsfonts}
\usepackage{array}
\usepackage{bm}

\usepackage{graphics}
\graphicspath{{./fig/}}

\usepackage{mathrsfs}
\usepackage{xcolor}
\usepackage{cancel}
\usepackage[normalem]{ulem}

\usepackage[unicode, pdfprintscaling=None, colorlinks]{hyperref}
\hypersetup{
    colorlinks,
    allcolors=blue!50!black,
}
\usepackage{inconsolata}

\usepackage[capitalise]{cleveref}

\usepackage{relsize}

\usepackage{microtype}

\usepackage{xfrac}
\usepackage{orcidlink}

\usepackage{ulem} 
\usepackage{amssymb}

\newcommand\del\partial

%% file: main.bbl
%apsrev4-2.bst 2019-01-14 (MD) hand-edited version of apsrev4-1.bst
%Control: key (0)
%Control: author (8) initials jnrlst
%Control: editor formatted (1) identically to author
%Control: production of article title (0) allowed
%Control: page (0) single
%Control: year (1) truncated
%Control: production of eprint (0) enabled
\begin{thebibliography}{180}%
\makeatletter
\providecommand \@ifxundefined [1]{%
 \@ifx{#1\undefined}
}%
\providecommand \@ifnum [1]{%
 \ifnum #1\expandafter \@firstoftwo
 \else \expandafter \@secondoftwo
 \fi
}%
\providecommand \@ifx [1]{%
 \ifx #1\expandafter \@firstoftwo
 \else \expandafter \@secondoftwo
 \fi
}%
\providecommand \natexlab [1]{#1}%
\providecommand \enquote  [1]{``#1''}%
\providecommand \bibnamefont  [1]{#1}%
\providecommand \bibfnamefont [1]{#1}%
\providecommand \citenamefont [1]{#1}%
\providecommand \href@noop [0]{\@secondoftwo}%
\providecommand \href [0]{\begingroup \@sanitize@url \@href}%
\providecommand \@href[1]{\@@startlink{#1}\@@href}%
\providecommand \@@href[1]{\endgroup#1\@@endlink}%
\providecommand \@sanitize@url [0]{\catcode `\\12\catcode `\$12\catcode
  `\&12\catcode `\#12\catcode `\^12\catcode `\_12\catcode `\%12\relax}%
\providecommand \@@startlink[1]{}%
\providecommand \@@endlink[0]{}%
\providecommand \url  [0]{\begingroup\@sanitize@url \@url }%
\providecommand \@url [1]{\endgroup\@href {#1}{\urlprefix }}%
\providecommand \urlprefix  [0]{URL }%
\providecommand \Eprint [0]{\href }%
\providecommand \doibase [0]{https://doi.org/}%
\providecommand \selectlanguage [0]{\@gobble}%
\providecommand \bibinfo  [0]{\@secondoftwo}%
\providecommand \bibfield  [0]{\@secondoftwo}%
\providecommand \translation [1]{[#1]}%
\providecommand \BibitemOpen [0]{}%
\providecommand \bibitemStop [0]{}%
\providecommand \bibitemNoStop [0]{.\EOS\space}%
\providecommand \EOS [0]{\spacefactor3000\relax}%
\providecommand \BibitemShut  [1]{\csname bibitem#1\endcsname}%
\let\auto@bib@innerbib\@empty
%</preamble>
\bibitem [{\citenamefont {Jordan}\ \emph {et~al.}(2019)\citenamefont {Jordan},
  \citenamefont {Lee},\ and\ \citenamefont {Preskill}}]{jordan2019quantum}%
  \BibitemOpen
  \bibfield  {author} {\bibinfo {author} {\bibfnamefont {S.~P.}\ \bibnamefont
  {Jordan}}, \bibinfo {author} {\bibfnamefont {K.~S.~M.}\ \bibnamefont {Lee}},\
  and\ \bibinfo {author} {\bibfnamefont {J.}~\bibnamefont {Preskill}},\
  }\href@noop {} {\bibinfo {title} {Quantum computation of scattering in scalar
  quantum field theories}} (\bibinfo {year} {2019}),\ \Eprint
  {https://arxiv.org/abs/1112.4833} {arXiv:1112.4833 [hep-th]} \BibitemShut
  {NoStop}%
\bibitem [{\citenamefont {Jordan}\ \emph {et~al.}(2014)\citenamefont {Jordan},
  \citenamefont {Lee},\ and\ \citenamefont {Preskill}}]{jordan2014quantum}%
  \BibitemOpen
  \bibfield  {author} {\bibinfo {author} {\bibfnamefont {S.~P.}\ \bibnamefont
  {Jordan}}, \bibinfo {author} {\bibfnamefont {K.~S.~M.}\ \bibnamefont {Lee}},\
  and\ \bibinfo {author} {\bibfnamefont {J.}~\bibnamefont {Preskill}},\
  }\href@noop {} {\bibinfo {title} {Quantum algorithms for fermionic quantum
  field theories}} (\bibinfo {year} {2014}),\ \Eprint
  {https://arxiv.org/abs/1404.7115} {arXiv:1404.7115 [hep-th]} \BibitemShut
  {NoStop}%
\bibitem [{\citenamefont {Preskill}(2018)}]{preskill2018simulating}%
  \BibitemOpen
  \bibfield  {author} {\bibinfo {author} {\bibfnamefont {J.}~\bibnamefont
  {Preskill}},\ }\href@noop {} {\bibinfo {title} {Simulating quantum field
  theory with a quantum computer}} (\bibinfo {year} {2018}),\ \Eprint
  {https://arxiv.org/abs/1811.10085} {arXiv:1811.10085 [hep-lat]} \BibitemShut
  {NoStop}%
\bibitem [{\citenamefont {Kurkcuoglu}\ \emph {et~al.}(2021)\citenamefont
  {Kurkcuoglu}, \citenamefont {Alam}, \citenamefont {Job}, \citenamefont {Li},
  \citenamefont {Macridin}, \citenamefont {Perdue},\ and\ \citenamefont
  {Providence}}]{https://doi.org/10.48550/arxiv.2108.13357}%
  \BibitemOpen
  \bibfield  {author} {\bibinfo {author} {\bibfnamefont {D.~M.}\ \bibnamefont
  {Kurkcuoglu}}, \bibinfo {author} {\bibfnamefont {M.~S.}\ \bibnamefont
  {Alam}}, \bibinfo {author} {\bibfnamefont {J.~A.}\ \bibnamefont {Job}},
  \bibinfo {author} {\bibfnamefont {A.~C.~Y.}\ \bibnamefont {Li}}, \bibinfo
  {author} {\bibfnamefont {A.}~\bibnamefont {Macridin}}, \bibinfo {author}
  {\bibfnamefont {G.~N.}\ \bibnamefont {Perdue}},\ and\ \bibinfo {author}
  {\bibfnamefont {S.}~\bibnamefont {Providence}},\ }\href
  {https://doi.org/10.48550/ARXIV.2108.13357} {\bibinfo {title} {Quantum
  simulation of $\phi^4$ theories in qudit systems}} (\bibinfo {year}
  {2021})\BibitemShut {NoStop}%
\bibitem [{\citenamefont {Lamm}\ \emph {et~al.}(2020)\citenamefont {Lamm},
  \citenamefont {Lawrence},\ and\ \citenamefont {Yamauchi}}]{Lamm_2020}%
  \BibitemOpen
  \bibfield  {author} {\bibinfo {author} {\bibfnamefont {H.}~\bibnamefont
  {Lamm}}, \bibinfo {author} {\bibfnamefont {S.}~\bibnamefont {Lawrence}},\
  and\ \bibinfo {author} {\bibfnamefont {Y.}~\bibnamefont {Yamauchi}},\
  }\bibfield  {title} {\bibinfo {title} {Parton physics on a quantum
  computer},\ }\bibfield  {journal} {\bibinfo  {journal} {Physical Review
  Research}\ }\textbf {\bibinfo {volume} {2}},\ \href
  {https://doi.org/10.1103/physrevresearch.2.013272}
  {10.1103/physrevresearch.2.013272} (\bibinfo {year} {2020})\BibitemShut
  {NoStop}%
\bibitem [{\citenamefont {Mazza}\ \emph {et~al.}(2012)\citenamefont {Mazza},
  \citenamefont {Bermudez}, \citenamefont {Goldman}, \citenamefont {Rizzi},
  \citenamefont {Martin-Delgado},\ and\ \citenamefont
  {Lewenstein}}]{Mazza_2012}%
  \BibitemOpen
  \bibfield  {author} {\bibinfo {author} {\bibfnamefont {L.}~\bibnamefont
  {Mazza}}, \bibinfo {author} {\bibfnamefont {A.}~\bibnamefont {Bermudez}},
  \bibinfo {author} {\bibfnamefont {N.}~\bibnamefont {Goldman}}, \bibinfo
  {author} {\bibfnamefont {M.}~\bibnamefont {Rizzi}}, \bibinfo {author}
  {\bibfnamefont {M.~A.}\ \bibnamefont {Martin-Delgado}},\ and\ \bibinfo
  {author} {\bibfnamefont {M.}~\bibnamefont {Lewenstein}},\ }\bibfield  {title}
  {\bibinfo {title} {An optical-lattice-based quantum simulator for
  relativistic field theories and topological insulators},\ }\href
  {https://doi.org/10.1088/1367-2630/14/1/015007} {\bibfield  {journal}
  {\bibinfo  {journal} {New Journal of Physics}\ }\textbf {\bibinfo {volume}
  {14}},\ \bibinfo {pages} {015007} (\bibinfo {year} {2012})}\BibitemShut
  {NoStop}%
\bibitem [{\citenamefont {Yeter-Aydeniz}\ \emph {et~al.}(2019)\citenamefont
  {Yeter-Aydeniz}, \citenamefont {Dumitrescu}, \citenamefont {McCaskey},
  \citenamefont {Bennink}, \citenamefont {Pooser},\ and\ \citenamefont
  {Siopsis}}]{Yeter_Aydeniz_2019}%
  \BibitemOpen
  \bibfield  {author} {\bibinfo {author} {\bibfnamefont {K.}~\bibnamefont
  {Yeter-Aydeniz}}, \bibinfo {author} {\bibfnamefont {E.~F.}\ \bibnamefont
  {Dumitrescu}}, \bibinfo {author} {\bibfnamefont {A.~J.}\ \bibnamefont
  {McCaskey}}, \bibinfo {author} {\bibfnamefont {R.~S.}\ \bibnamefont
  {Bennink}}, \bibinfo {author} {\bibfnamefont {R.~C.}\ \bibnamefont
  {Pooser}},\ and\ \bibinfo {author} {\bibfnamefont {G.}~\bibnamefont
  {Siopsis}},\ }\bibfield  {title} {\bibinfo {title} {Scalar quantum field
  theories as a benchmark for near-term quantum computers},\ }\bibfield
  {journal} {\bibinfo  {journal} {Physical Review A}\ }\textbf {\bibinfo
  {volume} {99}},\ \href {https://doi.org/10.1103/physreva.99.032306}
  {10.1103/physreva.99.032306} (\bibinfo {year} {2019})\BibitemShut {NoStop}%
\bibitem [{\citenamefont {Kokail}\ \emph {et~al.}(2019)\citenamefont {Kokail},
  \citenamefont {Maier}, \citenamefont {van Bijnen}, \citenamefont {Brydges},
  \citenamefont {Joshi}, \citenamefont {Jurcevic}, \citenamefont {Muschik},
  \citenamefont {Silvi}, \citenamefont {Blatt}, \citenamefont {Roos},\ and\
  \citenamefont {et~al.}}]{Kokail_2019}%
  \BibitemOpen
  \bibfield  {author} {\bibinfo {author} {\bibfnamefont {C.}~\bibnamefont
  {Kokail}}, \bibinfo {author} {\bibfnamefont {C.}~\bibnamefont {Maier}},
  \bibinfo {author} {\bibfnamefont {R.}~\bibnamefont {van Bijnen}}, \bibinfo
  {author} {\bibfnamefont {T.}~\bibnamefont {Brydges}}, \bibinfo {author}
  {\bibfnamefont {M.~K.}\ \bibnamefont {Joshi}}, \bibinfo {author}
  {\bibfnamefont {P.}~\bibnamefont {Jurcevic}}, \bibinfo {author}
  {\bibfnamefont {C.~A.}\ \bibnamefont {Muschik}}, \bibinfo {author}
  {\bibfnamefont {P.}~\bibnamefont {Silvi}}, \bibinfo {author} {\bibfnamefont
  {R.}~\bibnamefont {Blatt}}, \bibinfo {author} {\bibfnamefont {C.~F.}\
  \bibnamefont {Roos}},\ and\ \bibinfo {author} {\bibnamefont {et~al.}},\
  }\bibfield  {title} {\bibinfo {title} {Self-verifying variational quantum
  simulation of lattice models},\ }\href
  {https://doi.org/10.1038/s41586-019-1177-4} {\bibfield  {journal} {\bibinfo
  {journal} {Nature}\ }\textbf {\bibinfo {volume} {569}},\ \bibinfo {pages}
  {355–360} (\bibinfo {year} {2019})}\BibitemShut {NoStop}%
\bibitem [{\citenamefont {Kharzeev}\ and\ \citenamefont
  {Kikuchi}(2020)}]{Kharzeev_2020}%
  \BibitemOpen
  \bibfield  {author} {\bibinfo {author} {\bibfnamefont {D.~E.}\ \bibnamefont
  {Kharzeev}}\ and\ \bibinfo {author} {\bibfnamefont {Y.}~\bibnamefont
  {Kikuchi}},\ }\bibfield  {title} {\bibinfo {title} {Real-time chiral dynamics
  from a digital quantum simulation},\ }\bibfield  {journal} {\bibinfo
  {journal} {Physical Review Research}\ }\textbf {\bibinfo {volume} {2}},\
  \href {https://doi.org/10.1103/physrevresearch.2.023342}
  {10.1103/physrevresearch.2.023342} (\bibinfo {year} {2020})\BibitemShut
  {NoStop}%
\bibitem [{\citenamefont {Zohar}\ \emph {et~al.}(2012)\citenamefont {Zohar},
  \citenamefont {Cirac},\ and\ \citenamefont {Reznik}}]{Zohar:2012ay}%
  \BibitemOpen
  \bibfield  {author} {\bibinfo {author} {\bibfnamefont {E.}~\bibnamefont
  {Zohar}}, \bibinfo {author} {\bibfnamefont {J.~I.}\ \bibnamefont {Cirac}},\
  and\ \bibinfo {author} {\bibfnamefont {B.}~\bibnamefont {Reznik}},\
  }\bibfield  {title} {\bibinfo {title} {Simulating compact quantum
  electrodynamics with ultracold atoms: Probing confinement and nonperturbative
  effects},\ }\href {https://doi.org/10.1103/PhysRevLett.109.125302} {\bibfield
   {journal} {\bibinfo  {journal} {Phys. Rev. Lett.}\ }\textbf {\bibinfo
  {volume} {109}},\ \bibinfo {pages} {125302} (\bibinfo {year} {2012})},\
  \Eprint {https://arxiv.org/abs/1204.6574} {arXiv:1204.6574 [quant-ph]}
  \BibitemShut {NoStop}%
%%CITATION = ARXIV:1204.6574;%%
\bibitem [{\citenamefont {Zohar}\ \emph {et~al.}(2013)\citenamefont {Zohar},
  \citenamefont {Cirac},\ and\ \citenamefont {Reznik}}]{Zohar:2012xf}%
  \BibitemOpen
  \bibfield  {author} {\bibinfo {author} {\bibfnamefont {E.}~\bibnamefont
  {Zohar}}, \bibinfo {author} {\bibfnamefont {J.~I.}\ \bibnamefont {Cirac}},\
  and\ \bibinfo {author} {\bibfnamefont {B.}~\bibnamefont {Reznik}},\
  }\bibfield  {title} {\bibinfo {title} {Cold-atom quantum simulator for su(2)
  yang-mills lattice gauge theory},\ }\href
  {https://doi.org/10.1103/PhysRevLett.110.125304} {\bibfield  {journal}
  {\bibinfo  {journal} {Phys. Rev. Lett.}\ }\textbf {\bibinfo {volume} {110}},\
  \bibinfo {pages} {125304} (\bibinfo {year} {2013})},\ \Eprint
  {https://arxiv.org/abs/1211.2241} {arXiv:1211.2241 [quant-ph]} \BibitemShut
  {NoStop}%
%%CITATION = ARXIV:1211.2241;%%
\bibitem [{\citenamefont {Banerjee}\ \emph {et~al.}(2013)\citenamefont
  {Banerjee}, \citenamefont {B\"{o}gli}, \citenamefont {Dalmonte},
  \citenamefont {Rico}, \citenamefont {Stebler}, \citenamefont {Wiese},\ and\
  \citenamefont {Zoller}}]{Banerjee:2012xg}%
  \BibitemOpen
  \bibfield  {author} {\bibinfo {author} {\bibfnamefont {D.}~\bibnamefont
  {Banerjee}}, \bibinfo {author} {\bibfnamefont {M.}~\bibnamefont {B\"{o}gli}},
  \bibinfo {author} {\bibfnamefont {M.}~\bibnamefont {Dalmonte}}, \bibinfo
  {author} {\bibfnamefont {E.}~\bibnamefont {Rico}}, \bibinfo {author}
  {\bibfnamefont {P.}~\bibnamefont {Stebler}}, \bibinfo {author} {\bibfnamefont
  {U.~J.}\ \bibnamefont {Wiese}},\ and\ \bibinfo {author} {\bibfnamefont
  {P.}~\bibnamefont {Zoller}},\ }\bibfield  {title} {\bibinfo {title} {Atomic
  quantum simulation of u(n) and su(n) non-abelian lattice gauge theories},\
  }\href {https://doi.org/10.1103/PhysRevLett.110.125303} {\bibfield  {journal}
  {\bibinfo  {journal} {Phys. Rev. Lett.}\ }\textbf {\bibinfo {volume} {110}},\
  \bibinfo {pages} {125303} (\bibinfo {year} {2013})},\ \Eprint
  {https://arxiv.org/abs/1211.2242} {arXiv:1211.2242 [cond-mat.quant-gas]}
  \BibitemShut {NoStop}%
%%CITATION = ARXIV:1211.2242;%%
\bibitem [{\citenamefont {Banerjee}\ \emph {et~al.}(2012)\citenamefont
  {Banerjee}, \citenamefont {Dalmonte}, \citenamefont {Muller}, \citenamefont
  {Rico}, \citenamefont {Stebler}, \citenamefont {Wiese},\ and\ \citenamefont
  {Zoller}}]{Banerjee:2012pg}%
  \BibitemOpen
  \bibfield  {author} {\bibinfo {author} {\bibfnamefont {D.}~\bibnamefont
  {Banerjee}}, \bibinfo {author} {\bibfnamefont {M.}~\bibnamefont {Dalmonte}},
  \bibinfo {author} {\bibfnamefont {M.}~\bibnamefont {Muller}}, \bibinfo
  {author} {\bibfnamefont {E.}~\bibnamefont {Rico}}, \bibinfo {author}
  {\bibfnamefont {P.}~\bibnamefont {Stebler}}, \bibinfo {author} {\bibfnamefont
  {U.~J.}\ \bibnamefont {Wiese}},\ and\ \bibinfo {author} {\bibfnamefont
  {P.}~\bibnamefont {Zoller}},\ }\bibfield  {title} {\bibinfo {title} {Atomic
  quantum simulation of dynamical gauge fields coupled to fermionic matter:
  From string breaking to evolution after a quench},\ }\href
  {https://doi.org/10.1103/PhysRevLett.109.175302} {\bibfield  {journal}
  {\bibinfo  {journal} {Phys. Rev. Lett.}\ }\textbf {\bibinfo {volume} {109}},\
  \bibinfo {pages} {175302} (\bibinfo {year} {2012})},\ \Eprint
  {https://arxiv.org/abs/1205.6366} {arXiv:1205.6366 [cond-mat.quant-gas]}
  \BibitemShut {NoStop}%
%%CITATION = ARXIV:1205.6366;%%
\bibitem [{\citenamefont {Martinez}\ \emph {et~al.}(2016)\citenamefont
  {Martinez}, \citenamefont {Muschik}, \citenamefont {Schindler}, \citenamefont
  {Nigg}, \citenamefont {Erhard}, \citenamefont {Heyl}, \citenamefont {Hauke},
  \citenamefont {Dalmonte}, \citenamefont {Monz}, \citenamefont {Zoller},\ and\
  \citenamefont {Blatt}}]{Martinez2016}%
  \BibitemOpen
  \bibfield  {author} {\bibinfo {author} {\bibfnamefont {E.~A.}\ \bibnamefont
  {Martinez}}, \bibinfo {author} {\bibfnamefont {C.~A.}\ \bibnamefont
  {Muschik}}, \bibinfo {author} {\bibfnamefont {P.}~\bibnamefont {Schindler}},
  \bibinfo {author} {\bibfnamefont {D.}~\bibnamefont {Nigg}}, \bibinfo {author}
  {\bibfnamefont {A.}~\bibnamefont {Erhard}}, \bibinfo {author} {\bibfnamefont
  {M.}~\bibnamefont {Heyl}}, \bibinfo {author} {\bibfnamefont {P.}~\bibnamefont
  {Hauke}}, \bibinfo {author} {\bibfnamefont {M.}~\bibnamefont {Dalmonte}},
  \bibinfo {author} {\bibfnamefont {T.}~\bibnamefont {Monz}}, \bibinfo {author}
  {\bibfnamefont {P.}~\bibnamefont {Zoller}},\ and\ \bibinfo {author}
  {\bibfnamefont {R.}~\bibnamefont {Blatt}},\ }\bibfield  {title} {\bibinfo
  {title} {Real-time dynamics of lattice gauge theories with a few-qubit
  quantum computer},\ }\href {http://dx.doi.org/10.1038/nature18318} {\bibfield
   {journal} {\bibinfo  {journal} {Nature}\ }\textbf {\bibinfo {volume}
  {534}},\ \bibinfo {pages} {516 EP } (\bibinfo {year} {2016})}\BibitemShut
  {NoStop}%
\bibitem [{\citenamefont {Muschik}\ \emph {et~al.}(2017)\citenamefont
  {Muschik}, \citenamefont {Heyl}, \citenamefont {Martinez}, \citenamefont
  {Monz}, \citenamefont {Schindler}, \citenamefont {Vogell}, \citenamefont
  {Dalmonte}, \citenamefont {Hauke}, \citenamefont {Blatt},\ and\ \citenamefont
  {Zoller}}]{Muschik:2016tws}%
  \BibitemOpen
  \bibfield  {author} {\bibinfo {author} {\bibfnamefont {C.}~\bibnamefont
  {Muschik}}, \bibinfo {author} {\bibfnamefont {M.}~\bibnamefont {Heyl}},
  \bibinfo {author} {\bibfnamefont {E.}~\bibnamefont {Martinez}}, \bibinfo
  {author} {\bibfnamefont {T.}~\bibnamefont {Monz}}, \bibinfo {author}
  {\bibfnamefont {P.}~\bibnamefont {Schindler}}, \bibinfo {author}
  {\bibfnamefont {B.}~\bibnamefont {Vogell}}, \bibinfo {author} {\bibfnamefont
  {M.}~\bibnamefont {Dalmonte}}, \bibinfo {author} {\bibfnamefont
  {P.}~\bibnamefont {Hauke}}, \bibinfo {author} {\bibfnamefont
  {R.}~\bibnamefont {Blatt}},\ and\ \bibinfo {author} {\bibfnamefont
  {P.}~\bibnamefont {Zoller}},\ }\bibfield  {title} {\bibinfo {title} {U(1)
  wilson lattice gauge theories in digital quantum simulators},\ }\href
  {https://doi.org/10.1088/1367-2630/aa89ab} {\bibfield  {journal} {\bibinfo
  {journal} {New J. Phys.}\ }\textbf {\bibinfo {volume} {19}},\ \bibinfo
  {pages} {103020} (\bibinfo {year} {2017})},\ \Eprint
  {https://arxiv.org/abs/1612.08653} {arXiv:1612.08653 [quant-ph]} \BibitemShut
  {NoStop}%
%%CITATION = ARXIV:1612.08653;%%
\bibitem [{\citenamefont {Zohar}\ \emph {et~al.}(2017)\citenamefont {Zohar},
  \citenamefont {Farace}, \citenamefont {Reznik},\ and\ \citenamefont
  {Cirac}}]{Zohar:2016iic}%
  \BibitemOpen
  \bibfield  {author} {\bibinfo {author} {\bibfnamefont {E.}~\bibnamefont
  {Zohar}}, \bibinfo {author} {\bibfnamefont {A.}~\bibnamefont {Farace}},
  \bibinfo {author} {\bibfnamefont {B.}~\bibnamefont {Reznik}},\ and\ \bibinfo
  {author} {\bibfnamefont {J.~I.}\ \bibnamefont {Cirac}},\ }\bibfield  {title}
  {\bibinfo {title} {Digital lattice gauge theories},\ }\href
  {https://doi.org/10.1103/PhysRevA.95.023604} {\bibfield  {journal} {\bibinfo
  {journal} {Phys. Rev.}\ }\textbf {\bibinfo {volume} {A95}},\ \bibinfo {pages}
  {023604} (\bibinfo {year} {2017})},\ \Eprint
  {https://arxiv.org/abs/1607.08121} {arXiv:1607.08121 [quant-ph]} \BibitemShut
  {NoStop}%
%%CITATION = ARXIV:1607.08121;%%
\bibitem [{\citenamefont {Banuls}\ \emph {et~al.}(2017)\citenamefont {Banuls},
  \citenamefont {Cichy}, \citenamefont {Cirac}, \citenamefont {Jansen},\ and\
  \citenamefont {Kuhn}}]{Banuls:2017ena}%
  \BibitemOpen
  \bibfield  {author} {\bibinfo {author} {\bibfnamefont {M.~C.}\ \bibnamefont
  {Banuls}}, \bibinfo {author} {\bibfnamefont {K.}~\bibnamefont {Cichy}},
  \bibinfo {author} {\bibfnamefont {J.~I.}\ \bibnamefont {Cirac}}, \bibinfo
  {author} {\bibfnamefont {K.}~\bibnamefont {Jansen}},\ and\ \bibinfo {author}
  {\bibfnamefont {S.}~\bibnamefont {Kuhn}},\ }\bibfield  {title} {\bibinfo
  {title} {Efficient basis formulation for 1+1 dimensional su(2) lattice gauge
  theory: Spectral calculations with matrix product states},\ }\href
  {https://doi.org/10.1103/PhysRevX.7.041046} {\bibfield  {journal} {\bibinfo
  {journal} {Phys. Rev.}\ }\textbf {\bibinfo {volume} {X7}},\ \bibinfo {pages}
  {041046} (\bibinfo {year} {2017})},\ \Eprint
  {https://arxiv.org/abs/1707.06434} {arXiv:1707.06434 [hep-lat]} \BibitemShut
  {NoStop}%
%%CITATION = ARXIV:1707.06434;%%
\bibitem [{\citenamefont {Kaplan}\ and\ \citenamefont
  {Stryker}(2020)}]{Kaplan:2018vnj}%
  \BibitemOpen
  \bibfield  {author} {\bibinfo {author} {\bibfnamefont {D.~B.}\ \bibnamefont
  {Kaplan}}\ and\ \bibinfo {author} {\bibfnamefont {J.~R.}\ \bibnamefont
  {Stryker}},\ }\bibfield  {title} {\bibinfo {title} {Gauss's law, duality, and
  the hamiltonian formulation of u(1) lattice gauge theory},\ }\href@noop {}
  {\bibfield  {journal} {\bibinfo  {journal} {Phys. Rev. D}\ }\textbf {\bibinfo
  {volume} {102}},\ \bibinfo {pages} {094515} (\bibinfo {year} {2020})},\
  \Eprint {https://arxiv.org/abs/1806.08797} {arXiv:1806.08797 [hep-lat]}
  \BibitemShut {NoStop}%
%%CITATION = ARXIV:1806.08797;%%
\bibitem [{\citenamefont {Zache}\ \emph {et~al.}(2018)\citenamefont {Zache},
  \citenamefont {Hebenstreit}, \citenamefont {Jendrzejewski}, \citenamefont
  {Oberthaler}, \citenamefont {Berges},\ and\ \citenamefont
  {Hauke}}]{Zache:2018jbt}%
  \BibitemOpen
  \bibfield  {author} {\bibinfo {author} {\bibfnamefont {T.~V.}\ \bibnamefont
  {Zache}}, \bibinfo {author} {\bibfnamefont {F.}~\bibnamefont {Hebenstreit}},
  \bibinfo {author} {\bibfnamefont {F.}~\bibnamefont {Jendrzejewski}}, \bibinfo
  {author} {\bibfnamefont {M.~K.}\ \bibnamefont {Oberthaler}}, \bibinfo
  {author} {\bibfnamefont {J.}~\bibnamefont {Berges}},\ and\ \bibinfo {author}
  {\bibfnamefont {P.}~\bibnamefont {Hauke}},\ }\bibfield  {title} {\bibinfo
  {title} {Quantum simulation of lattice gauge theories using wilson
  fermions},\ }\href {https://doi.org/10.1088/2058-9565/aac33b} {\bibfield
  {journal} {\bibinfo  {journal} {Sci. Technol.}\ }\textbf {\bibinfo {volume}
  {3}},\ \bibinfo {pages} {034010} (\bibinfo {year} {2018})},\ \Eprint
  {https://arxiv.org/abs/1802.06704} {arXiv:1802.06704 [cond-mat.quant-gas]}
  \BibitemShut {NoStop}%
%%CITATION = ARXIV:1802.06704;%%
\bibitem [{\citenamefont {Stryker}(2019)}]{Stryker:2018efp}%
  \BibitemOpen
  \bibfield  {author} {\bibinfo {author} {\bibfnamefont {J.~R.}\ \bibnamefont
  {Stryker}},\ }\bibfield  {title} {\bibinfo {title} {Oracles for gauss's law
  on digital quantum computers},\ }\href
  {https://doi.org/10.1103/PhysRevA.99.042301} {\bibfield  {journal} {\bibinfo
  {journal} {Phys. Rev.}\ }\textbf {\bibinfo {volume} {A99}},\ \bibinfo {pages}
  {042301} (\bibinfo {year} {2019})},\ \Eprint
  {https://arxiv.org/abs/1812.01617} {arXiv:1812.01617 [quant-ph]} \BibitemShut
  {NoStop}%
%%CITATION = ARXIV:1812.01617;%%
\bibitem [{\citenamefont {Raychowdhury}(2019)}]{Raychowdhury:2018tfj}%
  \BibitemOpen
  \bibfield  {author} {\bibinfo {author} {\bibfnamefont {I.}~\bibnamefont
  {Raychowdhury}},\ }\bibfield  {title} {\bibinfo {title} {Low energy spectrum
  of su(2) lattice gauge theory},\ }\href
  {https://doi.org/10.1140/epjc/s10052-019-6753-0} {\bibfield  {journal}
  {\bibinfo  {journal} {Eur. Phys. J.}\ }\textbf {\bibinfo {volume} {C79}},\
  \bibinfo {pages} {235} (\bibinfo {year} {2019})},\ \Eprint
  {https://arxiv.org/abs/1804.01304} {arXiv:1804.01304 [hep-lat]} \BibitemShut
  {NoStop}%
%%CITATION = ARXIV:1804.01304;%%
\bibitem [{\citenamefont {Davoudi}\ \emph {et~al.}(2020)\citenamefont
  {Davoudi}, \citenamefont {Hafezi}, \citenamefont {Monroe}, \citenamefont
  {Pagano}, \citenamefont {Seif},\ and\ \citenamefont
  {Shaw}}]{Davoudi:2019bhy}%
  \BibitemOpen
  \bibfield  {author} {\bibinfo {author} {\bibfnamefont {Z.}~\bibnamefont
  {Davoudi}}, \bibinfo {author} {\bibfnamefont {M.}~\bibnamefont {Hafezi}},
  \bibinfo {author} {\bibfnamefont {C.}~\bibnamefont {Monroe}}, \bibinfo
  {author} {\bibfnamefont {G.}~\bibnamefont {Pagano}}, \bibinfo {author}
  {\bibfnamefont {A.}~\bibnamefont {Seif}},\ and\ \bibinfo {author}
  {\bibfnamefont {A.}~\bibnamefont {Shaw}},\ }\bibfield  {title} {\bibinfo
  {title} {Towards analog quantum simulations of lattice gauge theories with
  trapped ions},\ }\href@noop {} {\bibfield  {journal} {\bibinfo  {journal}
  {Phys. Rev. Research}\ }\textbf {\bibinfo {volume} {2}} (\bibinfo {year}
  {2020})},\ \Eprint {https://arxiv.org/abs/1908.03210} {arXiv:1908.03210
  [quant-ph]} \BibitemShut {NoStop}%
%%CITATION = ARXIV:1908.03210;%%
\bibitem [{\citenamefont {Haase}\ \emph {et~al.}(2021)\citenamefont {Haase},
  \citenamefont {Dellantonio}, \citenamefont {Celi}, \citenamefont {Paulson},
  \citenamefont {Kan}, \citenamefont {Jansen},\ and\ \citenamefont
  {Muschik}}]{Haase:2020kaj}%
  \BibitemOpen
  \bibfield  {author} {\bibinfo {author} {\bibfnamefont {J.~F.}\ \bibnamefont
  {Haase}}, \bibinfo {author} {\bibfnamefont {L.}~\bibnamefont {Dellantonio}},
  \bibinfo {author} {\bibfnamefont {A.}~\bibnamefont {Celi}}, \bibinfo {author}
  {\bibfnamefont {D.}~\bibnamefont {Paulson}}, \bibinfo {author} {\bibfnamefont
  {A.}~\bibnamefont {Kan}}, \bibinfo {author} {\bibfnamefont {K.}~\bibnamefont
  {Jansen}},\ and\ \bibinfo {author} {\bibfnamefont {C.~A.}\ \bibnamefont
  {Muschik}},\ }\bibfield  {title} {\bibinfo {title} {A resource efficient
  approach for quantum and classical simulations of gauge theories in particle
  physics},\ }\href@noop {} {\bibfield  {journal} {\bibinfo  {journal}
  {Quantum}\ }\textbf {\bibinfo {volume} {5}} (\bibinfo {year} {2021})},\
  \Eprint {https://arxiv.org/abs/2006.14160} {arXiv:2006.14160 [quant-ph]}
  \BibitemShut {NoStop}%
\bibitem [{\citenamefont {Paulson}\ \emph {et~al.}(2021)\citenamefont
  {Paulson}, \citenamefont {Dellantonio}, \citenamefont {Haase}, \citenamefont
  {Celi}, \citenamefont {Kan}, \citenamefont {Jena}, \citenamefont {Kokail},
  \citenamefont {van Bijnen}, \citenamefont {Jansen}, \citenamefont {Zoller},\
  and\ \citenamefont {et~al.}}]{Paulson_2021}%
  \BibitemOpen
  \bibfield  {author} {\bibinfo {author} {\bibfnamefont {D.}~\bibnamefont
  {Paulson}}, \bibinfo {author} {\bibfnamefont {L.}~\bibnamefont
  {Dellantonio}}, \bibinfo {author} {\bibfnamefont {J.~F.}\ \bibnamefont
  {Haase}}, \bibinfo {author} {\bibfnamefont {A.}~\bibnamefont {Celi}},
  \bibinfo {author} {\bibfnamefont {A.}~\bibnamefont {Kan}}, \bibinfo {author}
  {\bibfnamefont {A.}~\bibnamefont {Jena}}, \bibinfo {author} {\bibfnamefont
  {C.}~\bibnamefont {Kokail}}, \bibinfo {author} {\bibfnamefont
  {R.}~\bibnamefont {van Bijnen}}, \bibinfo {author} {\bibfnamefont
  {K.}~\bibnamefont {Jansen}}, \bibinfo {author} {\bibfnamefont
  {P.}~\bibnamefont {Zoller}},\ and\ \bibinfo {author} {\bibnamefont
  {et~al.}},\ }\bibfield  {title} {\bibinfo {title} {Simulating 2d effects in
  lattice gauge theories on a quantum computer},\ }\bibfield  {journal}
  {\bibinfo  {journal} {PRX Quantum}\ }\textbf {\bibinfo {volume} {2}},\ \href
  {https://doi.org/10.1103/prxquantum.2.030334} {10.1103/prxquantum.2.030334}
  (\bibinfo {year} {2021})\BibitemShut {NoStop}%
\bibitem [{\citenamefont {Davoudi}\ \emph
  {et~al.}(2021{\natexlab{a}})\citenamefont {Davoudi}, \citenamefont
  {Raychowdhury},\ and\ \citenamefont {Shaw}}]{Davoudi:2020yln}%
  \BibitemOpen
  \bibfield  {author} {\bibinfo {author} {\bibfnamefont {Z.}~\bibnamefont
  {Davoudi}}, \bibinfo {author} {\bibfnamefont {I.}~\bibnamefont
  {Raychowdhury}},\ and\ \bibinfo {author} {\bibfnamefont {A.}~\bibnamefont
  {Shaw}},\ }\bibfield  {title} {\bibinfo {title} {Search for efficient
  formulations for hamiltonian simulation of non-abelian lattice gauge
  theories},\ }\href@noop {} {\bibfield  {journal} {\bibinfo  {journal} {Phys.
  Rev. D}\ }\textbf {\bibinfo {volume} {104}} (\bibinfo {year}
  {2021}{\natexlab{a}})},\ \Eprint {https://arxiv.org/abs/2009.11802}
  {arXiv:2009.11802 [hep-lat]} \BibitemShut {NoStop}%
\bibitem [{\citenamefont {Barata}\ \emph {et~al.}(2021)\citenamefont {Barata},
  \citenamefont {Mueller}, \citenamefont {Tarasov},\ and\ \citenamefont
  {Venugopalan}}]{PhysRevA.103.042410}%
  \BibitemOpen
  \bibfield  {author} {\bibinfo {author} {\bibfnamefont {J.~a.}\ \bibnamefont
  {Barata}}, \bibinfo {author} {\bibfnamefont {N.}~\bibnamefont {Mueller}},
  \bibinfo {author} {\bibfnamefont {A.}~\bibnamefont {Tarasov}},\ and\ \bibinfo
  {author} {\bibfnamefont {R.}~\bibnamefont {Venugopalan}},\ }\bibfield
  {title} {\bibinfo {title} {Single-particle digitization strategy for quantum
  computation of a ${\ensuremath{\phi}}^{4}$ scalar field theory},\ }\href
  {https://doi.org/10.1103/PhysRevA.103.042410} {\bibfield  {journal} {\bibinfo
   {journal} {Phys. Rev. A}\ }\textbf {\bibinfo {volume} {103}},\ \bibinfo
  {pages} {042410} (\bibinfo {year} {2021})}\BibitemShut {NoStop}%
\bibitem [{\citenamefont {Buser}\ \emph {et~al.}(2020)\citenamefont {Buser},
  \citenamefont {Gharibyan}, \citenamefont {Hanada}, \citenamefont {Honda},\
  and\ \citenamefont {Liu}}]{Buser:2020cvn}%
  \BibitemOpen
  \bibfield  {author} {\bibinfo {author} {\bibfnamefont {A.}~\bibnamefont
  {Buser}}, \bibinfo {author} {\bibfnamefont {H.}~\bibnamefont {Gharibyan}},
  \bibinfo {author} {\bibfnamefont {M.}~\bibnamefont {Hanada}}, \bibinfo
  {author} {\bibfnamefont {M.}~\bibnamefont {Honda}},\ and\ \bibinfo {author}
  {\bibfnamefont {J.}~\bibnamefont {Liu}},\ }\href@noop {} {\bibinfo {title}
  {Quantum simulation of gauge theory via orbifold lattice}} (\bibinfo {year}
  {2020}),\ \Eprint {https://arxiv.org/abs/2011.06576} {arXiv:2011.06576
  [hep-th]} \BibitemShut {NoStop}%
\bibitem [{\citenamefont {Raychowdhury}\ and\ \citenamefont
  {Stryker}(2018)}]{Raychowdhury:2018osk}%
  \BibitemOpen
  \bibfield  {author} {\bibinfo {author} {\bibfnamefont {I.}~\bibnamefont
  {Raychowdhury}}\ and\ \bibinfo {author} {\bibfnamefont {J.~R.}\ \bibnamefont
  {Stryker}},\ }\href@noop {} {\bibinfo {title} {Tailoring non-abelian lattice
  gauge theory for quantum simulation}} (\bibinfo {year} {2018}),\ \Eprint
  {https://arxiv.org/abs/1812.07554} {arXiv:1812.07554 [hep-lat]} \BibitemShut
  {NoStop}%
%%CITATION = ARXIV:1812.07554;%%
\bibitem [{\citenamefont {Raychowdhury}\ and\ \citenamefont
  {Stryker}(2020)}]{Raychowdhury:2019iki}%
  \BibitemOpen
  \bibfield  {author} {\bibinfo {author} {\bibfnamefont {I.}~\bibnamefont
  {Raychowdhury}}\ and\ \bibinfo {author} {\bibfnamefont {J.~R.}\ \bibnamefont
  {Stryker}},\ }\bibfield  {title} {\bibinfo {title} {Loop, string, and hadron
  dynamics in su(2) hamiltonian lattice gauge theories},\ }\href
  {https://doi.org/10.1103/PhysRevD.101.114502} {\bibfield  {journal} {\bibinfo
   {journal} {Phys. Rev. D}\ }\textbf {\bibinfo {volume} {101}},\ \bibinfo
  {pages} {114502} (\bibinfo {year} {2020})},\ \Eprint
  {https://arxiv.org/abs/1912.06133} {arXiv:1912.06133 [hep-lat]} \BibitemShut
  {NoStop}%
\bibitem [{\citenamefont {Kreshchuk}\ \emph {et~al.}(2021)\citenamefont
  {Kreshchuk}, \citenamefont {Jia}, \citenamefont {Kirby}, \citenamefont
  {Goldstein}, \citenamefont {Vary},\ and\ \citenamefont
  {Love}}]{Kreshchuk:2020kcz}%
  \BibitemOpen
  \bibfield  {author} {\bibinfo {author} {\bibfnamefont {M.}~\bibnamefont
  {Kreshchuk}}, \bibinfo {author} {\bibfnamefont {S.}~\bibnamefont {Jia}},
  \bibinfo {author} {\bibfnamefont {W.~M.}\ \bibnamefont {Kirby}}, \bibinfo
  {author} {\bibfnamefont {G.}~\bibnamefont {Goldstein}}, \bibinfo {author}
  {\bibfnamefont {J.~P.}\ \bibnamefont {Vary}},\ and\ \bibinfo {author}
  {\bibfnamefont {P.~J.}\ \bibnamefont {Love}},\ }\bibfield  {title} {\bibinfo
  {title} {Light-front field theory on current quantum computers},\ }\href
  {https://doi.org/10.3390/e23050597} {\bibfield  {journal} {\bibinfo
  {journal} {Entropy}\ }\textbf {\bibinfo {volume} {23}},\ \bibinfo {pages}
  {597} (\bibinfo {year} {2021})}\BibitemShut {NoStop}%
\bibitem [{\citenamefont {Tagliacozzo}\ \emph {et~al.}(2013)\citenamefont
  {Tagliacozzo}, \citenamefont {Celi}, \citenamefont {Orland},\ and\
  \citenamefont {Lewenstein}}]{Tagliacozzo:2012df}%
  \BibitemOpen
  \bibfield  {author} {\bibinfo {author} {\bibfnamefont {L.}~\bibnamefont
  {Tagliacozzo}}, \bibinfo {author} {\bibfnamefont {A.}~\bibnamefont {Celi}},
  \bibinfo {author} {\bibfnamefont {P.}~\bibnamefont {Orland}},\ and\ \bibinfo
  {author} {\bibfnamefont {M.}~\bibnamefont {Lewenstein}},\ }\bibfield  {title}
  {\bibinfo {title} {Simulations of non-abelian gauge theories with optical
  lattices},\ }\href {https://doi.org/10.1038/ncomms3615} {\bibfield  {journal}
  {\bibinfo  {journal} {Nature Commun.}\ }\textbf {\bibinfo {volume} {4}},\
  \bibinfo {pages} {2615} (\bibinfo {year} {2013})},\ \Eprint
  {https://arxiv.org/abs/1211.2704} {arXiv:1211.2704 [cond-mat.quant-gas]}
  \BibitemShut {NoStop}%
%%CITATION = ARXIV:1211.2704;%%
\bibitem [{\citenamefont {Ji}\ \emph {et~al.}(2020)\citenamefont {Ji},
  \citenamefont {Lamm},\ and\ \citenamefont {Zhu}}]{Ji:2020kjk}%
  \BibitemOpen
  \bibfield  {author} {\bibinfo {author} {\bibfnamefont {Y.}~\bibnamefont
  {Ji}}, \bibinfo {author} {\bibfnamefont {H.}~\bibnamefont {Lamm}},\ and\
  \bibinfo {author} {\bibfnamefont {S.}~\bibnamefont {Zhu}} (\bibinfo
  {collaboration} {NuQS}),\ }\bibfield  {title} {\bibinfo {title} {Gluon field
  digitization via group space decimation for quantum computers},\ }\href
  {https://doi.org/10.1103/PhysRevD.102.114513} {\bibfield  {journal} {\bibinfo
   {journal} {Phys. Rev. D}\ }\textbf {\bibinfo {volume} {102}},\ \bibinfo
  {pages} {114513} (\bibinfo {year} {2020})},\ \Eprint
  {https://arxiv.org/abs/2005.14221} {arXiv:2005.14221 [hep-lat]} \BibitemShut
  {NoStop}%
\bibitem [{\citenamefont {Bender}\ \emph {et~al.}(2018)\citenamefont {Bender},
  \citenamefont {Zohar}, \citenamefont {Farace},\ and\ \citenamefont
  {Cirac}}]{Bender_2018}%
  \BibitemOpen
  \bibfield  {author} {\bibinfo {author} {\bibfnamefont {J.}~\bibnamefont
  {Bender}}, \bibinfo {author} {\bibfnamefont {E.}~\bibnamefont {Zohar}},
  \bibinfo {author} {\bibfnamefont {A.}~\bibnamefont {Farace}},\ and\ \bibinfo
  {author} {\bibfnamefont {J.~I.}\ \bibnamefont {Cirac}},\ }\bibfield  {title}
  {\bibinfo {title} {Digital quantum simulation of lattice gauge theories in
  three spatial dimensions},\ }\href {https://doi.org/10.1088/1367-2630/aadb71}
  {\bibfield  {journal} {\bibinfo  {journal} {New Journal of Physics}\ }\textbf
  {\bibinfo {volume} {20}},\ \bibinfo {pages} {093001} (\bibinfo {year}
  {2018})}\BibitemShut {NoStop}%
\bibitem [{\citenamefont {Klco}\ and\ \citenamefont
  {Savage}(2019)}]{Klco:2018zqz}%
  \BibitemOpen
  \bibfield  {author} {\bibinfo {author} {\bibfnamefont {N.}~\bibnamefont
  {Klco}}\ and\ \bibinfo {author} {\bibfnamefont {M.~J.}\ \bibnamefont
  {Savage}},\ }\bibfield  {title} {\bibinfo {title} {Digitization of scalar
  fields for quantum computing},\ }\href
  {https://doi.org/10.1103/PhysRevA.99.052335} {\bibfield  {journal} {\bibinfo
  {journal} {Phys. Rev.}\ }\textbf {\bibinfo {volume} {A99}},\ \bibinfo {pages}
  {052335} (\bibinfo {year} {2019})},\ \Eprint
  {https://arxiv.org/abs/1808.10378} {arXiv:1808.10378 [quant-ph]} \BibitemShut
  {NoStop}%
%%CITATION = ARXIV:1808.10378;%%
\bibitem [{\citenamefont {Somma}(2016)}]{10.5555/3179430.3179434}%
  \BibitemOpen
  \bibfield  {author} {\bibinfo {author} {\bibfnamefont {R.~D.}\ \bibnamefont
  {Somma}},\ }\bibfield  {title} {\bibinfo {title} {Quantum simulations of one
  dimensional quantum systems},\ }\href@noop {} {\bibfield  {journal} {\bibinfo
   {journal} {Quantum Info. Comput.}\ }\textbf {\bibinfo {volume} {16}},\
  \bibinfo {pages} {1125–1168} (\bibinfo {year} {2016})}\BibitemShut
  {NoStop}%
\bibitem [{\citenamefont {Macridin}\ \emph {et~al.}(2018)\citenamefont
  {Macridin}, \citenamefont {Spentzouris}, \citenamefont {Amundson},\ and\
  \citenamefont {Harnik}}]{PhysRevA.98.042312}%
  \BibitemOpen
  \bibfield  {author} {\bibinfo {author} {\bibfnamefont {A.}~\bibnamefont
  {Macridin}}, \bibinfo {author} {\bibfnamefont {P.}~\bibnamefont
  {Spentzouris}}, \bibinfo {author} {\bibfnamefont {J.}~\bibnamefont
  {Amundson}},\ and\ \bibinfo {author} {\bibfnamefont {R.}~\bibnamefont
  {Harnik}},\ }\bibfield  {title} {\bibinfo {title} {Digital quantum
  computation of fermion-boson interacting systems},\ }\href
  {https://doi.org/10.1103/PhysRevA.98.042312} {\bibfield  {journal} {\bibinfo
  {journal} {Phys. Rev. A}\ }\textbf {\bibinfo {volume} {98}},\ \bibinfo
  {pages} {042312} (\bibinfo {year} {2018})}\BibitemShut {NoStop}%
\bibitem [{\citenamefont {Gharibyan}\ \emph {et~al.}(2021)\citenamefont
  {Gharibyan}, \citenamefont {Hanada}, \citenamefont {Honda},\ and\
  \citenamefont {Liu}}]{Gharibyan_2021}%
  \BibitemOpen
  \bibfield  {author} {\bibinfo {author} {\bibfnamefont {H.}~\bibnamefont
  {Gharibyan}}, \bibinfo {author} {\bibfnamefont {M.}~\bibnamefont {Hanada}},
  \bibinfo {author} {\bibfnamefont {M.}~\bibnamefont {Honda}},\ and\ \bibinfo
  {author} {\bibfnamefont {J.}~\bibnamefont {Liu}},\ }\bibfield  {title}
  {\bibinfo {title} {Toward simulating superstring/m-theory on a quantum
  computer},\ }\bibfield  {journal} {\bibinfo  {journal} {Journal of High
  Energy Physics}\ }\textbf {\bibinfo {volume} {2021}},\ \href
  {https://doi.org/10.1007/jhep07(2021)140} {10.1007/jhep07(2021)140} (\bibinfo
  {year} {2021})\BibitemShut {NoStop}%
\bibitem [{\citenamefont {Ciavarella}\ \emph {et~al.}(2021)\citenamefont
  {Ciavarella}, \citenamefont {Klco},\ and\ \citenamefont
  {Savage}}]{Ciavarella_2021}%
  \BibitemOpen
  \bibfield  {author} {\bibinfo {author} {\bibfnamefont {A.}~\bibnamefont
  {Ciavarella}}, \bibinfo {author} {\bibfnamefont {N.}~\bibnamefont {Klco}},\
  and\ \bibinfo {author} {\bibfnamefont {M.~J.}\ \bibnamefont {Savage}},\
  }\bibfield  {title} {\bibinfo {title} {Trailhead for quantum simulation of
  su(3) yang-mills lattice gauge theory in the local multiplet basis},\
  }\bibfield  {journal} {\bibinfo  {journal} {Physical Review D}\ }\textbf
  {\bibinfo {volume} {103}},\ \href
  {https://doi.org/10.1103/physrevd.103.094501} {10.1103/physrevd.103.094501}
  (\bibinfo {year} {2021})\BibitemShut {NoStop}%
\bibitem [{\citenamefont {Klco}\ \emph {et~al.}(2021)\citenamefont {Klco},
  \citenamefont {Roggero},\ and\ \citenamefont {Savage}}]{klco2021standard}%
  \BibitemOpen
  \bibfield  {author} {\bibinfo {author} {\bibfnamefont {N.}~\bibnamefont
  {Klco}}, \bibinfo {author} {\bibfnamefont {A.}~\bibnamefont {Roggero}},\ and\
  \bibinfo {author} {\bibfnamefont {M.~J.}\ \bibnamefont {Savage}},\
  }\href@noop {} {\bibinfo {title} {Standard model physics and the digital
  quantum revolution: Thoughts about the interface}} (\bibinfo {year} {2021}),\
  \Eprint {https://arxiv.org/abs/2107.04769} {arXiv:2107.04769 [quant-ph]}
  \BibitemShut {NoStop}%
\bibitem [{\citenamefont {de~Jong}\ \emph {et~al.}(2021)\citenamefont
  {de~Jong}, \citenamefont {Lee}, \citenamefont {Mulligan}, \citenamefont
  {Płoskoń}, \citenamefont {Ringer},\ and\ \citenamefont
  {Yao}}]{dejong2021quantum}%
  \BibitemOpen
  \bibfield  {author} {\bibinfo {author} {\bibfnamefont {W.~A.}\ \bibnamefont
  {de~Jong}}, \bibinfo {author} {\bibfnamefont {K.}~\bibnamefont {Lee}},
  \bibinfo {author} {\bibfnamefont {J.}~\bibnamefont {Mulligan}}, \bibinfo
  {author} {\bibfnamefont {M.}~\bibnamefont {Płoskoń}}, \bibinfo {author}
  {\bibfnamefont {F.}~\bibnamefont {Ringer}},\ and\ \bibinfo {author}
  {\bibfnamefont {X.}~\bibnamefont {Yao}},\ }\href@noop {} {\bibinfo {title}
  {Quantum simulation of non-equilibrium dynamics and thermalization in the
  schwinger model}} (\bibinfo {year} {2021}),\ \Eprint
  {https://arxiv.org/abs/2106.08394} {arXiv:2106.08394 [quant-ph]} \BibitemShut
  {NoStop}%
\bibitem [{\citenamefont {Klco}\ and\ \citenamefont
  {Savage}(2020)}]{icons2020}%
  \BibitemOpen
  \bibfield  {author} {\bibinfo {author} {\bibfnamefont {N.}~\bibnamefont
  {Klco}}\ and\ \bibinfo {author} {\bibfnamefont {M.~J.}\ \bibnamefont
  {Savage}},\ }\bibfield  {title} {\bibinfo {title} {Minimally entangled state
  preparation of localized wave functions on quantum computers},\ }\bibfield
  {journal} {\bibinfo  {journal} {Physical Review A}\ }\textbf {\bibinfo
  {volume} {102}},\ \href {https://doi.org/10.1103/physreva.102.012612}
  {10.1103/physreva.102.012612} (\bibinfo {year} {2020})\BibitemShut {NoStop}%
\bibitem [{\citenamefont {Meurice}(2021)}]{meurice2021theoretical}%
  \BibitemOpen
  \bibfield  {author} {\bibinfo {author} {\bibfnamefont {Y.}~\bibnamefont
  {Meurice}},\ }\href@noop {} {\bibinfo {title} {Theoretical methods to design
  and test quantum simulators for the compact abelian higgs model}} (\bibinfo
  {year} {2021}),\ \Eprint {https://arxiv.org/abs/2107.11366} {arXiv:2107.11366
  [quant-ph]} \BibitemShut {NoStop}%
\bibitem [{\citenamefont {Schlittgen}\ and\ \citenamefont
  {Wiese}(2001)}]{PhysRevD.63.085007}%
  \BibitemOpen
  \bibfield  {author} {\bibinfo {author} {\bibfnamefont {B.}~\bibnamefont
  {Schlittgen}}\ and\ \bibinfo {author} {\bibfnamefont {U.-J.}\ \bibnamefont
  {Wiese}},\ }\bibfield  {title} {\bibinfo {title} {Low-energy effective
  theories of quantum spin and quantum link models},\ }\href
  {https://doi.org/10.1103/PhysRevD.63.085007} {\bibfield  {journal} {\bibinfo
  {journal} {Phys. Rev. D}\ }\textbf {\bibinfo {volume} {63}},\ \bibinfo
  {pages} {085007} (\bibinfo {year} {2001})}\BibitemShut {NoStop}%
\bibitem [{\citenamefont {Brower}\ \emph {et~al.}(1999)\citenamefont {Brower},
  \citenamefont {Chandrasekharan},\ and\ \citenamefont
  {Wiese}}]{PhysRevD.60.094502}%
  \BibitemOpen
  \bibfield  {author} {\bibinfo {author} {\bibfnamefont {R.}~\bibnamefont
  {Brower}}, \bibinfo {author} {\bibfnamefont {S.}~\bibnamefont
  {Chandrasekharan}},\ and\ \bibinfo {author} {\bibfnamefont {U.-J.}\
  \bibnamefont {Wiese}},\ }\bibfield  {title} {\bibinfo {title} {Qcd as a
  quantum link model},\ }\href {https://doi.org/10.1103/PhysRevD.60.094502}
  {\bibfield  {journal} {\bibinfo  {journal} {Phys. Rev. D}\ }\textbf {\bibinfo
  {volume} {60}},\ \bibinfo {pages} {094502} (\bibinfo {year}
  {1999})}\BibitemShut {NoStop}%
\bibitem [{\citenamefont {Alam}\ \emph {et~al.}(2022)\citenamefont {Alam},
  \citenamefont {Hadfield}, \citenamefont {Lamm},\ and\ \citenamefont
  {and}}]{Alam_2022}%
  \BibitemOpen
  \bibfield  {author} {\bibinfo {author} {\bibfnamefont {M.~S.}\ \bibnamefont
  {Alam}}, \bibinfo {author} {\bibfnamefont {S.}~\bibnamefont {Hadfield}},
  \bibinfo {author} {\bibfnamefont {H.}~\bibnamefont {Lamm}},\ and\ \bibinfo
  {author} {\bibfnamefont {A.~C.~L.}\ \bibnamefont {and}},\ }\bibfield  {title}
  {\bibinfo {title} {Primitive quantum gates for dihedral gauge theories},\
  }\bibfield  {journal} {\bibinfo  {journal} {Physical Review D}\ }\textbf
  {\bibinfo {volume} {105}},\ \href
  {https://doi.org/10.1103/physrevd.105.114501} {10.1103/physrevd.105.114501}
  (\bibinfo {year} {2022})\BibitemShut {NoStop}%
\bibitem [{\citenamefont {Alexandru}\ \emph {et~al.}(2022)\citenamefont
  {Alexandru}, \citenamefont {Bedaque}, \citenamefont {Brett},\ and\
  \citenamefont {Lamm}}]{PhysRevD.105.114508}%
  \BibitemOpen
  \bibfield  {author} {\bibinfo {author} {\bibfnamefont {A.}~\bibnamefont
  {Alexandru}}, \bibinfo {author} {\bibfnamefont {P.~F.}\ \bibnamefont
  {Bedaque}}, \bibinfo {author} {\bibfnamefont {R.}~\bibnamefont {Brett}},\
  and\ \bibinfo {author} {\bibfnamefont {H.}~\bibnamefont {Lamm}},\ }\bibfield
  {title} {\bibinfo {title} {Spectrum of digitized qcd: Glueballs in a
  $s(1080)$ gauge theory},\ }\href
  {https://doi.org/10.1103/PhysRevD.105.114508} {\bibfield  {journal} {\bibinfo
   {journal} {Phys. Rev. D}\ }\textbf {\bibinfo {volume} {105}},\ \bibinfo
  {pages} {114508} (\bibinfo {year} {2022})}\BibitemShut {NoStop}%
\bibitem [{\citenamefont {Zohar}(2021)}]{zohar2021quantum}%
  \BibitemOpen
  \bibfield  {author} {\bibinfo {author} {\bibfnamefont {E.}~\bibnamefont
  {Zohar}},\ }\href@noop {} {\bibinfo {title} {Quantum simulation of lattice
  gauge theories in more than one space dimension -- requirements, challenges,
  methods}} (\bibinfo {year} {2021}),\ \Eprint
  {https://arxiv.org/abs/2106.04609} {arXiv:2106.04609 [quant-ph]} \BibitemShut
  {NoStop}%
\bibitem [{\citenamefont {Armon}\ \emph {et~al.}(2021)\citenamefont {Armon},
  \citenamefont {Ashkenazi}, \citenamefont {García-Moreno}, \citenamefont
  {González-Tudela},\ and\ \citenamefont {Zohar}}]{armon2021photonmediated}%
  \BibitemOpen
  \bibfield  {author} {\bibinfo {author} {\bibfnamefont {T.}~\bibnamefont
  {Armon}}, \bibinfo {author} {\bibfnamefont {S.}~\bibnamefont {Ashkenazi}},
  \bibinfo {author} {\bibfnamefont {G.}~\bibnamefont {García-Moreno}},
  \bibinfo {author} {\bibfnamefont {A.}~\bibnamefont {González-Tudela}},\ and\
  \bibinfo {author} {\bibfnamefont {E.}~\bibnamefont {Zohar}},\ }\href@noop {}
  {\bibinfo {title} {Photon-mediated stroboscopic quantum simulation of a
  $\mathbb{Z}_{2}$ lattice gauge theory}} (\bibinfo {year} {2021}),\ \Eprint
  {https://arxiv.org/abs/2107.13024} {arXiv:2107.13024 [quant-ph]} \BibitemShut
  {NoStop}%
\bibitem [{\citenamefont {Klco}\ and\ \citenamefont
  {Savage}(2021)}]{klco2021hierarchical}%
  \BibitemOpen
  \bibfield  {author} {\bibinfo {author} {\bibfnamefont {N.}~\bibnamefont
  {Klco}}\ and\ \bibinfo {author} {\bibfnamefont {M.~J.}\ \bibnamefont
  {Savage}},\ }\href@noop {} {\bibinfo {title} {Hierarchical qubit maps and
  hierarchical quantum error correction}} (\bibinfo {year} {2021}),\ \Eprint
  {https://arxiv.org/abs/2109.01953} {arXiv:2109.01953 [quant-ph]} \BibitemShut
  {NoStop}%
\bibitem [{\citenamefont {Hostetler}\ \emph {et~al.}(2021)\citenamefont
  {Hostetler}, \citenamefont {Zhang}, \citenamefont {Sakai}, \citenamefont
  {Unmuth-Yockey}, \citenamefont {Bazavov},\ and\ \citenamefont
  {Meurice}}]{Hostetler_2021}%
  \BibitemOpen
  \bibfield  {author} {\bibinfo {author} {\bibfnamefont {L.}~\bibnamefont
  {Hostetler}}, \bibinfo {author} {\bibfnamefont {J.}~\bibnamefont {Zhang}},
  \bibinfo {author} {\bibfnamefont {R.}~\bibnamefont {Sakai}}, \bibinfo
  {author} {\bibfnamefont {J.}~\bibnamefont {Unmuth-Yockey}}, \bibinfo {author}
  {\bibfnamefont {A.}~\bibnamefont {Bazavov}},\ and\ \bibinfo {author}
  {\bibfnamefont {Y.}~\bibnamefont {Meurice}},\ }\bibfield  {title} {\bibinfo
  {title} {Clock model interpolation and symmetry breaking in o(2) models},\
  }\bibfield  {journal} {\bibinfo  {journal} {Physical Review D}\ }\textbf
  {\bibinfo {volume} {104}},\ \href
  {https://doi.org/10.1103/physrevd.104.054505} {10.1103/physrevd.104.054505}
  (\bibinfo {year} {2021})\BibitemShut {NoStop}%
\bibitem [{\citenamefont {Andrade}\ \emph {et~al.}(2022)\citenamefont
  {Andrade}, \citenamefont {Davoudi}, \citenamefont {Gra{\ss}}, \citenamefont
  {Hafezi}, \citenamefont {Pagano},\ and\ \citenamefont {Seif}}]{Andrade_2022}%
  \BibitemOpen
  \bibfield  {author} {\bibinfo {author} {\bibfnamefont {B.}~\bibnamefont
  {Andrade}}, \bibinfo {author} {\bibfnamefont {Z.}~\bibnamefont {Davoudi}},
  \bibinfo {author} {\bibfnamefont {T.}~\bibnamefont {Gra{\ss}}}, \bibinfo
  {author} {\bibfnamefont {M.}~\bibnamefont {Hafezi}}, \bibinfo {author}
  {\bibfnamefont {G.}~\bibnamefont {Pagano}},\ and\ \bibinfo {author}
  {\bibfnamefont {A.}~\bibnamefont {Seif}},\ }\bibfield  {title} {\bibinfo
  {title} {Engineering an effective three-spin hamiltonian in trapped-ion
  systems for applications in quantum simulation},\ }\href
  {https://doi.org/10.1088/2058-9565/ac5f5b} {\bibfield  {journal} {\bibinfo
  {journal} {Quantum Science and Technology}\ }\textbf {\bibinfo {volume}
  {7}},\ \bibinfo {pages} {034001} (\bibinfo {year} {2022})}\BibitemShut
  {NoStop}%
\bibitem [{\citenamefont {Bermudez}\ \emph {et~al.}(2010)\citenamefont
  {Bermudez}, \citenamefont {Mazza}, \citenamefont {Rizzi}, \citenamefont
  {Goldman}, \citenamefont {Lewenstein},\ and\ \citenamefont
  {Martin-Delgado}}]{PhysRevLett.105.190404}%
  \BibitemOpen
  \bibfield  {author} {\bibinfo {author} {\bibfnamefont {A.}~\bibnamefont
  {Bermudez}}, \bibinfo {author} {\bibfnamefont {L.}~\bibnamefont {Mazza}},
  \bibinfo {author} {\bibfnamefont {M.}~\bibnamefont {Rizzi}}, \bibinfo
  {author} {\bibfnamefont {N.}~\bibnamefont {Goldman}}, \bibinfo {author}
  {\bibfnamefont {M.}~\bibnamefont {Lewenstein}},\ and\ \bibinfo {author}
  {\bibfnamefont {M.~A.}\ \bibnamefont {Martin-Delgado}},\ }\bibfield  {title}
  {\bibinfo {title} {Wilson fermions and axion electrodynamics in optical
  lattices},\ }\href {https://doi.org/10.1103/PhysRevLett.105.190404}
  {\bibfield  {journal} {\bibinfo  {journal} {Phys. Rev. Lett.}\ }\textbf
  {\bibinfo {volume} {105}},\ \bibinfo {pages} {190404} (\bibinfo {year}
  {2010})}\BibitemShut {NoStop}%
\bibitem [{\citenamefont {Byrnes}\ and\ \citenamefont
  {Yamamoto}(2006)}]{Byrnes_2006}%
  \BibitemOpen
  \bibfield  {author} {\bibinfo {author} {\bibfnamefont {T.}~\bibnamefont
  {Byrnes}}\ and\ \bibinfo {author} {\bibfnamefont {Y.}~\bibnamefont
  {Yamamoto}},\ }\bibfield  {title} {\bibinfo {title} {Simulating lattice gauge
  theories on a quantum computer},\ }\bibfield  {journal} {\bibinfo  {journal}
  {Physical Review A}\ }\textbf {\bibinfo {volume} {73}},\ \href
  {https://doi.org/10.1103/physreva.73.022328} {10.1103/physreva.73.022328}
  (\bibinfo {year} {2006})\BibitemShut {NoStop}%
\bibitem [{\citenamefont {Kan}\ and\ \citenamefont
  {Nam}(2021)}]{kan2021lattice}%
  \BibitemOpen
  \bibfield  {author} {\bibinfo {author} {\bibfnamefont {A.}~\bibnamefont
  {Kan}}\ and\ \bibinfo {author} {\bibfnamefont {Y.}~\bibnamefont {Nam}},\
  }\href@noop {} {\bibinfo {title} {Lattice quantum chromodynamics and
  electrodynamics on a universal quantum computer}} (\bibinfo {year} {2021}),\
  \Eprint {https://arxiv.org/abs/2107.12769} {arXiv:2107.12769 [quant-ph]}
  \BibitemShut {NoStop}%
\bibitem [{\citenamefont {Stryker}(2021)}]{stryker2021shearing}%
  \BibitemOpen
  \bibfield  {author} {\bibinfo {author} {\bibfnamefont {J.~R.}\ \bibnamefont
  {Stryker}},\ }\href@noop {} {\bibinfo {title} {Shearing approach to gauge
  invariant trotterization}} (\bibinfo {year} {2021}),\ \Eprint
  {https://arxiv.org/abs/2105.11548} {arXiv:2105.11548 [hep-lat]} \BibitemShut
  {NoStop}%
\bibitem [{\citenamefont {Stetina}\ \emph {et~al.}(2022)\citenamefont
  {Stetina}, \citenamefont {Ciavarella}, \citenamefont {Li},\ and\
  \citenamefont {Wiebe}}]{stetina2022simulating}%
  \BibitemOpen
  \bibfield  {author} {\bibinfo {author} {\bibfnamefont {T.~F.}\ \bibnamefont
  {Stetina}}, \bibinfo {author} {\bibfnamefont {A.}~\bibnamefont {Ciavarella}},
  \bibinfo {author} {\bibfnamefont {X.}~\bibnamefont {Li}},\ and\ \bibinfo
  {author} {\bibfnamefont {N.}~\bibnamefont {Wiebe}},\ }\bibfield  {title}
  {\bibinfo {title} {Simulating effective qed on quantum computers},\
  }\href@noop {} {\bibfield  {journal} {\bibinfo  {journal} {Quantum}\ }\textbf
  {\bibinfo {volume} {6}},\ \bibinfo {pages} {622} (\bibinfo {year}
  {2022})}\BibitemShut {NoStop}%
\bibitem [{\citenamefont {Farrell}\ \emph {et~al.}(2022)\citenamefont
  {Farrell}, \citenamefont {Chernyshev}, \citenamefont {Powell}, \citenamefont
  {Zemlevskiy}, \citenamefont {Illa},\ and\ \citenamefont
  {Savage}}]{https://doi.org/10.48550/arxiv.2207.01731}%
  \BibitemOpen
  \bibfield  {author} {\bibinfo {author} {\bibfnamefont {R.~C.}\ \bibnamefont
  {Farrell}}, \bibinfo {author} {\bibfnamefont {I.~A.}\ \bibnamefont
  {Chernyshev}}, \bibinfo {author} {\bibfnamefont {S.~J.~M.}\ \bibnamefont
  {Powell}}, \bibinfo {author} {\bibfnamefont {N.~A.}\ \bibnamefont
  {Zemlevskiy}}, \bibinfo {author} {\bibfnamefont {M.}~\bibnamefont {Illa}},\
  and\ \bibinfo {author} {\bibfnamefont {M.~J.}\ \bibnamefont {Savage}},\
  }\href {https://doi.org/10.48550/ARXIV.2207.01731} {\bibinfo {title}
  {Preparations for quantum simulations of quantum chromodynamics in 1+1
  dimensions: (i) axial gauge}} (\bibinfo {year} {2022})\BibitemShut {NoStop}%
\bibitem [{\citenamefont {Davoudi}\ \emph
  {et~al.}(2021{\natexlab{b}})\citenamefont {Davoudi}, \citenamefont {Linke},\
  and\ \citenamefont {Pagano}}]{davoudi2021simulating}%
  \BibitemOpen
  \bibfield  {author} {\bibinfo {author} {\bibfnamefont {Z.}~\bibnamefont
  {Davoudi}}, \bibinfo {author} {\bibfnamefont {N.~M.}\ \bibnamefont {Linke}},\
  and\ \bibinfo {author} {\bibfnamefont {G.}~\bibnamefont {Pagano}},\
  }\href@noop {} {\bibinfo {title} {Toward simulating quantum field theories
  with controlled phonon-ion dynamics: A hybrid analog-digital approach}}
  (\bibinfo {year} {2021}{\natexlab{b}}),\ \Eprint
  {https://arxiv.org/abs/2104.09346} {arXiv:2104.09346 [quant-ph]} \BibitemShut
  {NoStop}%
\bibitem [{\citenamefont {Singh}\ and\ \citenamefont
  {Chandrasekharan}(2019)}]{PhysRevD.100.054505}%
  \BibitemOpen
  \bibfield  {author} {\bibinfo {author} {\bibfnamefont {H.}~\bibnamefont
  {Singh}}\ and\ \bibinfo {author} {\bibfnamefont {S.}~\bibnamefont
  {Chandrasekharan}},\ }\bibfield  {title} {\bibinfo {title} {Qubit
  regularization of the $o(3)$ sigma model},\ }\href
  {https://doi.org/10.1103/PhysRevD.100.054505} {\bibfield  {journal} {\bibinfo
   {journal} {Phys. Rev. D}\ }\textbf {\bibinfo {volume} {100}},\ \bibinfo
  {pages} {054505} (\bibinfo {year} {2019})}\BibitemShut {NoStop}%
\bibitem [{\citenamefont {Bhattacharya}\ \emph {et~al.}(2021)\citenamefont
  {Bhattacharya}, \citenamefont {Buser}, \citenamefont {Chandrasekharan},
  \citenamefont {Gupta},\ and\ \citenamefont {Singh}}]{PhysRevLett.126.172001}%
  \BibitemOpen
  \bibfield  {author} {\bibinfo {author} {\bibfnamefont {T.}~\bibnamefont
  {Bhattacharya}}, \bibinfo {author} {\bibfnamefont {A.~J.}\ \bibnamefont
  {Buser}}, \bibinfo {author} {\bibfnamefont {S.}~\bibnamefont
  {Chandrasekharan}}, \bibinfo {author} {\bibfnamefont {R.}~\bibnamefont
  {Gupta}},\ and\ \bibinfo {author} {\bibfnamefont {H.}~\bibnamefont {Singh}},\
  }\bibfield  {title} {\bibinfo {title} {Qubit regularization of asymptotic
  freedom},\ }\href {https://doi.org/10.1103/PhysRevLett.126.172001} {\bibfield
   {journal} {\bibinfo  {journal} {Phys. Rev. Lett.}\ }\textbf {\bibinfo
  {volume} {126}},\ \bibinfo {pages} {172001} (\bibinfo {year}
  {2021})}\BibitemShut {NoStop}%
\bibitem [{\citenamefont
  {Singh}(2019)}]{https://doi.org/10.48550/arxiv.1911.12353}%
  \BibitemOpen
  \bibfield  {author} {\bibinfo {author} {\bibfnamefont {H.}~\bibnamefont
  {Singh}},\ }\href {https://doi.org/10.48550/ARXIV.1911.12353} {\bibinfo
  {title} {Qubit regularized $o(n)$ nonlinear sigma models}} (\bibinfo {year}
  {2019})\BibitemShut {NoStop}%
\bibitem [{\citenamefont
  {Singh}(2022)}]{https://doi.org/10.48550/arxiv.2203.00059}%
  \BibitemOpen
  \bibfield  {author} {\bibinfo {author} {\bibfnamefont {H.}~\bibnamefont
  {Singh}},\ }\href {https://doi.org/10.48550/ARXIV.2203.00059} {\bibinfo
  {title} {Large-charge conformal dimensions at the $o(n)$ wilson-fisher fixed
  point}} (\bibinfo {year} {2022})\BibitemShut {NoStop}%
\bibitem [{\citenamefont {Caspar}\ and\ \citenamefont
  {Singh}(2022)}]{https://doi.org/10.48550/arxiv.2203.15766}%
  \BibitemOpen
  \bibfield  {author} {\bibinfo {author} {\bibfnamefont {S.}~\bibnamefont
  {Caspar}}\ and\ \bibinfo {author} {\bibfnamefont {H.}~\bibnamefont {Singh}},\
  }\href {https://doi.org/10.48550/ARXIV.2203.15766} {\bibinfo {title} {From
  asymptotic freedom to $θ$ vacua: Qubit embeddings of the o(3) nonlinear $σ$
  model}} (\bibinfo {year} {2022})\BibitemShut {NoStop}%
\bibitem [{\citenamefont {Gustafson}(2021)}]{PhysRevD.103.114505}%
  \BibitemOpen
  \bibfield  {author} {\bibinfo {author} {\bibfnamefont {E.~J.}\ \bibnamefont
  {Gustafson}},\ }\bibfield  {title} {\bibinfo {title} {Prospects for
  simulating a qudit-based model of $(1+1)\mathrm{D}$ scalar qed},\ }\href
  {https://doi.org/10.1103/PhysRevD.103.114505} {\bibfield  {journal} {\bibinfo
   {journal} {Phys. Rev. D}\ }\textbf {\bibinfo {volume} {103}},\ \bibinfo
  {pages} {114505} (\bibinfo {year} {2021})}\BibitemShut {NoStop}%
\bibitem [{\citenamefont {Unmuth-Yockey}(2019)}]{PhysRevD.99.074502}%
  \BibitemOpen
  \bibfield  {author} {\bibinfo {author} {\bibfnamefont {J.~F.}\ \bibnamefont
  {Unmuth-Yockey}},\ }\bibfield  {title} {\bibinfo {title} {Gauge-invariant
  rotor hamiltonian from dual variables of 3d $u\mathbf{(}1\mathbf{)}$ gauge
  theory},\ }\href {https://doi.org/10.1103/PhysRevD.99.074502} {\bibfield
  {journal} {\bibinfo  {journal} {Phys. Rev. D}\ }\textbf {\bibinfo {volume}
  {99}},\ \bibinfo {pages} {074502} (\bibinfo {year} {2019})}\BibitemShut
  {NoStop}%
\bibitem [{\citenamefont {Zhang}\ \emph {et~al.}(2021)\citenamefont {Zhang},
  \citenamefont {Ferguson}, \citenamefont {Kühn}, \citenamefont {Haase},
  \citenamefont {Wilson}, \citenamefont {Jansen},\ and\ \citenamefont
  {Muschik}}]{https://doi.org/10.48550/arxiv.2108.08248}%
  \BibitemOpen
  \bibfield  {author} {\bibinfo {author} {\bibfnamefont {J.}~\bibnamefont
  {Zhang}}, \bibinfo {author} {\bibfnamefont {R.}~\bibnamefont {Ferguson}},
  \bibinfo {author} {\bibfnamefont {S.}~\bibnamefont {Kühn}}, \bibinfo
  {author} {\bibfnamefont {J.~F.}\ \bibnamefont {Haase}}, \bibinfo {author}
  {\bibfnamefont {C.~M.}\ \bibnamefont {Wilson}}, \bibinfo {author}
  {\bibfnamefont {K.}~\bibnamefont {Jansen}},\ and\ \bibinfo {author}
  {\bibfnamefont {C.~A.}\ \bibnamefont {Muschik}},\ }\href
  {https://doi.org/10.48550/ARXIV.2108.08248} {\bibinfo {title} {Simulating
  gauge theories with variational quantum eigensolvers in superconducting
  microwave cavities}} (\bibinfo {year} {2021})\BibitemShut {NoStop}%
\bibitem [{\citenamefont {\"Unsal}(2012)}]{PhysRevD.86.105012}%
  \BibitemOpen
  \bibfield  {author} {\bibinfo {author} {\bibfnamefont {M.}~\bibnamefont
  {\"Unsal}},\ }\bibfield  {title} {\bibinfo {title} {Theta dependence, sign
  problems, and topological interference},\ }\href
  {https://doi.org/10.1103/PhysRevD.86.105012} {\bibfield  {journal} {\bibinfo
  {journal} {Phys. Rev. D}\ }\textbf {\bibinfo {volume} {86}},\ \bibinfo
  {pages} {105012} (\bibinfo {year} {2012})}\BibitemShut {NoStop}%
\bibitem [{\citenamefont {Allton}\ \emph {et~al.}(2002)\citenamefont {Allton},
  \citenamefont {Ejiri}, \citenamefont {Hands}, \citenamefont {Kaczmarek},
  \citenamefont {Karsch}, \citenamefont {Laermann}, \citenamefont {Schmidt},\
  and\ \citenamefont {Scorzato}}]{PhysRevD.66.074507}%
  \BibitemOpen
  \bibfield  {author} {\bibinfo {author} {\bibfnamefont {C.~R.}\ \bibnamefont
  {Allton}}, \bibinfo {author} {\bibfnamefont {S.}~\bibnamefont {Ejiri}},
  \bibinfo {author} {\bibfnamefont {S.~J.}\ \bibnamefont {Hands}}, \bibinfo
  {author} {\bibfnamefont {O.}~\bibnamefont {Kaczmarek}}, \bibinfo {author}
  {\bibfnamefont {F.}~\bibnamefont {Karsch}}, \bibinfo {author} {\bibfnamefont
  {E.}~\bibnamefont {Laermann}}, \bibinfo {author} {\bibfnamefont
  {C.}~\bibnamefont {Schmidt}},\ and\ \bibinfo {author} {\bibfnamefont
  {L.}~\bibnamefont {Scorzato}},\ }\bibfield  {title} {\bibinfo {title} {Qcd
  thermal phase transition in the presence of a small chemical potential},\
  }\href {https://doi.org/10.1103/PhysRevD.66.074507} {\bibfield  {journal}
  {\bibinfo  {journal} {Phys. Rev. D}\ }\textbf {\bibinfo {volume} {66}},\
  \bibinfo {pages} {074507} (\bibinfo {year} {2002})}\BibitemShut {NoStop}%
\bibitem [{\citenamefont {de~Forcrand}\ and\ \citenamefont
  {Philipsen}(2002)}]{DEFORCRAND2002290}%
  \BibitemOpen
  \bibfield  {author} {\bibinfo {author} {\bibfnamefont {P.}~\bibnamefont
  {de~Forcrand}}\ and\ \bibinfo {author} {\bibfnamefont {O.}~\bibnamefont
  {Philipsen}},\ }\bibfield  {title} {\bibinfo {title} {The qcd phase diagram
  for small densities from imaginary chemical potential},\ }\href
  {https://doi.org/https://doi.org/10.1016/S0550-3213(02)00626-0} {\bibfield
  {journal} {\bibinfo  {journal} {Nuclear Physics B}\ }\textbf {\bibinfo
  {volume} {642}},\ \bibinfo {pages} {290} (\bibinfo {year}
  {2002})}\BibitemShut {NoStop}%
\bibitem [{\citenamefont {Feynman}(1982)}]{Feynman1982}%
  \BibitemOpen
  \bibfield  {author} {\bibinfo {author} {\bibfnamefont {R.~P.}\ \bibnamefont
  {Feynman}},\ }\bibfield  {title} {\bibinfo {title} {Simulating physics with
  computers},\ }\href {https://doi.org/10.1007/BF02650179} {\bibfield
  {journal} {\bibinfo  {journal} {International Journal of Theoretical
  Physics}\ }\textbf {\bibinfo {volume} {21}},\ \bibinfo {pages} {467}
  (\bibinfo {year} {1982})}\BibitemShut {NoStop}%
\bibitem [{\citenamefont {Benioff}(1980)}]{Benioff:1980}%
  \BibitemOpen
  \bibfield  {author} {\bibinfo {author} {\bibfnamefont {P.}~\bibnamefont
  {Benioff}},\ }\bibfield  {title} {\bibinfo {title} {The computer as a
  physical system: A microscopic quantum mechanical hamiltonian model of
  computers as represented by turing machines},\ }\href@noop {} {\bibfield
  {journal} {\bibinfo  {journal} {Journal of statistical physics}\ }\textbf
  {\bibinfo {volume} {22}},\ \bibinfo {pages} {563} (\bibinfo {year}
  {1980})}\BibitemShut {NoStop}%
\bibitem [{\citenamefont {Huang}\ \emph {et~al.}(2020)\citenamefont {Huang},
  \citenamefont {Wu}, \citenamefont {Fan},\ and\ \citenamefont
  {Zhu}}]{huang2020superconducting}%
  \BibitemOpen
  \bibfield  {author} {\bibinfo {author} {\bibfnamefont {H.-L.}\ \bibnamefont
  {Huang}}, \bibinfo {author} {\bibfnamefont {D.}~\bibnamefont {Wu}}, \bibinfo
  {author} {\bibfnamefont {D.}~\bibnamefont {Fan}},\ and\ \bibinfo {author}
  {\bibfnamefont {X.}~\bibnamefont {Zhu}},\ }\href@noop {} {\bibinfo {title}
  {Superconducting quantum computing: A review}} (\bibinfo {year} {2020}),\
  \Eprint {https://arxiv.org/abs/2006.10433} {arXiv:2006.10433 [quant-ph]}
  \BibitemShut {NoStop}%
\bibitem [{\citenamefont {Slussarenko}\ and\ \citenamefont
  {Pryde}(2019)}]{2019Photon}%
  \BibitemOpen
  \bibfield  {author} {\bibinfo {author} {\bibfnamefont {S.}~\bibnamefont
  {Slussarenko}}\ and\ \bibinfo {author} {\bibfnamefont {G.~J.}\ \bibnamefont
  {Pryde}},\ }\bibfield  {title} {\bibinfo {title} {Photonic quantum
  information processing: A concise review},\ }\href
  {https://doi.org/10.1063/1.5115814} {\bibfield  {journal} {\bibinfo
  {journal} {Applied Physics Reviews}\ }\textbf {\bibinfo {volume} {6}},\
  \bibinfo {pages} {041303} (\bibinfo {year} {2019})}\BibitemShut {NoStop}%
\bibitem [{\citenamefont {Barthelemy}\ and\ \citenamefont
  {Vandersypen}(2013)}]{barthelemy2013quantum}%
  \BibitemOpen
  \bibfield  {author} {\bibinfo {author} {\bibfnamefont {P.}~\bibnamefont
  {Barthelemy}}\ and\ \bibinfo {author} {\bibfnamefont {L.~M.}\ \bibnamefont
  {Vandersypen}},\ }\bibfield  {title} {\bibinfo {title} {Quantum dot systems:
  a versatile platform for quantum simulations},\ }\href@noop {} {\bibfield
  {journal} {\bibinfo  {journal} {Annalen der Physik}\ }\textbf {\bibinfo
  {volume} {525}},\ \bibinfo {pages} {808} (\bibinfo {year}
  {2013})}\BibitemShut {NoStop}%
\bibitem [{\citenamefont {Bruzewicz}\ \emph {et~al.}(2019)\citenamefont
  {Bruzewicz}, \citenamefont {Chiaverini}, \citenamefont {McConnell},\ and\
  \citenamefont {Sage}}]{2019Trap}%
  \BibitemOpen
  \bibfield  {author} {\bibinfo {author} {\bibfnamefont {C.~D.}\ \bibnamefont
  {Bruzewicz}}, \bibinfo {author} {\bibfnamefont {J.}~\bibnamefont
  {Chiaverini}}, \bibinfo {author} {\bibfnamefont {R.}~\bibnamefont
  {McConnell}},\ and\ \bibinfo {author} {\bibfnamefont {J.~M.}\ \bibnamefont
  {Sage}},\ }\bibfield  {title} {\bibinfo {title} {Trapped-ion quantum
  computing: Progress and challenges},\ }\href
  {https://doi.org/10.1063/1.5088164} {\bibfield  {journal} {\bibinfo
  {journal} {Applied Physics Reviews}\ }\textbf {\bibinfo {volume} {6}},\
  \bibinfo {pages} {021314} (\bibinfo {year} {2019})}\BibitemShut {NoStop}%
\bibitem [{\citenamefont {Cohen}\ and\ \citenamefont
  {Thompson}(2021)}]{PRXQuantum.2.030322}%
  \BibitemOpen
  \bibfield  {author} {\bibinfo {author} {\bibfnamefont {S.~R.}\ \bibnamefont
  {Cohen}}\ and\ \bibinfo {author} {\bibfnamefont {J.~D.}\ \bibnamefont
  {Thompson}},\ }\bibfield  {title} {\bibinfo {title} {Quantum computing with
  circular rydberg atoms},\ }\href
  {https://doi.org/10.1103/PRXQuantum.2.030322} {\bibfield  {journal} {\bibinfo
   {journal} {PRX Quantum}\ }\textbf {\bibinfo {volume} {2}},\ \bibinfo {pages}
  {030322} (\bibinfo {year} {2021})}\BibitemShut {NoStop}%
\bibitem [{\citenamefont {Weimer}\ \emph {et~al.}(2011)\citenamefont {Weimer},
  \citenamefont {Müller}, \citenamefont {Büchler},\ and\ \citenamefont
  {Lesanovsky}}]{Weimer_2011}%
  \BibitemOpen
  \bibfield  {author} {\bibinfo {author} {\bibfnamefont {H.}~\bibnamefont
  {Weimer}}, \bibinfo {author} {\bibfnamefont {M.}~\bibnamefont {Müller}},
  \bibinfo {author} {\bibfnamefont {H.~P.}\ \bibnamefont {Büchler}},\ and\
  \bibinfo {author} {\bibfnamefont {I.}~\bibnamefont {Lesanovsky}},\ }\bibfield
   {title} {\bibinfo {title} {Digital quantum simulation with rydberg atoms},\
  }\href {https://doi.org/10.1007/s11128-011-0303-5} {\bibfield  {journal}
  {\bibinfo  {journal} {Quantum Information Processing}\ }\textbf {\bibinfo
  {volume} {10}},\ \bibinfo {pages} {885} (\bibinfo {year} {2011})}\BibitemShut
  {NoStop}%
\bibitem [{\citenamefont {Heeres}\ \emph {et~al.}(2015)\citenamefont {Heeres},
  \citenamefont {Vlastakis}, \citenamefont {Holland}, \citenamefont
  {Krastanov}, \citenamefont {Albert}, \citenamefont {Frunzio}, \citenamefont
  {Jiang},\ and\ \citenamefont {Schoelkopf}}]{PhysRevLett.115.137002}%
  \BibitemOpen
  \bibfield  {author} {\bibinfo {author} {\bibfnamefont {R.~W.}\ \bibnamefont
  {Heeres}}, \bibinfo {author} {\bibfnamefont {B.}~\bibnamefont {Vlastakis}},
  \bibinfo {author} {\bibfnamefont {E.}~\bibnamefont {Holland}}, \bibinfo
  {author} {\bibfnamefont {S.}~\bibnamefont {Krastanov}}, \bibinfo {author}
  {\bibfnamefont {V.~V.}\ \bibnamefont {Albert}}, \bibinfo {author}
  {\bibfnamefont {L.}~\bibnamefont {Frunzio}}, \bibinfo {author} {\bibfnamefont
  {L.}~\bibnamefont {Jiang}},\ and\ \bibinfo {author} {\bibfnamefont {R.~J.}\
  \bibnamefont {Schoelkopf}},\ }\bibfield  {title} {\bibinfo {title} {Cavity
  state manipulation using photon-number selective phase gates},\ }\href
  {https://doi.org/10.1103/PhysRevLett.115.137002} {\bibfield  {journal}
  {\bibinfo  {journal} {Phys. Rev. Lett.}\ }\textbf {\bibinfo {volume} {115}},\
  \bibinfo {pages} {137002} (\bibinfo {year} {2015})}\BibitemShut {NoStop}%
\bibitem [{\citenamefont {Bianchetti}\ \emph {et~al.}(2010)\citenamefont
  {Bianchetti}, \citenamefont {Filipp}, \citenamefont {Baur}, \citenamefont
  {Fink}, \citenamefont {Lang}, \citenamefont {Steffen}, \citenamefont
  {Boissonneault}, \citenamefont {Blais},\ and\ \citenamefont
  {Wallraff}}]{PhysRevLett.105.223601}%
  \BibitemOpen
  \bibfield  {author} {\bibinfo {author} {\bibfnamefont {R.}~\bibnamefont
  {Bianchetti}}, \bibinfo {author} {\bibfnamefont {S.}~\bibnamefont {Filipp}},
  \bibinfo {author} {\bibfnamefont {M.}~\bibnamefont {Baur}}, \bibinfo {author}
  {\bibfnamefont {J.~M.}\ \bibnamefont {Fink}}, \bibinfo {author}
  {\bibfnamefont {C.}~\bibnamefont {Lang}}, \bibinfo {author} {\bibfnamefont
  {L.}~\bibnamefont {Steffen}}, \bibinfo {author} {\bibfnamefont
  {M.}~\bibnamefont {Boissonneault}}, \bibinfo {author} {\bibfnamefont
  {A.}~\bibnamefont {Blais}},\ and\ \bibinfo {author} {\bibfnamefont
  {A.}~\bibnamefont {Wallraff}},\ }\bibfield  {title} {\bibinfo {title}
  {Control and tomography of a three level superconducting artificial atom},\
  }\href {https://doi.org/10.1103/PhysRevLett.105.223601} {\bibfield  {journal}
  {\bibinfo  {journal} {Phys. Rev. Lett.}\ }\textbf {\bibinfo {volume} {105}},\
  \bibinfo {pages} {223601} (\bibinfo {year} {2010})}\BibitemShut {NoStop}%
\bibitem [{\citenamefont {Wang}\ \emph {et~al.}(2022)\citenamefont {Wang},
  \citenamefont {Gonin}, \citenamefont {Grassellino}, \citenamefont {Kazakov},
  \citenamefont {Romanenko}, \citenamefont {Yakovlev},\ and\ \citenamefont
  {Zorzetti}}]{https://doi.org/10.48550/arxiv.2206.15467}%
  \BibitemOpen
  \bibfield  {author} {\bibinfo {author} {\bibfnamefont {C.}~\bibnamefont
  {Wang}}, \bibinfo {author} {\bibfnamefont {I.}~\bibnamefont {Gonin}},
  \bibinfo {author} {\bibfnamefont {A.}~\bibnamefont {Grassellino}}, \bibinfo
  {author} {\bibfnamefont {S.}~\bibnamefont {Kazakov}}, \bibinfo {author}
  {\bibfnamefont {A.}~\bibnamefont {Romanenko}}, \bibinfo {author}
  {\bibfnamefont {V.~P.}\ \bibnamefont {Yakovlev}},\ and\ \bibinfo {author}
  {\bibfnamefont {S.}~\bibnamefont {Zorzetti}},\ }\href
  {https://doi.org/10.48550/ARXIV.2206.15467} {\bibinfo {title}
  {High-efficiency microwave-optical quantum transduction based on a cavity
  electro-optic superconducting system with long coherence time}} (\bibinfo
  {year} {2022})\BibitemShut {NoStop}%
\bibitem [{\citenamefont {Wallraff}\ \emph {et~al.}(2004)\citenamefont
  {Wallraff}, \citenamefont {Schuster}, \citenamefont {Blais}, \citenamefont
  {Frunzio}, \citenamefont {Huang}, \citenamefont {Majer}, \citenamefont
  {Kumar}, \citenamefont {Girvin},\ and\ \citenamefont
  {Schoelkopf}}]{wallraff2004strong}%
  \BibitemOpen
  \bibfield  {author} {\bibinfo {author} {\bibfnamefont {A.}~\bibnamefont
  {Wallraff}}, \bibinfo {author} {\bibfnamefont {D.~I.}\ \bibnamefont
  {Schuster}}, \bibinfo {author} {\bibfnamefont {A.}~\bibnamefont {Blais}},
  \bibinfo {author} {\bibfnamefont {L.}~\bibnamefont {Frunzio}}, \bibinfo
  {author} {\bibfnamefont {R.-S.}\ \bibnamefont {Huang}}, \bibinfo {author}
  {\bibfnamefont {J.}~\bibnamefont {Majer}}, \bibinfo {author} {\bibfnamefont
  {S.}~\bibnamefont {Kumar}}, \bibinfo {author} {\bibfnamefont {S.~M.}\
  \bibnamefont {Girvin}},\ and\ \bibinfo {author} {\bibfnamefont {R.~J.}\
  \bibnamefont {Schoelkopf}},\ }\bibfield  {title} {\bibinfo {title} {Strong
  coupling of a single photon to a superconducting qubit using circuit quantum
  electrodynamics},\ }\href@noop {} {\bibfield  {journal} {\bibinfo  {journal}
  {Nature}\ }\textbf {\bibinfo {volume} {431}},\ \bibinfo {pages} {162}
  (\bibinfo {year} {2004})}\BibitemShut {NoStop}%
\bibitem [{\citenamefont {Chiorescu}\ \emph {et~al.}(2004)\citenamefont
  {Chiorescu}, \citenamefont {Bertet}, \citenamefont {Semba}, \citenamefont
  {Nakamura}, \citenamefont {Harmans},\ and\ \citenamefont
  {Mooij}}]{chiorescu2004coherent}%
  \BibitemOpen
  \bibfield  {author} {\bibinfo {author} {\bibfnamefont {I.}~\bibnamefont
  {Chiorescu}}, \bibinfo {author} {\bibfnamefont {P.}~\bibnamefont {Bertet}},
  \bibinfo {author} {\bibfnamefont {K.}~\bibnamefont {Semba}}, \bibinfo
  {author} {\bibfnamefont {Y.}~\bibnamefont {Nakamura}}, \bibinfo {author}
  {\bibfnamefont {C.}~\bibnamefont {Harmans}},\ and\ \bibinfo {author}
  {\bibfnamefont {J.}~\bibnamefont {Mooij}},\ }\bibfield  {title} {\bibinfo
  {title} {Coherent dynamics of a flux qubit coupled to a harmonic
  oscillator},\ }\href@noop {} {\bibfield  {journal} {\bibinfo  {journal}
  {Nature}\ }\textbf {\bibinfo {volume} {431}},\ \bibinfo {pages} {159}
  (\bibinfo {year} {2004})}\BibitemShut {NoStop}%
\bibitem [{\citenamefont {Flannigan}\ \emph {et~al.}(2022)\citenamefont
  {Flannigan}, \citenamefont {Pearson}, \citenamefont {Low}, \citenamefont
  {Buyskikh}, \citenamefont {Bloch}, \citenamefont {Zoller}, \citenamefont
  {Troyer},\ and\ \citenamefont
  {Daley}}]{https://doi.org/10.48550/arxiv.2204.13644}%
  \BibitemOpen
  \bibfield  {author} {\bibinfo {author} {\bibfnamefont {S.}~\bibnamefont
  {Flannigan}}, \bibinfo {author} {\bibfnamefont {N.}~\bibnamefont {Pearson}},
  \bibinfo {author} {\bibfnamefont {G.~H.}\ \bibnamefont {Low}}, \bibinfo
  {author} {\bibfnamefont {A.}~\bibnamefont {Buyskikh}}, \bibinfo {author}
  {\bibfnamefont {I.}~\bibnamefont {Bloch}}, \bibinfo {author} {\bibfnamefont
  {P.}~\bibnamefont {Zoller}}, \bibinfo {author} {\bibfnamefont
  {M.}~\bibnamefont {Troyer}},\ and\ \bibinfo {author} {\bibfnamefont {A.~J.}\
  \bibnamefont {Daley}},\ }\href {https://doi.org/10.48550/ARXIV.2204.13644}
  {\bibinfo {title} {Propagation of errors and quantitative quantum simulation
  with quantum advantage}} (\bibinfo {year} {2022})\BibitemShut {NoStop}%
\bibitem [{\citenamefont {Chandrasekharan}\ \emph {et~al.}(2002)\citenamefont
  {Chandrasekharan}, \citenamefont {Scarlet},\ and\ \citenamefont
  {Wiese}}]{CHANDRASEKHARAN2002388}%
  \BibitemOpen
  \bibfield  {author} {\bibinfo {author} {\bibfnamefont {S.}~\bibnamefont
  {Chandrasekharan}}, \bibinfo {author} {\bibfnamefont {B.}~\bibnamefont
  {Scarlet}},\ and\ \bibinfo {author} {\bibfnamefont {U.-J.}\ \bibnamefont
  {Wiese}},\ }\bibfield  {title} {\bibinfo {title} {From spin ladders to the 2d
  o(3) model at non-zero density},\ }\href
  {https://doi.org/https://doi.org/10.1016/S0010-4655(02)00311-9} {\bibfield
  {journal} {\bibinfo  {journal} {Computer Physics Communications}\ }\textbf
  {\bibinfo {volume} {147}},\ \bibinfo {pages} {388} (\bibinfo {year}
  {2002})},\ \bibinfo {note} {proceedings of the Europhysics Conference on
  Computational Physics Computational Modeling and Simulation of Complex
  Systems}\BibitemShut {NoStop}%
\bibitem [{\citenamefont {Brower}\ \emph {et~al.}(2004)\citenamefont {Brower},
  \citenamefont {Chandrasekharan}, \citenamefont {Riederer},\ and\
  \citenamefont {Wiese}}]{BROWER2004149}%
  \BibitemOpen
  \bibfield  {author} {\bibinfo {author} {\bibfnamefont {R.}~\bibnamefont
  {Brower}}, \bibinfo {author} {\bibfnamefont {S.}~\bibnamefont
  {Chandrasekharan}}, \bibinfo {author} {\bibfnamefont {S.}~\bibnamefont
  {Riederer}},\ and\ \bibinfo {author} {\bibfnamefont {U.-J.}\ \bibnamefont
  {Wiese}},\ }\bibfield  {title} {\bibinfo {title} {D-theory: field
  quantization by dimensional reduction of discrete variables},\ }\href
  {https://doi.org/https://doi.org/10.1016/j.nuclphysb.2004.06.007} {\bibfield
  {journal} {\bibinfo  {journal} {Nuclear Physics B}\ }\textbf {\bibinfo
  {volume} {693}},\ \bibinfo {pages} {149} (\bibinfo {year}
  {2004})}\BibitemShut {NoStop}%
\bibitem [{\citenamefont {Bruckmann}\ \emph {et~al.}(2019)\citenamefont
  {Bruckmann}, \citenamefont {Jansen},\ and\ \citenamefont
  {K\"uhn}}]{PhysRevD.99.074501}%
  \BibitemOpen
  \bibfield  {author} {\bibinfo {author} {\bibfnamefont {F.}~\bibnamefont
  {Bruckmann}}, \bibinfo {author} {\bibfnamefont {K.}~\bibnamefont {Jansen}},\
  and\ \bibinfo {author} {\bibfnamefont {S.}~\bibnamefont {K\"uhn}},\
  }\bibfield  {title} {\bibinfo {title} {O(3) nonlinear sigma model in $1+1$
  dimensions with matrix product states},\ }\href
  {https://doi.org/10.1103/PhysRevD.99.074501} {\bibfield  {journal} {\bibinfo
  {journal} {Phys. Rev. D}\ }\textbf {\bibinfo {volume} {99}},\ \bibinfo
  {pages} {074501} (\bibinfo {year} {2019})}\BibitemShut {NoStop}%
\bibitem [{\citenamefont {Duan}\ \emph {et~al.}(2006)\citenamefont {Duan},
  \citenamefont {Fuller},\ and\ \citenamefont {Qian}}]{Duan:2005cp}%
  \BibitemOpen
  \bibfield  {author} {\bibinfo {author} {\bibfnamefont {H.}~\bibnamefont
  {Duan}}, \bibinfo {author} {\bibfnamefont {G.~M.}\ \bibnamefont {Fuller}},\
  and\ \bibinfo {author} {\bibfnamefont {Y.-Z.}\ \bibnamefont {Qian}},\
  }\bibfield  {title} {\bibinfo {title} {{Collective neutrino flavor
  transformation in supernovae}},\ }\href
  {https://doi.org/10.1103/PhysRevD.74.123004} {\bibfield  {journal} {\bibinfo
  {journal} {Phys. Rev. D}\ }\textbf {\bibinfo {volume} {74}},\ \bibinfo
  {pages} {123004} (\bibinfo {year} {2006})},\ \Eprint
  {https://arxiv.org/abs/astro-ph/0511275} {arXiv:astro-ph/0511275}
  \BibitemShut {NoStop}%
\bibitem [{\citenamefont {Balantekin}\ and\ \citenamefont
  {Pehlivan}(2006)}]{Balantekin_2006}%
  \BibitemOpen
  \bibfield  {author} {\bibinfo {author} {\bibfnamefont {A.~B.}\ \bibnamefont
  {Balantekin}}\ and\ \bibinfo {author} {\bibfnamefont {Y.}~\bibnamefont
  {Pehlivan}},\ }\bibfield  {title} {\bibinfo {title}
  {Neutrino{\textendash}neutrino interactions and flavour mixing in dense
  matter},\ }\href {https://doi.org/10.1088/0954-3899/34/1/004} {\bibfield
  {journal} {\bibinfo  {journal} {Journal of Physics G: Nuclear and Particle
  Physics}\ }\textbf {\bibinfo {volume} {34}},\ \bibinfo {pages} {47} (\bibinfo
  {year} {2006})}\BibitemShut {NoStop}%
\bibitem [{\citenamefont {Patwardhan}\ \emph {et~al.}(2019)\citenamefont
  {Patwardhan}, \citenamefont {Cervia},\ and\ \citenamefont
  {Balantekin}}]{PhysRevD.99.123013}%
  \BibitemOpen
  \bibfield  {author} {\bibinfo {author} {\bibfnamefont {A.~V.}\ \bibnamefont
  {Patwardhan}}, \bibinfo {author} {\bibfnamefont {M.~J.}\ \bibnamefont
  {Cervia}},\ and\ \bibinfo {author} {\bibfnamefont {A.~B.}\ \bibnamefont
  {Balantekin}},\ }\bibfield  {title} {\bibinfo {title} {Eigenvalues and
  eigenstates of the many-body collective neutrino oscillation problem},\
  }\href {https://doi.org/10.1103/PhysRevD.99.123013} {\bibfield  {journal}
  {\bibinfo  {journal} {Phys. Rev. D}\ }\textbf {\bibinfo {volume} {99}},\
  \bibinfo {pages} {123013} (\bibinfo {year} {2019})}\BibitemShut {NoStop}%
\bibitem [{\citenamefont {Patwardhan}\ \emph {et~al.}(2021)\citenamefont
  {Patwardhan}, \citenamefont {Cervia},\ and\ \citenamefont
  {Balantekin}}]{PhysRevD.104.123035}%
  \BibitemOpen
  \bibfield  {author} {\bibinfo {author} {\bibfnamefont {A.~V.}\ \bibnamefont
  {Patwardhan}}, \bibinfo {author} {\bibfnamefont {M.~J.}\ \bibnamefont
  {Cervia}},\ and\ \bibinfo {author} {\bibfnamefont {A.~B.}\ \bibnamefont
  {Balantekin}},\ }\bibfield  {title} {\bibinfo {title} {Spectral splits and
  entanglement entropy in collective neutrino oscillations},\ }\href
  {https://doi.org/10.1103/PhysRevD.104.123035} {\bibfield  {journal} {\bibinfo
   {journal} {Phys. Rev. D}\ }\textbf {\bibinfo {volume} {104}},\ \bibinfo
  {pages} {123035} (\bibinfo {year} {2021})}\BibitemShut {NoStop}%
\bibitem [{\citenamefont {Rrapaj}(2020)}]{PhysRevC.101.065805}%
  \BibitemOpen
  \bibfield  {author} {\bibinfo {author} {\bibfnamefont {E.}~\bibnamefont
  {Rrapaj}},\ }\bibfield  {title} {\bibinfo {title} {Exact solution of
  multiangle quantum many-body collective neutrino-flavor oscillations},\
  }\href {https://doi.org/10.1103/PhysRevC.101.065805} {\bibfield  {journal}
  {\bibinfo  {journal} {Phys. Rev. C}\ }\textbf {\bibinfo {volume} {101}},\
  \bibinfo {pages} {065805} (\bibinfo {year} {2020})}\BibitemShut {NoStop}%
\bibitem [{\citenamefont {Martin}\ \emph {et~al.}(2020)\citenamefont {Martin},
  \citenamefont {Carlson},\ and\ \citenamefont {Duan}}]{Martin:2019dof}%
  \BibitemOpen
  \bibfield  {author} {\bibinfo {author} {\bibfnamefont {J.~D.}\ \bibnamefont
  {Martin}}, \bibinfo {author} {\bibfnamefont {J.}~\bibnamefont {Carlson}},\
  and\ \bibinfo {author} {\bibfnamefont {H.}~\bibnamefont {Duan}},\ }\bibfield
  {title} {\bibinfo {title} {{Spectral swaps in a two-dimensional neutrino ring
  model}},\ }\href {https://doi.org/10.1103/PhysRevD.101.023007} {\bibfield
  {journal} {\bibinfo  {journal} {Phys. Rev. D}\ }\textbf {\bibinfo {volume}
  {101}},\ \bibinfo {pages} {023007} (\bibinfo {year} {2020})},\ \Eprint
  {https://arxiv.org/abs/1911.09772} {arXiv:1911.09772 [hep-ph]} \BibitemShut
  {NoStop}%
\bibitem [{\citenamefont {Roggero}(2021{\natexlab{a}})}]{PhysRevD.104.103016}%
  \BibitemOpen
  \bibfield  {author} {\bibinfo {author} {\bibfnamefont {A.}~\bibnamefont
  {Roggero}},\ }\bibfield  {title} {\bibinfo {title} {Entanglement and
  many-body effects in collective neutrino oscillations},\ }\href
  {https://doi.org/10.1103/PhysRevD.104.103016} {\bibfield  {journal} {\bibinfo
   {journal} {Phys. Rev. D}\ }\textbf {\bibinfo {volume} {104}},\ \bibinfo
  {pages} {103016} (\bibinfo {year} {2021}{\natexlab{a}})}\BibitemShut
  {NoStop}%
\bibitem [{\citenamefont {Martin}\ \emph {et~al.}(2021)\citenamefont {Martin},
  \citenamefont {Carlson}, \citenamefont {Cirigliano},\ and\ \citenamefont
  {Duan}}]{Martin:2021xyl}%
  \BibitemOpen
  \bibfield  {author} {\bibinfo {author} {\bibfnamefont {J.~D.}\ \bibnamefont
  {Martin}}, \bibinfo {author} {\bibfnamefont {J.}~\bibnamefont {Carlson}},
  \bibinfo {author} {\bibfnamefont {V.}~\bibnamefont {Cirigliano}},\ and\
  \bibinfo {author} {\bibfnamefont {H.}~\bibnamefont {Duan}},\ }\bibfield
  {title} {\bibinfo {title} {{Fast flavor oscillations in dense neutrino media
  with collisions}},\ }\href {https://doi.org/10.1103/PhysRevD.103.063001}
  {\bibfield  {journal} {\bibinfo  {journal} {Phys. Rev. D}\ }\textbf {\bibinfo
  {volume} {103}},\ \bibinfo {pages} {063001} (\bibinfo {year} {2021})},\
  \Eprint {https://arxiv.org/abs/2101.01278} {arXiv:2101.01278 [hep-ph]}
  \BibitemShut {NoStop}%
\bibitem [{\citenamefont {Roggero}\ \emph {et~al.}(2022)\citenamefont
  {Roggero}, \citenamefont {Rrapaj},\ and\ \citenamefont
  {Xiong}}]{https://doi.org/10.48550/arxiv.2203.02783}%
  \BibitemOpen
  \bibfield  {author} {\bibinfo {author} {\bibfnamefont {A.}~\bibnamefont
  {Roggero}}, \bibinfo {author} {\bibfnamefont {E.}~\bibnamefont {Rrapaj}},\
  and\ \bibinfo {author} {\bibfnamefont {Z.}~\bibnamefont {Xiong}},\ }\href
  {https://doi.org/10.48550/ARXIV.2203.02783} {\bibinfo {title} {Entanglement
  and correlations in fast collective neutrino flavor oscillations}} (\bibinfo
  {year} {2022})\BibitemShut {NoStop}%
\bibitem [{\citenamefont {Illa}\ and\ \citenamefont
  {Savage}(2023)}]{Illa:2022zgu}%
  \BibitemOpen
  \bibfield  {author} {\bibinfo {author} {\bibfnamefont {M.}~\bibnamefont
  {Illa}}\ and\ \bibinfo {author} {\bibfnamefont {M.~J.}\ \bibnamefont
  {Savage}},\ }\bibfield  {title} {\bibinfo {title} {{Multi-Neutrino
  Entanglement and Correlations in Dense Neutrino Systems}},\ }\href
  {https://doi.org/10.1103/PhysRevLett.130.221003} {\bibfield  {journal}
  {\bibinfo  {journal} {Phys. Rev. Lett.}\ }\textbf {\bibinfo {volume} {130}},\
  \bibinfo {pages} {221003} (\bibinfo {year} {2023})},\ \Eprint
  {https://arxiv.org/abs/2210.08656} {arXiv:2210.08656 [nucl-th]} \BibitemShut
  {NoStop}%
\bibitem [{\citenamefont {Martin}\ \emph {et~al.}(2023)\citenamefont {Martin},
  \citenamefont {Roggero}, \citenamefont {Duan},\ and\ \citenamefont
  {Carlson}}]{Martin:2023ljq}%
  \BibitemOpen
  \bibfield  {author} {\bibinfo {author} {\bibfnamefont {J.~D.}\ \bibnamefont
  {Martin}}, \bibinfo {author} {\bibfnamefont {A.}~\bibnamefont {Roggero}},
  \bibinfo {author} {\bibfnamefont {H.}~\bibnamefont {Duan}},\ and\ \bibinfo
  {author} {\bibfnamefont {J.}~\bibnamefont {Carlson}},\ }\href@noop {}
  {\bibinfo {title} {{Many-body neutrino flavor entanglement in a simple
  dynamic model}}} (\bibinfo {year} {2023}),\ \Eprint
  {https://arxiv.org/abs/2301.07049} {arXiv:2301.07049 [hep-ph]} \BibitemShut
  {NoStop}%
\bibitem [{\citenamefont {Hall}\ \emph {et~al.}(2021)\citenamefont {Hall},
  \citenamefont {Roggero}, \citenamefont {Baroni},\ and\ \citenamefont
  {Carlson}}]{PhysRevD.104.063009}%
  \BibitemOpen
  \bibfield  {author} {\bibinfo {author} {\bibfnamefont {B.}~\bibnamefont
  {Hall}}, \bibinfo {author} {\bibfnamefont {A.}~\bibnamefont {Roggero}},
  \bibinfo {author} {\bibfnamefont {A.}~\bibnamefont {Baroni}},\ and\ \bibinfo
  {author} {\bibfnamefont {J.}~\bibnamefont {Carlson}},\ }\bibfield  {title}
  {\bibinfo {title} {Simulation of collective neutrino oscillations on a
  quantum computer},\ }\href {https://doi.org/10.1103/PhysRevD.104.063009}
  {\bibfield  {journal} {\bibinfo  {journal} {Phys. Rev. D}\ }\textbf {\bibinfo
  {volume} {104}},\ \bibinfo {pages} {063009} (\bibinfo {year}
  {2021})}\BibitemShut {NoStop}%
\bibitem [{\citenamefont {Roggero}(2021{\natexlab{b}})}]{PhysRevD.104.123023}%
  \BibitemOpen
  \bibfield  {author} {\bibinfo {author} {\bibfnamefont {A.}~\bibnamefont
  {Roggero}},\ }\bibfield  {title} {\bibinfo {title} {Dynamical phase
  transitions in models of collective neutrino oscillations},\ }\href
  {https://doi.org/10.1103/PhysRevD.104.123023} {\bibfield  {journal} {\bibinfo
   {journal} {Phys. Rev. D}\ }\textbf {\bibinfo {volume} {104}},\ \bibinfo
  {pages} {123023} (\bibinfo {year} {2021}{\natexlab{b}})}\BibitemShut
  {NoStop}%
\bibitem [{\citenamefont {Martin}\ \emph {et~al.}(2022)\citenamefont {Martin},
  \citenamefont {Roggero}, \citenamefont {Duan}, \citenamefont {Carlson},\ and\
  \citenamefont {Cirigliano}}]{PhysRevD.105.083020}%
  \BibitemOpen
  \bibfield  {author} {\bibinfo {author} {\bibfnamefont {J.~D.}\ \bibnamefont
  {Martin}}, \bibinfo {author} {\bibfnamefont {A.}~\bibnamefont {Roggero}},
  \bibinfo {author} {\bibfnamefont {H.}~\bibnamefont {Duan}}, \bibinfo {author}
  {\bibfnamefont {J.}~\bibnamefont {Carlson}},\ and\ \bibinfo {author}
  {\bibfnamefont {V.}~\bibnamefont {Cirigliano}},\ }\bibfield  {title}
  {\bibinfo {title} {Classical and quantum evolution in a simple coherent
  neutrino problem},\ }\href {https://doi.org/10.1103/PhysRevD.105.083020}
  {\bibfield  {journal} {\bibinfo  {journal} {Phys. Rev. D}\ }\textbf {\bibinfo
  {volume} {105}},\ \bibinfo {pages} {083020} (\bibinfo {year}
  {2022})}\BibitemShut {NoStop}%
\bibitem [{\citenamefont {Yeter-Aydeniz}\ \emph {et~al.}(2022)\citenamefont
  {Yeter-Aydeniz}, \citenamefont {Bangar}, \citenamefont {Siopsis},\ and\
  \citenamefont {Pooser}}]{Yeter-Aydeniz:2021olz}%
  \BibitemOpen
  \bibfield  {author} {\bibinfo {author} {\bibfnamefont {K.}~\bibnamefont
  {Yeter-Aydeniz}}, \bibinfo {author} {\bibfnamefont {S.}~\bibnamefont
  {Bangar}}, \bibinfo {author} {\bibfnamefont {G.}~\bibnamefont {Siopsis}},\
  and\ \bibinfo {author} {\bibfnamefont {R.~C.}\ \bibnamefont {Pooser}},\
  }\bibfield  {title} {\bibinfo {title} {{Collective neutrino oscillations on a
  quantum computer}},\ }\href {https://doi.org/10.1007/s11128-021-03348-x}
  {\bibfield  {journal} {\bibinfo  {journal} {Quant. Inf. Proc.}\ }\textbf
  {\bibinfo {volume} {21}},\ \bibinfo {pages} {84} (\bibinfo {year} {2022})},\
  \Eprint {https://arxiv.org/abs/2104.03273} {arXiv:2104.03273 [quant-ph]}
  \BibitemShut {NoStop}%
\bibitem [{\citenamefont {Amitrano}\ \emph {et~al.}(2023)\citenamefont
  {Amitrano}, \citenamefont {Roggero}, \citenamefont {Luchi}, \citenamefont
  {Turro}, \citenamefont {Vespucci},\ and\ \citenamefont
  {Pederiva}}]{Amitrano:2022yyn}%
  \BibitemOpen
  \bibfield  {author} {\bibinfo {author} {\bibfnamefont {V.}~\bibnamefont
  {Amitrano}}, \bibinfo {author} {\bibfnamefont {A.}~\bibnamefont {Roggero}},
  \bibinfo {author} {\bibfnamefont {P.}~\bibnamefont {Luchi}}, \bibinfo
  {author} {\bibfnamefont {F.}~\bibnamefont {Turro}}, \bibinfo {author}
  {\bibfnamefont {L.}~\bibnamefont {Vespucci}},\ and\ \bibinfo {author}
  {\bibfnamefont {F.}~\bibnamefont {Pederiva}},\ }\bibfield  {title} {\bibinfo
  {title} {{Trapped-ion quantum simulation of collective neutrino
  oscillations}},\ }\href {https://doi.org/10.1103/PhysRevD.107.023007}
  {\bibfield  {journal} {\bibinfo  {journal} {Phys. Rev. D}\ }\textbf {\bibinfo
  {volume} {107}},\ \bibinfo {pages} {023007} (\bibinfo {year} {2023})},\
  \Eprint {https://arxiv.org/abs/2207.03189} {arXiv:2207.03189 [quant-ph]}
  \BibitemShut {NoStop}%
\bibitem [{\citenamefont {Childs}\ \emph {et~al.}(2018)\citenamefont {Childs},
  \citenamefont {Maslov}, \citenamefont {Nam}, \citenamefont {Ross},\ and\
  \citenamefont {Su}}]{Childs_2018}%
  \BibitemOpen
  \bibfield  {author} {\bibinfo {author} {\bibfnamefont {A.~M.}\ \bibnamefont
  {Childs}}, \bibinfo {author} {\bibfnamefont {D.}~\bibnamefont {Maslov}},
  \bibinfo {author} {\bibfnamefont {Y.}~\bibnamefont {Nam}}, \bibinfo {author}
  {\bibfnamefont {N.~J.}\ \bibnamefont {Ross}},\ and\ \bibinfo {author}
  {\bibfnamefont {Y.}~\bibnamefont {Su}},\ }\bibfield  {title} {\bibinfo
  {title} {Toward the first quantum simulation with quantum speedup},\ }\href
  {https://doi.org/10.1073/pnas.1801723115} {\bibfield  {journal} {\bibinfo
  {journal} {Proceedings of the National Academy of Sciences}\ }\textbf
  {\bibinfo {volume} {115}},\ \bibinfo {pages} {9456} (\bibinfo {year}
  {2018})}\BibitemShut {NoStop}%
\bibitem [{\citenamefont {Childs}\ \emph {et~al.}(2021)\citenamefont {Childs},
  \citenamefont {Su}, \citenamefont {Tran}, \citenamefont {Wiebe},\ and\
  \citenamefont {Zhu}}]{Childs_2021}%
  \BibitemOpen
  \bibfield  {author} {\bibinfo {author} {\bibfnamefont {A.~M.}\ \bibnamefont
  {Childs}}, \bibinfo {author} {\bibfnamefont {Y.}~\bibnamefont {Su}}, \bibinfo
  {author} {\bibfnamefont {M.~C.}\ \bibnamefont {Tran}}, \bibinfo {author}
  {\bibfnamefont {N.}~\bibnamefont {Wiebe}},\ and\ \bibinfo {author}
  {\bibfnamefont {S.}~\bibnamefont {Zhu}},\ }\bibfield  {title} {\bibinfo
  {title} {Theory of trotter error with commutator scaling},\ }\bibfield
  {journal} {\bibinfo  {journal} {Physical Review X}\ }\textbf {\bibinfo
  {volume} {11}},\ \href {https://doi.org/10.1103/physrevx.11.011020}
  {10.1103/physrevx.11.011020} (\bibinfo {year} {2021})\BibitemShut {NoStop}%
\bibitem [{\citenamefont {Gonzalez-Raya}\ \emph {et~al.}(2021)\citenamefont
  {Gonzalez-Raya}, \citenamefont {Asensio-Perea}, \citenamefont {Martin},
  \citenamefont {C\'eleri}, \citenamefont {Sanz}, \citenamefont {Lougovski},\
  and\ \citenamefont {Dumitrescu}}]{PRXQuantum.2.020328}%
  \BibitemOpen
  \bibfield  {author} {\bibinfo {author} {\bibfnamefont {T.}~\bibnamefont
  {Gonzalez-Raya}}, \bibinfo {author} {\bibfnamefont {R.}~\bibnamefont
  {Asensio-Perea}}, \bibinfo {author} {\bibfnamefont {A.}~\bibnamefont
  {Martin}}, \bibinfo {author} {\bibfnamefont {L.~C.}\ \bibnamefont
  {C\'eleri}}, \bibinfo {author} {\bibfnamefont {M.}~\bibnamefont {Sanz}},
  \bibinfo {author} {\bibfnamefont {P.}~\bibnamefont {Lougovski}},\ and\
  \bibinfo {author} {\bibfnamefont {E.~F.}\ \bibnamefont {Dumitrescu}},\
  }\bibfield  {title} {\bibinfo {title} {Digital-analog quantum simulations
  using the cross-resonance effect},\ }\href
  {https://doi.org/10.1103/PRXQuantum.2.020328} {\bibfield  {journal} {\bibinfo
   {journal} {PRX Quantum}\ }\textbf {\bibinfo {volume} {2}},\ \bibinfo {pages}
  {020328} (\bibinfo {year} {2021})}\BibitemShut {NoStop}%
\bibitem [{\citenamefont {Bermudez}\ \emph {et~al.}(2017)\citenamefont
  {Bermudez}, \citenamefont {Tagliacozzo}, \citenamefont {Sierra},\ and\
  \citenamefont {Richerme}}]{PhysRevB.95.024431}%
  \BibitemOpen
  \bibfield  {author} {\bibinfo {author} {\bibfnamefont {A.}~\bibnamefont
  {Bermudez}}, \bibinfo {author} {\bibfnamefont {L.}~\bibnamefont
  {Tagliacozzo}}, \bibinfo {author} {\bibfnamefont {G.}~\bibnamefont
  {Sierra}},\ and\ \bibinfo {author} {\bibfnamefont {P.}~\bibnamefont
  {Richerme}},\ }\bibfield  {title} {\bibinfo {title} {Long-range heisenberg
  models in quasiperiodically driven crystals of trapped ions},\ }\href
  {https://doi.org/10.1103/PhysRevB.95.024431} {\bibfield  {journal} {\bibinfo
  {journal} {Phys. Rev. B}\ }\textbf {\bibinfo {volume} {95}},\ \bibinfo
  {pages} {024431} (\bibinfo {year} {2017})}\BibitemShut {NoStop}%
\bibitem [{\citenamefont {Scholl}\ \emph {et~al.}(2022)\citenamefont {Scholl},
  \citenamefont {Williams}, \citenamefont {Bornet}, \citenamefont {Wallner},
  \citenamefont {Barredo}, \citenamefont {Henriet}, \citenamefont {Signoles},
  \citenamefont {Hainaut}, \citenamefont {Franz}, \citenamefont {Geier},
  \citenamefont {Tebben}, \citenamefont {Salzinger}, \citenamefont {Z\"urn},
  \citenamefont {Lahaye}, \citenamefont {Weidem\"uller},\ and\ \citenamefont
  {Browaeys}}]{PRXQuantum.3.020303}%
  \BibitemOpen
  \bibfield  {author} {\bibinfo {author} {\bibfnamefont {P.}~\bibnamefont
  {Scholl}}, \bibinfo {author} {\bibfnamefont {H.~J.}\ \bibnamefont
  {Williams}}, \bibinfo {author} {\bibfnamefont {G.}~\bibnamefont {Bornet}},
  \bibinfo {author} {\bibfnamefont {F.}~\bibnamefont {Wallner}}, \bibinfo
  {author} {\bibfnamefont {D.}~\bibnamefont {Barredo}}, \bibinfo {author}
  {\bibfnamefont {L.}~\bibnamefont {Henriet}}, \bibinfo {author} {\bibfnamefont
  {A.}~\bibnamefont {Signoles}}, \bibinfo {author} {\bibfnamefont
  {C.}~\bibnamefont {Hainaut}}, \bibinfo {author} {\bibfnamefont
  {T.}~\bibnamefont {Franz}}, \bibinfo {author} {\bibfnamefont
  {S.}~\bibnamefont {Geier}}, \bibinfo {author} {\bibfnamefont
  {A.}~\bibnamefont {Tebben}}, \bibinfo {author} {\bibfnamefont
  {A.}~\bibnamefont {Salzinger}}, \bibinfo {author} {\bibfnamefont
  {G.}~\bibnamefont {Z\"urn}}, \bibinfo {author} {\bibfnamefont
  {T.}~\bibnamefont {Lahaye}}, \bibinfo {author} {\bibfnamefont
  {M.}~\bibnamefont {Weidem\"uller}},\ and\ \bibinfo {author} {\bibfnamefont
  {A.}~\bibnamefont {Browaeys}},\ }\bibfield  {title} {\bibinfo {title}
  {Microwave engineering of programmable $xxz$ hamiltonians in arrays of
  rydberg atoms},\ }\href {https://doi.org/10.1103/PRXQuantum.3.020303}
  {\bibfield  {journal} {\bibinfo  {journal} {PRX Quantum}\ }\textbf {\bibinfo
  {volume} {3}},\ \bibinfo {pages} {020303} (\bibinfo {year}
  {2022})}\BibitemShut {NoStop}%
\bibitem [{\citenamefont {Mádi}\ \emph {et~al.}(1997)\citenamefont {Mádi},
  \citenamefont {Brutscher}, \citenamefont {Schulte-Herbrüggen}, \citenamefont
  {Brüschweiler},\ and\ \citenamefont {Ernst}}]{MADI1997300}%
  \BibitemOpen
  \bibfield  {author} {\bibinfo {author} {\bibfnamefont {Z.}~\bibnamefont
  {Mádi}}, \bibinfo {author} {\bibfnamefont {B.}~\bibnamefont {Brutscher}},
  \bibinfo {author} {\bibfnamefont {T.}~\bibnamefont {Schulte-Herbrüggen}},
  \bibinfo {author} {\bibfnamefont {R.}~\bibnamefont {Brüschweiler}},\ and\
  \bibinfo {author} {\bibfnamefont {R.}~\bibnamefont {Ernst}},\ }\bibfield
  {title} {\bibinfo {title} {Time-resolved observation of spin waves in a
  linear chain of nuclear spins},\ }\href
  {https://doi.org/https://doi.org/10.1016/S0009-2614(97)00194-2} {\bibfield
  {journal} {\bibinfo  {journal} {Chemical Physics Letters}\ }\textbf {\bibinfo
  {volume} {268}},\ \bibinfo {pages} {300} (\bibinfo {year}
  {1997})}\BibitemShut {NoStop}%
\bibitem [{\citenamefont {S\'anchez}\ \emph {et~al.}(2020)\citenamefont
  {S\'anchez}, \citenamefont {Chattah}, \citenamefont {Wei}, \citenamefont
  {Buljubasich}, \citenamefont {Cappellaro},\ and\ \citenamefont
  {Pastawski}}]{PhysRevLett.124.030601}%
  \BibitemOpen
  \bibfield  {author} {\bibinfo {author} {\bibfnamefont {C.~M.}\ \bibnamefont
  {S\'anchez}}, \bibinfo {author} {\bibfnamefont {A.~K.}\ \bibnamefont
  {Chattah}}, \bibinfo {author} {\bibfnamefont {K.~X.}\ \bibnamefont {Wei}},
  \bibinfo {author} {\bibfnamefont {L.}~\bibnamefont {Buljubasich}}, \bibinfo
  {author} {\bibfnamefont {P.}~\bibnamefont {Cappellaro}},\ and\ \bibinfo
  {author} {\bibfnamefont {H.~M.}\ \bibnamefont {Pastawski}},\ }\bibfield
  {title} {\bibinfo {title} {Perturbation independent decay of the loschmidt
  echo in a many-body system},\ }\href
  {https://doi.org/10.1103/PhysRevLett.124.030601} {\bibfield  {journal}
  {\bibinfo  {journal} {Phys. Rev. Lett.}\ }\textbf {\bibinfo {volume} {124}},\
  \bibinfo {pages} {030601} (\bibinfo {year} {2020})}\BibitemShut {NoStop}%
\bibitem [{\citenamefont {Rost}\ \emph {et~al.}(2021)\citenamefont {Rost},
  \citenamefont {Re}, \citenamefont {Earnest}, \citenamefont {Kemper},
  \citenamefont {Jones},\ and\ \citenamefont
  {Freericks}}]{rost2021demonstrating}%
  \BibitemOpen
  \bibfield  {author} {\bibinfo {author} {\bibfnamefont {B.}~\bibnamefont
  {Rost}}, \bibinfo {author} {\bibfnamefont {L.~D.}\ \bibnamefont {Re}},
  \bibinfo {author} {\bibfnamefont {N.}~\bibnamefont {Earnest}}, \bibinfo
  {author} {\bibfnamefont {A.~F.}\ \bibnamefont {Kemper}}, \bibinfo {author}
  {\bibfnamefont {B.}~\bibnamefont {Jones}},\ and\ \bibinfo {author}
  {\bibfnamefont {J.~K.}\ \bibnamefont {Freericks}},\ }\href@noop {} {\bibinfo
  {title} {Demonstrating robust simulation of driven-dissipative problems on
  near-term quantum computers}} (\bibinfo {year} {2021}),\ \Eprint
  {https://arxiv.org/abs/2108.01183} {arXiv:2108.01183 [quant-ph]} \BibitemShut
  {NoStop}%
\bibitem [{\citenamefont {Re}\ \emph {et~al.}(2022)\citenamefont {Re},
  \citenamefont {Rost}, \citenamefont {Foss-Feig}, \citenamefont {Kemper},\
  and\ \citenamefont {Freericks}}]{delre2022robust}%
  \BibitemOpen
  \bibfield  {author} {\bibinfo {author} {\bibfnamefont {L.~D.}\ \bibnamefont
  {Re}}, \bibinfo {author} {\bibfnamefont {B.}~\bibnamefont {Rost}}, \bibinfo
  {author} {\bibfnamefont {M.}~\bibnamefont {Foss-Feig}}, \bibinfo {author}
  {\bibfnamefont {A.~F.}\ \bibnamefont {Kemper}},\ and\ \bibinfo {author}
  {\bibfnamefont {J.~K.}\ \bibnamefont {Freericks}},\ }\href@noop {} {\bibinfo
  {title} {Robust measurements of n-point correlation functions of
  driven-dissipative quantum systems on a digital quantum computer}} (\bibinfo
  {year} {2022}),\ \Eprint {https://arxiv.org/abs/2204.12400} {arXiv:2204.12400
  [quant-ph]} \BibitemShut {NoStop}%
\bibitem [{\citenamefont {Ebadi}\ \emph {et~al.}(2021)\citenamefont {Ebadi},
  \citenamefont {Wang}, \citenamefont {Levine}, \citenamefont {Keesling},
  \citenamefont {Semeghini}, \citenamefont {Omran}, \citenamefont {Bluvstein},
  \citenamefont {Samajdar}, \citenamefont {Pichler}, \citenamefont {Ho} \emph
  {et~al.}}]{ebadi2021quantum}%
  \BibitemOpen
  \bibfield  {author} {\bibinfo {author} {\bibfnamefont {S.}~\bibnamefont
  {Ebadi}}, \bibinfo {author} {\bibfnamefont {T.~T.}\ \bibnamefont {Wang}},
  \bibinfo {author} {\bibfnamefont {H.}~\bibnamefont {Levine}}, \bibinfo
  {author} {\bibfnamefont {A.}~\bibnamefont {Keesling}}, \bibinfo {author}
  {\bibfnamefont {G.}~\bibnamefont {Semeghini}}, \bibinfo {author}
  {\bibfnamefont {A.}~\bibnamefont {Omran}}, \bibinfo {author} {\bibfnamefont
  {D.}~\bibnamefont {Bluvstein}}, \bibinfo {author} {\bibfnamefont
  {R.}~\bibnamefont {Samajdar}}, \bibinfo {author} {\bibfnamefont
  {H.}~\bibnamefont {Pichler}}, \bibinfo {author} {\bibfnamefont {W.~W.}\
  \bibnamefont {Ho}}, \emph {et~al.},\ }\bibfield  {title} {\bibinfo {title}
  {Quantum phases of matter on a 256-atom programmable quantum simulator},\
  }\href@noop {} {\bibfield  {journal} {\bibinfo  {journal} {Nature}\ }\textbf
  {\bibinfo {volume} {595}},\ \bibinfo {pages} {227} (\bibinfo {year}
  {2021})}\BibitemShut {NoStop}%
\bibitem [{\citenamefont {Semeghini}\ \emph {et~al.}(2021)\citenamefont
  {Semeghini}, \citenamefont {Levine}, \citenamefont {Keesling}, \citenamefont
  {Ebadi}, \citenamefont {Wang}, \citenamefont {Bluvstein}, \citenamefont
  {Verresen}, \citenamefont {Pichler}, \citenamefont {Kalinowski},
  \citenamefont {Samajdar}, \citenamefont {Omran}, \citenamefont {Sachdev},
  \citenamefont {Vishwanath}, \citenamefont {Greiner}, \citenamefont
  {Vuletić},\ and\ \citenamefont {Lukin}}]{doi:10.1126/science.abi8794}%
  \BibitemOpen
  \bibfield  {author} {\bibinfo {author} {\bibfnamefont {G.}~\bibnamefont
  {Semeghini}}, \bibinfo {author} {\bibfnamefont {H.}~\bibnamefont {Levine}},
  \bibinfo {author} {\bibfnamefont {A.}~\bibnamefont {Keesling}}, \bibinfo
  {author} {\bibfnamefont {S.}~\bibnamefont {Ebadi}}, \bibinfo {author}
  {\bibfnamefont {T.~T.}\ \bibnamefont {Wang}}, \bibinfo {author}
  {\bibfnamefont {D.}~\bibnamefont {Bluvstein}}, \bibinfo {author}
  {\bibfnamefont {R.}~\bibnamefont {Verresen}}, \bibinfo {author}
  {\bibfnamefont {H.}~\bibnamefont {Pichler}}, \bibinfo {author} {\bibfnamefont
  {M.}~\bibnamefont {Kalinowski}}, \bibinfo {author} {\bibfnamefont
  {R.}~\bibnamefont {Samajdar}}, \bibinfo {author} {\bibfnamefont
  {A.}~\bibnamefont {Omran}}, \bibinfo {author} {\bibfnamefont
  {S.}~\bibnamefont {Sachdev}}, \bibinfo {author} {\bibfnamefont
  {A.}~\bibnamefont {Vishwanath}}, \bibinfo {author} {\bibfnamefont
  {M.}~\bibnamefont {Greiner}}, \bibinfo {author} {\bibfnamefont
  {V.}~\bibnamefont {Vuletić}},\ and\ \bibinfo {author} {\bibfnamefont
  {M.~D.}\ \bibnamefont {Lukin}},\ }\bibfield  {title} {\bibinfo {title}
  {Probing topological spin liquids on a programmable quantum simulator},\
  }\href {https://doi.org/10.1126/science.abi8794} {\bibfield  {journal}
  {\bibinfo  {journal} {Science}\ }\textbf {\bibinfo {volume} {374}},\ \bibinfo
  {pages} {1242} (\bibinfo {year} {2021})},\ \Eprint
  {https://arxiv.org/abs/https://www.science.org/doi/pdf/10.1126/science.abi8794}
  {https://www.science.org/doi/pdf/10.1126/science.abi8794} \BibitemShut
  {NoStop}%
\bibitem [{\citenamefont {Bluvstein}\ \emph {et~al.}(2022)\citenamefont
  {Bluvstein}, \citenamefont {Levine}, \citenamefont {Semeghini}, \citenamefont
  {Wang}, \citenamefont {Ebadi}, \citenamefont {Kalinowski}, \citenamefont
  {Keesling}, \citenamefont {Maskara}, \citenamefont {Pichler}, \citenamefont
  {Greiner} \emph {et~al.}}]{bluvstein2022quantum}%
  \BibitemOpen
  \bibfield  {author} {\bibinfo {author} {\bibfnamefont {D.}~\bibnamefont
  {Bluvstein}}, \bibinfo {author} {\bibfnamefont {H.}~\bibnamefont {Levine}},
  \bibinfo {author} {\bibfnamefont {G.}~\bibnamefont {Semeghini}}, \bibinfo
  {author} {\bibfnamefont {T.~T.}\ \bibnamefont {Wang}}, \bibinfo {author}
  {\bibfnamefont {S.}~\bibnamefont {Ebadi}}, \bibinfo {author} {\bibfnamefont
  {M.}~\bibnamefont {Kalinowski}}, \bibinfo {author} {\bibfnamefont
  {A.}~\bibnamefont {Keesling}}, \bibinfo {author} {\bibfnamefont
  {N.}~\bibnamefont {Maskara}}, \bibinfo {author} {\bibfnamefont
  {H.}~\bibnamefont {Pichler}}, \bibinfo {author} {\bibfnamefont
  {M.}~\bibnamefont {Greiner}}, \emph {et~al.},\ }\bibfield  {title} {\bibinfo
  {title} {A quantum processor based on coherent transport of entangled atom
  arrays},\ }\href@noop {} {\bibfield  {journal} {\bibinfo  {journal} {Nature}\
  }\textbf {\bibinfo {volume} {604}},\ \bibinfo {pages} {451} (\bibinfo {year}
  {2022})}\BibitemShut {NoStop}%
\bibitem [{\citenamefont {Porras}\ and\ \citenamefont
  {Cirac}(2004)}]{PhysRevLett.92.207901}%
  \BibitemOpen
  \bibfield  {author} {\bibinfo {author} {\bibfnamefont {D.}~\bibnamefont
  {Porras}}\ and\ \bibinfo {author} {\bibfnamefont {J.~I.}\ \bibnamefont
  {Cirac}},\ }\bibfield  {title} {\bibinfo {title} {Effective quantum spin
  systems with trapped ions},\ }\href
  {https://doi.org/10.1103/PhysRevLett.92.207901} {\bibfield  {journal}
  {\bibinfo  {journal} {Phys. Rev. Lett.}\ }\textbf {\bibinfo {volume} {92}},\
  \bibinfo {pages} {207901} (\bibinfo {year} {2004})}\BibitemShut {NoStop}%
\bibitem [{\citenamefont {Gra{\ss}}\ and\ \citenamefont
  {Lewenstein}(2014)}]{grass2014trapped}%
  \BibitemOpen
  \bibfield  {author} {\bibinfo {author} {\bibfnamefont {T.}~\bibnamefont
  {Gra{\ss}}}\ and\ \bibinfo {author} {\bibfnamefont {M.}~\bibnamefont
  {Lewenstein}},\ }\bibfield  {title} {\bibinfo {title} {Trapped-ion quantum
  simulation of tunable-range heisenberg chains},\ }\href@noop {} {\bibfield
  {journal} {\bibinfo  {journal} {EPJ Quantum Technology}\ }\textbf {\bibinfo
  {volume} {1}},\ \bibinfo {pages} {1} (\bibinfo {year} {2014})}\BibitemShut
  {NoStop}%
\bibitem [{\citenamefont {Monroe}\ \emph {et~al.}(2021)\citenamefont {Monroe},
  \citenamefont {Campbell}, \citenamefont {Duan}, \citenamefont {Gong},
  \citenamefont {Gorshkov}, \citenamefont {Hess}, \citenamefont {Islam},
  \citenamefont {Kim}, \citenamefont {Linke}, \citenamefont {Pagano},
  \citenamefont {Richerme}, \citenamefont {Senko},\ and\ \citenamefont
  {Yao}}]{RevModPhys.93.025001}%
  \BibitemOpen
  \bibfield  {author} {\bibinfo {author} {\bibfnamefont {C.}~\bibnamefont
  {Monroe}}, \bibinfo {author} {\bibfnamefont {W.~C.}\ \bibnamefont
  {Campbell}}, \bibinfo {author} {\bibfnamefont {L.-M.}\ \bibnamefont {Duan}},
  \bibinfo {author} {\bibfnamefont {Z.-X.}\ \bibnamefont {Gong}}, \bibinfo
  {author} {\bibfnamefont {A.~V.}\ \bibnamefont {Gorshkov}}, \bibinfo {author}
  {\bibfnamefont {P.~W.}\ \bibnamefont {Hess}}, \bibinfo {author}
  {\bibfnamefont {R.}~\bibnamefont {Islam}}, \bibinfo {author} {\bibfnamefont
  {K.}~\bibnamefont {Kim}}, \bibinfo {author} {\bibfnamefont {N.~M.}\
  \bibnamefont {Linke}}, \bibinfo {author} {\bibfnamefont {G.}~\bibnamefont
  {Pagano}}, \bibinfo {author} {\bibfnamefont {P.}~\bibnamefont {Richerme}},
  \bibinfo {author} {\bibfnamefont {C.}~\bibnamefont {Senko}},\ and\ \bibinfo
  {author} {\bibfnamefont {N.~Y.}\ \bibnamefont {Yao}},\ }\bibfield  {title}
  {\bibinfo {title} {Programmable quantum simulations of spin systems with
  trapped ions},\ }\href {https://doi.org/10.1103/RevModPhys.93.025001}
  {\bibfield  {journal} {\bibinfo  {journal} {Rev. Mod. Phys.}\ }\textbf
  {\bibinfo {volume} {93}},\ \bibinfo {pages} {025001} (\bibinfo {year}
  {2021})}\BibitemShut {NoStop}%
\bibitem [{\citenamefont {Johnson}\ \emph {et~al.}(2011)\citenamefont
  {Johnson}, \citenamefont {Amin}, \citenamefont {Gildert}, \citenamefont
  {Lanting}, \citenamefont {Hamze}, \citenamefont {Dickson}, \citenamefont
  {Harris}, \citenamefont {Berkley}, \citenamefont {Johansson}, \citenamefont
  {Bunyk} \emph {et~al.}}]{johnson2011quantum}%
  \BibitemOpen
  \bibfield  {author} {\bibinfo {author} {\bibfnamefont {M.~W.}\ \bibnamefont
  {Johnson}}, \bibinfo {author} {\bibfnamefont {M.~H.}\ \bibnamefont {Amin}},
  \bibinfo {author} {\bibfnamefont {S.}~\bibnamefont {Gildert}}, \bibinfo
  {author} {\bibfnamefont {T.}~\bibnamefont {Lanting}}, \bibinfo {author}
  {\bibfnamefont {F.}~\bibnamefont {Hamze}}, \bibinfo {author} {\bibfnamefont
  {N.}~\bibnamefont {Dickson}}, \bibinfo {author} {\bibfnamefont
  {R.}~\bibnamefont {Harris}}, \bibinfo {author} {\bibfnamefont {A.~J.}\
  \bibnamefont {Berkley}}, \bibinfo {author} {\bibfnamefont {J.}~\bibnamefont
  {Johansson}}, \bibinfo {author} {\bibfnamefont {P.}~\bibnamefont {Bunyk}},
  \emph {et~al.},\ }\bibfield  {title} {\bibinfo {title} {Quantum annealing
  with manufactured spins},\ }\href@noop {} {\bibfield  {journal} {\bibinfo
  {journal} {Nature}\ }\textbf {\bibinfo {volume} {473}},\ \bibinfo {pages}
  {194} (\bibinfo {year} {2011})}\BibitemShut {NoStop}%
\bibitem [{\citenamefont {Harris}\ \emph {et~al.}(2010)\citenamefont {Harris},
  \citenamefont {Johnson}, \citenamefont {Lanting}, \citenamefont {Berkley},
  \citenamefont {Johansson}, \citenamefont {Bunyk}, \citenamefont {Tolkacheva},
  \citenamefont {Ladizinsky}, \citenamefont {Ladizinsky}, \citenamefont {Oh},
  \citenamefont {Cioata}, \citenamefont {Perminov}, \citenamefont {Spear},
  \citenamefont {Enderud}, \citenamefont {Rich}, \citenamefont {Uchaikin},
  \citenamefont {Thom}, \citenamefont {Chapple}, \citenamefont {Wang},
  \citenamefont {Wilson}, \citenamefont {Amin}, \citenamefont {Dickson},
  \citenamefont {Karimi}, \citenamefont {Macready}, \citenamefont {Truncik},\
  and\ \citenamefont {Rose}}]{PhysRevB.82.024511}%
  \BibitemOpen
  \bibfield  {author} {\bibinfo {author} {\bibfnamefont {R.}~\bibnamefont
  {Harris}}, \bibinfo {author} {\bibfnamefont {M.~W.}\ \bibnamefont {Johnson}},
  \bibinfo {author} {\bibfnamefont {T.}~\bibnamefont {Lanting}}, \bibinfo
  {author} {\bibfnamefont {A.~J.}\ \bibnamefont {Berkley}}, \bibinfo {author}
  {\bibfnamefont {J.}~\bibnamefont {Johansson}}, \bibinfo {author}
  {\bibfnamefont {P.}~\bibnamefont {Bunyk}}, \bibinfo {author} {\bibfnamefont
  {E.}~\bibnamefont {Tolkacheva}}, \bibinfo {author} {\bibfnamefont
  {E.}~\bibnamefont {Ladizinsky}}, \bibinfo {author} {\bibfnamefont
  {N.}~\bibnamefont {Ladizinsky}}, \bibinfo {author} {\bibfnamefont
  {T.}~\bibnamefont {Oh}}, \bibinfo {author} {\bibfnamefont {F.}~\bibnamefont
  {Cioata}}, \bibinfo {author} {\bibfnamefont {I.}~\bibnamefont {Perminov}},
  \bibinfo {author} {\bibfnamefont {P.}~\bibnamefont {Spear}}, \bibinfo
  {author} {\bibfnamefont {C.}~\bibnamefont {Enderud}}, \bibinfo {author}
  {\bibfnamefont {C.}~\bibnamefont {Rich}}, \bibinfo {author} {\bibfnamefont
  {S.}~\bibnamefont {Uchaikin}}, \bibinfo {author} {\bibfnamefont {M.~C.}\
  \bibnamefont {Thom}}, \bibinfo {author} {\bibfnamefont {E.~M.}\ \bibnamefont
  {Chapple}}, \bibinfo {author} {\bibfnamefont {J.}~\bibnamefont {Wang}},
  \bibinfo {author} {\bibfnamefont {B.}~\bibnamefont {Wilson}}, \bibinfo
  {author} {\bibfnamefont {M.~H.~S.}\ \bibnamefont {Amin}}, \bibinfo {author}
  {\bibfnamefont {N.}~\bibnamefont {Dickson}}, \bibinfo {author} {\bibfnamefont
  {K.}~\bibnamefont {Karimi}}, \bibinfo {author} {\bibfnamefont
  {B.}~\bibnamefont {Macready}}, \bibinfo {author} {\bibfnamefont {C.~J.~S.}\
  \bibnamefont {Truncik}},\ and\ \bibinfo {author} {\bibfnamefont
  {G.}~\bibnamefont {Rose}},\ }\bibfield  {title} {\bibinfo {title}
  {Experimental investigation of an eight-qubit unit cell in a superconducting
  optimization processor},\ }\href {https://doi.org/10.1103/PhysRevB.82.024511}
  {\bibfield  {journal} {\bibinfo  {journal} {Phys. Rev. B}\ }\textbf {\bibinfo
  {volume} {82}},\ \bibinfo {pages} {024511} (\bibinfo {year}
  {2010})}\BibitemShut {NoStop}%
\bibitem [{\citenamefont {Bunyk}\ \emph {et~al.}(2014)\citenamefont {Bunyk},
  \citenamefont {Hoskinson}, \citenamefont {Johnson}, \citenamefont
  {Tolkacheva}, \citenamefont {Altomare}, \citenamefont {Berkley},
  \citenamefont {Harris}, \citenamefont {Hilton}, \citenamefont {Lanting},
  \citenamefont {Przybysz},\ and\ \citenamefont {Whittaker}}]{6802426}%
  \BibitemOpen
  \bibfield  {author} {\bibinfo {author} {\bibfnamefont {P.~I.}\ \bibnamefont
  {Bunyk}}, \bibinfo {author} {\bibfnamefont {E.~M.}\ \bibnamefont
  {Hoskinson}}, \bibinfo {author} {\bibfnamefont {M.~W.}\ \bibnamefont
  {Johnson}}, \bibinfo {author} {\bibfnamefont {E.}~\bibnamefont {Tolkacheva}},
  \bibinfo {author} {\bibfnamefont {F.}~\bibnamefont {Altomare}}, \bibinfo
  {author} {\bibfnamefont {A.~J.}\ \bibnamefont {Berkley}}, \bibinfo {author}
  {\bibfnamefont {R.}~\bibnamefont {Harris}}, \bibinfo {author} {\bibfnamefont
  {J.~P.}\ \bibnamefont {Hilton}}, \bibinfo {author} {\bibfnamefont
  {T.}~\bibnamefont {Lanting}}, \bibinfo {author} {\bibfnamefont {A.~J.}\
  \bibnamefont {Przybysz}},\ and\ \bibinfo {author} {\bibfnamefont
  {J.}~\bibnamefont {Whittaker}},\ }\bibfield  {title} {\bibinfo {title}
  {Architectural considerations in the design of a superconducting quantum
  annealing processor},\ }\href {https://doi.org/10.1109/TASC.2014.2318294}
  {\bibfield  {journal} {\bibinfo  {journal} {IEEE Transactions on Applied
  Superconductivity}\ }\textbf {\bibinfo {volume} {24}},\ \bibinfo {pages} {1}
  (\bibinfo {year} {2014})}\BibitemShut {NoStop}%
\bibitem [{\citenamefont {James}\ \emph {et~al.}(2019)\citenamefont {James},
  \citenamefont {Konik},\ and\ \citenamefont
  {Robinson}}]{PhysRevLett.122.130603}%
  \BibitemOpen
  \bibfield  {author} {\bibinfo {author} {\bibfnamefont {A.~J.~A.}\
  \bibnamefont {James}}, \bibinfo {author} {\bibfnamefont {R.~M.}\ \bibnamefont
  {Konik}},\ and\ \bibinfo {author} {\bibfnamefont {N.~J.}\ \bibnamefont
  {Robinson}},\ }\bibfield  {title} {\bibinfo {title} {Nonthermal states
  arising from confinement in one and two dimensions},\ }\href
  {https://doi.org/10.1103/PhysRevLett.122.130603} {\bibfield  {journal}
  {\bibinfo  {journal} {Phys. Rev. Lett.}\ }\textbf {\bibinfo {volume} {122}},\
  \bibinfo {pages} {130603} (\bibinfo {year} {2019})}\BibitemShut {NoStop}%
\bibitem [{\citenamefont {Kormos}\ \emph {et~al.}(2017)\citenamefont {Kormos},
  \citenamefont {Collura}, \citenamefont {Tak{\'a}cs},\ and\ \citenamefont
  {Calabrese}}]{kormos2017real}%
  \BibitemOpen
  \bibfield  {author} {\bibinfo {author} {\bibfnamefont {M.}~\bibnamefont
  {Kormos}}, \bibinfo {author} {\bibfnamefont {M.}~\bibnamefont {Collura}},
  \bibinfo {author} {\bibfnamefont {G.}~\bibnamefont {Tak{\'a}cs}},\ and\
  \bibinfo {author} {\bibfnamefont {P.}~\bibnamefont {Calabrese}},\ }\bibfield
  {title} {\bibinfo {title} {Real-time confinement following a quantum quench
  to a non-integrable model},\ }\href@noop {} {\bibfield  {journal} {\bibinfo
  {journal} {Nature Physics}\ }\textbf {\bibinfo {volume} {13}},\ \bibinfo
  {pages} {246} (\bibinfo {year} {2017})}\BibitemShut {NoStop}%
\bibitem [{\citenamefont {Richerme}\ \emph {et~al.}(2014)\citenamefont
  {Richerme}, \citenamefont {Gong}, \citenamefont {Lee}, \citenamefont {Senko},
  \citenamefont {Smith}, \citenamefont {Foss-Feig}, \citenamefont {Michalakis},
  \citenamefont {Gorshkov},\ and\ \citenamefont {Monroe}}]{Richerme_2014}%
  \BibitemOpen
  \bibfield  {author} {\bibinfo {author} {\bibfnamefont {P.}~\bibnamefont
  {Richerme}}, \bibinfo {author} {\bibfnamefont {Z.-X.}\ \bibnamefont {Gong}},
  \bibinfo {author} {\bibfnamefont {A.}~\bibnamefont {Lee}}, \bibinfo {author}
  {\bibfnamefont {C.}~\bibnamefont {Senko}}, \bibinfo {author} {\bibfnamefont
  {J.}~\bibnamefont {Smith}}, \bibinfo {author} {\bibfnamefont
  {M.}~\bibnamefont {Foss-Feig}}, \bibinfo {author} {\bibfnamefont
  {S.}~\bibnamefont {Michalakis}}, \bibinfo {author} {\bibfnamefont {A.~V.}\
  \bibnamefont {Gorshkov}},\ and\ \bibinfo {author} {\bibfnamefont
  {C.}~\bibnamefont {Monroe}},\ }\bibfield  {title} {\bibinfo {title}
  {Non-local propagation of correlations in quantum systems with long-range
  interactions},\ }\href {https://doi.org/10.1038/nature13450} {\bibfield
  {journal} {\bibinfo  {journal} {Nature}\ }\textbf {\bibinfo {volume} {511}},\
  \bibinfo {pages} {198} (\bibinfo {year} {2014})}\BibitemShut {NoStop}%
\bibitem [{\citenamefont {Jurcevic}\ \emph {et~al.}(2014)\citenamefont
  {Jurcevic}, \citenamefont {Lanyon}, \citenamefont {Hauke}, \citenamefont
  {Hempel}, \citenamefont {Zoller}, \citenamefont {Blatt},\ and\ \citenamefont
  {Roos}}]{Jurcevic_2014}%
  \BibitemOpen
  \bibfield  {author} {\bibinfo {author} {\bibfnamefont {P.}~\bibnamefont
  {Jurcevic}}, \bibinfo {author} {\bibfnamefont {B.~P.}\ \bibnamefont
  {Lanyon}}, \bibinfo {author} {\bibfnamefont {P.}~\bibnamefont {Hauke}},
  \bibinfo {author} {\bibfnamefont {C.}~\bibnamefont {Hempel}}, \bibinfo
  {author} {\bibfnamefont {P.}~\bibnamefont {Zoller}}, \bibinfo {author}
  {\bibfnamefont {R.}~\bibnamefont {Blatt}},\ and\ \bibinfo {author}
  {\bibfnamefont {C.~F.}\ \bibnamefont {Roos}},\ }\bibfield  {title} {\bibinfo
  {title} {Quasiparticle engineering and entanglement propagation in a quantum
  many-body system},\ }\href {https://doi.org/10.1038/nature13461} {\bibfield
  {journal} {\bibinfo  {journal} {Nature}\ }\textbf {\bibinfo {volume} {511}},\
  \bibinfo {pages} {202} (\bibinfo {year} {2014})}\BibitemShut {NoStop}%
\bibitem [{\citenamefont {Wall}\ \emph {et~al.}(2017)\citenamefont {Wall},
  \citenamefont {Safavi-Naini},\ and\ \citenamefont {Rey}}]{Wall_2017}%
  \BibitemOpen
  \bibfield  {author} {\bibinfo {author} {\bibfnamefont {M.~L.}\ \bibnamefont
  {Wall}}, \bibinfo {author} {\bibfnamefont {A.}~\bibnamefont {Safavi-Naini}},\
  and\ \bibinfo {author} {\bibfnamefont {A.~M.}\ \bibnamefont {Rey}},\
  }\bibfield  {title} {\bibinfo {title} {Boson-mediated quantum spin simulators
  in transverse fields: Xy model and spin-boson entanglement},\ }\bibfield
  {journal} {\bibinfo  {journal} {Physical Review A}\ }\textbf {\bibinfo
  {volume} {95}},\ \href {https://doi.org/10.1103/physreva.95.013602}
  {10.1103/physreva.95.013602} (\bibinfo {year} {2017})\BibitemShut {NoStop}%
\bibitem [{\citenamefont {Kiely}\ and\ \citenamefont
  {Freericks}(2018)}]{Kiely_2018}%
  \BibitemOpen
  \bibfield  {author} {\bibinfo {author} {\bibfnamefont {T.~G.}\ \bibnamefont
  {Kiely}}\ and\ \bibinfo {author} {\bibfnamefont {J.~K.}\ \bibnamefont
  {Freericks}},\ }\bibfield  {title} {\bibinfo {title} {Relationship between
  the transverse-field ising model and the xy model via the rotating-wave
  approximation},\ }\bibfield  {journal} {\bibinfo  {journal} {Physical Review
  A}\ }\textbf {\bibinfo {volume} {97}},\ \href
  {https://doi.org/10.1103/physreva.97.023611} {10.1103/physreva.97.023611}
  (\bibinfo {year} {2018})\BibitemShut {NoStop}%
\bibitem [{\citenamefont {Robinson}\ \emph {et~al.}(2021)\citenamefont
  {Robinson}, \citenamefont {Castillo},\ and\ \citenamefont
  {Guzm{\'{a}}n-Gonz{\'{a}}lez}}]{Robinson_2021}%
  \BibitemOpen
  \bibfield  {author} {\bibinfo {author} {\bibfnamefont {N.~J.}\ \bibnamefont
  {Robinson}}, \bibinfo {author} {\bibfnamefont {I.~P.}\ \bibnamefont
  {Castillo}},\ and\ \bibinfo {author} {\bibfnamefont {E.}~\bibnamefont
  {Guzm{\'{a}}n-Gonz{\'{a}}lez}},\ }\bibfield  {title} {\bibinfo {title}
  {Quantum quench in a driven ising chain},\ }\bibfield  {journal} {\bibinfo
  {journal} {Physical Review B}\ }\textbf {\bibinfo {volume} {103}},\ \href
  {https://doi.org/10.1103/physrevb.103.l140407} {10.1103/physrevb.103.l140407}
  (\bibinfo {year} {2021})\BibitemShut {NoStop}%
\bibitem [{\citenamefont {Heyl}\ \emph {et~al.}(2013)\citenamefont {Heyl},
  \citenamefont {Polkovnikov},\ and\ \citenamefont
  {Kehrein}}]{PhysRevLett.110.135704}%
  \BibitemOpen
  \bibfield  {author} {\bibinfo {author} {\bibfnamefont {M.}~\bibnamefont
  {Heyl}}, \bibinfo {author} {\bibfnamefont {A.}~\bibnamefont {Polkovnikov}},\
  and\ \bibinfo {author} {\bibfnamefont {S.}~\bibnamefont {Kehrein}},\
  }\bibfield  {title} {\bibinfo {title} {Dynamical quantum phase transitions in
  the transverse-field ising model},\ }\href
  {https://doi.org/10.1103/PhysRevLett.110.135704} {\bibfield  {journal}
  {\bibinfo  {journal} {Phys. Rev. Lett.}\ }\textbf {\bibinfo {volume} {110}},\
  \bibinfo {pages} {135704} (\bibinfo {year} {2013})}\BibitemShut {NoStop}%
\bibitem [{\citenamefont {Karrasch}\ and\ \citenamefont
  {Schuricht}(2013)}]{PhysRevB.87.195104}%
  \BibitemOpen
  \bibfield  {author} {\bibinfo {author} {\bibfnamefont {C.}~\bibnamefont
  {Karrasch}}\ and\ \bibinfo {author} {\bibfnamefont {D.}~\bibnamefont
  {Schuricht}},\ }\bibfield  {title} {\bibinfo {title} {Dynamical phase
  transitions after quenches in nonintegrable models},\ }\href
  {https://doi.org/10.1103/PhysRevB.87.195104} {\bibfield  {journal} {\bibinfo
  {journal} {Phys. Rev. B}\ }\textbf {\bibinfo {volume} {87}},\ \bibinfo
  {pages} {195104} (\bibinfo {year} {2013})}\BibitemShut {NoStop}%
\bibitem [{\citenamefont {De~Nicola}\ \emph {et~al.}(2021)\citenamefont
  {De~Nicola}, \citenamefont {Michailidis},\ and\ \citenamefont
  {Serbyn}}]{PhysRevLett.126.040602}%
  \BibitemOpen
  \bibfield  {author} {\bibinfo {author} {\bibfnamefont {S.}~\bibnamefont
  {De~Nicola}}, \bibinfo {author} {\bibfnamefont {A.~A.}\ \bibnamefont
  {Michailidis}},\ and\ \bibinfo {author} {\bibfnamefont {M.}~\bibnamefont
  {Serbyn}},\ }\bibfield  {title} {\bibinfo {title} {Entanglement view of
  dynamical quantum phase transitions},\ }\href
  {https://doi.org/10.1103/PhysRevLett.126.040602} {\bibfield  {journal}
  {\bibinfo  {journal} {Phys. Rev. Lett.}\ }\textbf {\bibinfo {volume} {126}},\
  \bibinfo {pages} {040602} (\bibinfo {year} {2021})}\BibitemShut {NoStop}%
\bibitem [{\citenamefont {Rakovszky}\ \emph {et~al.}(2016)\citenamefont
  {Rakovszky}, \citenamefont {Mestyán}, \citenamefont {Collura}, \citenamefont
  {Kormos},\ and\ \citenamefont {Takács}}]{RAKOVSZKY2016805}%
  \BibitemOpen
  \bibfield  {author} {\bibinfo {author} {\bibfnamefont {T.}~\bibnamefont
  {Rakovszky}}, \bibinfo {author} {\bibfnamefont {M.}~\bibnamefont {Mestyán}},
  \bibinfo {author} {\bibfnamefont {M.}~\bibnamefont {Collura}}, \bibinfo
  {author} {\bibfnamefont {M.}~\bibnamefont {Kormos}},\ and\ \bibinfo {author}
  {\bibfnamefont {G.}~\bibnamefont {Takács}},\ }\bibfield  {title} {\bibinfo
  {title} {Hamiltonian truncation approach to quenches in the ising field
  theory},\ }\href
  {https://doi.org/https://doi.org/10.1016/j.nuclphysb.2016.08.024} {\bibfield
  {journal} {\bibinfo  {journal} {Nuclear Physics B}\ }\textbf {\bibinfo
  {volume} {911}},\ \bibinfo {pages} {805} (\bibinfo {year}
  {2016})}\BibitemShut {NoStop}%
\bibitem [{\citenamefont {Heyl}(2015)}]{Heyl_2015}%
  \BibitemOpen
  \bibfield  {author} {\bibinfo {author} {\bibfnamefont {M.}~\bibnamefont
  {Heyl}},\ }\bibfield  {title} {\bibinfo {title} {Scaling and universality at
  dynamical quantum phase transitions},\ }\bibfield  {journal} {\bibinfo
  {journal} {Physical Review Letters}\ }\textbf {\bibinfo {volume} {115}},\
  \href {https://doi.org/10.1103/physrevlett.115.140602}
  {10.1103/physrevlett.115.140602} (\bibinfo {year} {2015})\BibitemShut
  {NoStop}%
\bibitem [{\citenamefont {Hashizume}\ \emph {et~al.}(2022)\citenamefont
  {Hashizume}, \citenamefont {McCulloch},\ and\ \citenamefont
  {Halimeh}}]{PhysRevResearch.4.013250}%
  \BibitemOpen
  \bibfield  {author} {\bibinfo {author} {\bibfnamefont {T.}~\bibnamefont
  {Hashizume}}, \bibinfo {author} {\bibfnamefont {I.~P.}\ \bibnamefont
  {McCulloch}},\ and\ \bibinfo {author} {\bibfnamefont {J.~C.}\ \bibnamefont
  {Halimeh}},\ }\bibfield  {title} {\bibinfo {title} {Dynamical phase
  transitions in the two-dimensional transverse-field ising model},\ }\href
  {https://doi.org/10.1103/PhysRevResearch.4.013250} {\bibfield  {journal}
  {\bibinfo  {journal} {Phys. Rev. Research}\ }\textbf {\bibinfo {volume}
  {4}},\ \bibinfo {pages} {013250} (\bibinfo {year} {2022})}\BibitemShut
  {NoStop}%
\bibitem [{\citenamefont {Yoshinaga}\ \emph {et~al.}(2021)\citenamefont
  {Yoshinaga}, \citenamefont {Hakoshima}, \citenamefont {Imoto}, \citenamefont
  {Matsuzaki},\ and\ \citenamefont
  {Hamazaki}}]{https://doi.org/10.48550/arxiv.2111.05586}%
  \BibitemOpen
  \bibfield  {author} {\bibinfo {author} {\bibfnamefont {A.}~\bibnamefont
  {Yoshinaga}}, \bibinfo {author} {\bibfnamefont {H.}~\bibnamefont
  {Hakoshima}}, \bibinfo {author} {\bibfnamefont {T.}~\bibnamefont {Imoto}},
  \bibinfo {author} {\bibfnamefont {Y.}~\bibnamefont {Matsuzaki}},\ and\
  \bibinfo {author} {\bibfnamefont {R.}~\bibnamefont {Hamazaki}},\ }\href
  {https://doi.org/10.48550/ARXIV.2111.05586} {\bibinfo {title} {Emergence of
  hilbert space fragmentation in ising models with a weak transverse field}}
  (\bibinfo {year} {2021})\BibitemShut {NoStop}%
\bibitem [{\citenamefont {Hart}\ and\ \citenamefont
  {Nandkishore}(2022)}]{https://doi.org/10.48550/arxiv.2203.06188}%
  \BibitemOpen
  \bibfield  {author} {\bibinfo {author} {\bibfnamefont {O.}~\bibnamefont
  {Hart}}\ and\ \bibinfo {author} {\bibfnamefont {R.}~\bibnamefont
  {Nandkishore}},\ }\href {https://doi.org/10.48550/ARXIV.2203.06188} {\bibinfo
  {title} {Hilbert space shattering and dynamical freezing in the quantum ising
  model}} (\bibinfo {year} {2022})\BibitemShut {NoStop}%
\bibitem [{\citenamefont {Sieberer}\ \emph {et~al.}(2019)\citenamefont
  {Sieberer}, \citenamefont {Olsacher}, \citenamefont {Elben}, \citenamefont
  {Heyl}, \citenamefont {Hauke}, \citenamefont {Haake},\ and\ \citenamefont
  {Zoller}}]{sieberer2019digital}%
  \BibitemOpen
  \bibfield  {author} {\bibinfo {author} {\bibfnamefont {L.~M.}\ \bibnamefont
  {Sieberer}}, \bibinfo {author} {\bibfnamefont {T.}~\bibnamefont {Olsacher}},
  \bibinfo {author} {\bibfnamefont {A.}~\bibnamefont {Elben}}, \bibinfo
  {author} {\bibfnamefont {M.}~\bibnamefont {Heyl}}, \bibinfo {author}
  {\bibfnamefont {P.}~\bibnamefont {Hauke}}, \bibinfo {author} {\bibfnamefont
  {F.}~\bibnamefont {Haake}},\ and\ \bibinfo {author} {\bibfnamefont
  {P.}~\bibnamefont {Zoller}},\ }\bibfield  {title} {\bibinfo {title} {Digital
  quantum simulation, trotter errors, and quantum chaos of the kicked top},\
  }\href@noop {} {\bibfield  {journal} {\bibinfo  {journal} {npj Quantum
  Information}\ }\textbf {\bibinfo {volume} {5}},\ \bibinfo {pages} {1}
  (\bibinfo {year} {2019})}\BibitemShut {NoStop}%
\bibitem [{\citenamefont {Heyl}\ \emph {et~al.}(2019)\citenamefont {Heyl},
  \citenamefont {Hauke},\ and\ \citenamefont {Zoller}}]{Heyl_2019}%
  \BibitemOpen
  \bibfield  {author} {\bibinfo {author} {\bibfnamefont {M.}~\bibnamefont
  {Heyl}}, \bibinfo {author} {\bibfnamefont {P.}~\bibnamefont {Hauke}},\ and\
  \bibinfo {author} {\bibfnamefont {P.}~\bibnamefont {Zoller}},\ }\bibfield
  {title} {\bibinfo {title} {Quantum localization bounds trotter errors in
  digital quantum simulation},\ }\bibfield  {journal} {\bibinfo  {journal}
  {Science Advances}\ }\textbf {\bibinfo {volume} {5}},\ \href
  {https://doi.org/10.1126/sciadv.aau8342} {10.1126/sciadv.aau8342} (\bibinfo
  {year} {2019})\BibitemShut {NoStop}%
\bibitem [{\citenamefont {Young}\ \emph {et~al.}(2022)\citenamefont {Young},
  \citenamefont {Muleady}, \citenamefont {Perlin}, \citenamefont {Kaufman},\
  and\ \citenamefont {Rey}}]{https://doi.org/10.48550/arxiv.2208.01869}%
  \BibitemOpen
  \bibfield  {author} {\bibinfo {author} {\bibfnamefont {J.~T.}\ \bibnamefont
  {Young}}, \bibinfo {author} {\bibfnamefont {S.~R.}\ \bibnamefont {Muleady}},
  \bibinfo {author} {\bibfnamefont {M.~A.}\ \bibnamefont {Perlin}}, \bibinfo
  {author} {\bibfnamefont {A.~M.}\ \bibnamefont {Kaufman}},\ and\ \bibinfo
  {author} {\bibfnamefont {A.~M.}\ \bibnamefont {Rey}},\ }\href
  {https://doi.org/10.48550/ARXIV.2208.01869} {\bibinfo {title} {Enhancing spin
  squeezing using soft-core interactions}} (\bibinfo {year} {2022})\BibitemShut
  {NoStop}%
\bibitem [{\citenamefont {Robinson}\ \emph {et~al.}(2019)\citenamefont
  {Robinson}, \citenamefont {James},\ and\ \citenamefont
  {Konik}}]{PhysRevB.99.195108}%
  \BibitemOpen
  \bibfield  {author} {\bibinfo {author} {\bibfnamefont {N.~J.}\ \bibnamefont
  {Robinson}}, \bibinfo {author} {\bibfnamefont {A.~J.~A.}\ \bibnamefont
  {James}},\ and\ \bibinfo {author} {\bibfnamefont {R.~M.}\ \bibnamefont
  {Konik}},\ }\bibfield  {title} {\bibinfo {title} {Signatures of rare states
  and thermalization in a theory with confinement},\ }\href
  {https://doi.org/10.1103/PhysRevB.99.195108} {\bibfield  {journal} {\bibinfo
  {journal} {Phys. Rev. B}\ }\textbf {\bibinfo {volume} {99}},\ \bibinfo
  {pages} {195108} (\bibinfo {year} {2019})}\BibitemShut {NoStop}%
\bibitem [{\citenamefont {Cubitt}\ \emph {et~al.}(2018)\citenamefont {Cubitt},
  \citenamefont {Montanaro},\ and\ \citenamefont {Piddock}}]{Cubitt_2018}%
  \BibitemOpen
  \bibfield  {author} {\bibinfo {author} {\bibfnamefont {T.~S.}\ \bibnamefont
  {Cubitt}}, \bibinfo {author} {\bibfnamefont {A.}~\bibnamefont {Montanaro}},\
  and\ \bibinfo {author} {\bibfnamefont {S.}~\bibnamefont {Piddock}},\
  }\bibfield  {title} {\bibinfo {title} {Universal quantum hamiltonians},\
  }\href {https://doi.org/10.1073/pnas.1804949115} {\bibfield  {journal}
  {\bibinfo  {journal} {Proceedings of the National Academy of Sciences}\
  }\textbf {\bibinfo {volume} {115}},\ \bibinfo {pages} {9497} (\bibinfo {year}
  {2018})}\BibitemShut {NoStop}%
\bibitem [{\citenamefont {Goldman}\ and\ \citenamefont
  {Dalibard}(2014)}]{PhysRevX.4.031027}%
  \BibitemOpen
  \bibfield  {author} {\bibinfo {author} {\bibfnamefont {N.}~\bibnamefont
  {Goldman}}\ and\ \bibinfo {author} {\bibfnamefont {J.}~\bibnamefont
  {Dalibard}},\ }\bibfield  {title} {\bibinfo {title} {Periodically driven
  quantum systems: Effective hamiltonians and engineered gauge fields},\ }\href
  {https://doi.org/10.1103/PhysRevX.4.031027} {\bibfield  {journal} {\bibinfo
  {journal} {Phys. Rev. X}\ }\textbf {\bibinfo {volume} {4}},\ \bibinfo {pages}
  {031027} (\bibinfo {year} {2014})}\BibitemShut {NoStop}%
\bibitem [{\citenamefont {Shirley}(1965)}]{PhysRev.138.B979}%
  \BibitemOpen
  \bibfield  {author} {\bibinfo {author} {\bibfnamefont {J.~H.}\ \bibnamefont
  {Shirley}},\ }\bibfield  {title} {\bibinfo {title} {Solution of the
  schr\"odinger equation with a hamiltonian periodic in time},\ }\href
  {https://doi.org/10.1103/PhysRev.138.B979} {\bibfield  {journal} {\bibinfo
  {journal} {Phys. Rev.}\ }\textbf {\bibinfo {volume} {138}},\ \bibinfo {pages}
  {B979} (\bibinfo {year} {1965})}\BibitemShut {NoStop}%
\bibitem [{\citenamefont {Vandersypen}\ and\ \citenamefont
  {Chuang}(2005)}]{RevModPhys.76.1037}%
  \BibitemOpen
  \bibfield  {author} {\bibinfo {author} {\bibfnamefont {L.~M.~K.}\
  \bibnamefont {Vandersypen}}\ and\ \bibinfo {author} {\bibfnamefont {I.~L.}\
  \bibnamefont {Chuang}},\ }\bibfield  {title} {\bibinfo {title} {Nmr
  techniques for quantum control and computation},\ }\href
  {https://doi.org/10.1103/RevModPhys.76.1037} {\bibfield  {journal} {\bibinfo
  {journal} {Rev. Mod. Phys.}\ }\textbf {\bibinfo {volume} {76}},\ \bibinfo
  {pages} {1037} (\bibinfo {year} {2005})}\BibitemShut {NoStop}%
\bibitem [{\citenamefont {Salath\'e}\ \emph {et~al.}(2015)\citenamefont
  {Salath\'e}, \citenamefont {Mondal}, \citenamefont {Oppliger}, \citenamefont
  {Heinsoo}, \citenamefont {Kurpiers}, \citenamefont
  {Poto\ifmmode~\check{c}\else \v{c}\fi{}nik}, \citenamefont {Mezzacapo},
  \citenamefont {Las~Heras}, \citenamefont {Lamata}, \citenamefont {Solano},
  \citenamefont {Filipp},\ and\ \citenamefont {Wallraff}}]{PhysRevX.5.021027}%
  \BibitemOpen
  \bibfield  {author} {\bibinfo {author} {\bibfnamefont {Y.}~\bibnamefont
  {Salath\'e}}, \bibinfo {author} {\bibfnamefont {M.}~\bibnamefont {Mondal}},
  \bibinfo {author} {\bibfnamefont {M.}~\bibnamefont {Oppliger}}, \bibinfo
  {author} {\bibfnamefont {J.}~\bibnamefont {Heinsoo}}, \bibinfo {author}
  {\bibfnamefont {P.}~\bibnamefont {Kurpiers}}, \bibinfo {author}
  {\bibfnamefont {A.}~\bibnamefont {Poto\ifmmode~\check{c}\else
  \v{c}\fi{}nik}}, \bibinfo {author} {\bibfnamefont {A.}~\bibnamefont
  {Mezzacapo}}, \bibinfo {author} {\bibfnamefont {U.}~\bibnamefont
  {Las~Heras}}, \bibinfo {author} {\bibfnamefont {L.}~\bibnamefont {Lamata}},
  \bibinfo {author} {\bibfnamefont {E.}~\bibnamefont {Solano}}, \bibinfo
  {author} {\bibfnamefont {S.}~\bibnamefont {Filipp}},\ and\ \bibinfo {author}
  {\bibfnamefont {A.}~\bibnamefont {Wallraff}},\ }\bibfield  {title} {\bibinfo
  {title} {Digital quantum simulation of spin models with circuit quantum
  electrodynamics},\ }\href {https://doi.org/10.1103/PhysRevX.5.021027}
  {\bibfield  {journal} {\bibinfo  {journal} {Phys. Rev. X}\ }\textbf {\bibinfo
  {volume} {5}},\ \bibinfo {pages} {021027} (\bibinfo {year}
  {2015})}\BibitemShut {NoStop}%
\bibitem [{\citenamefont {Aidelsburger}\ \emph {et~al.}(2013)\citenamefont
  {Aidelsburger}, \citenamefont {Atala}, \citenamefont {Lohse}, \citenamefont
  {Barreiro}, \citenamefont {Paredes},\ and\ \citenamefont
  {Bloch}}]{PhysRevLett.111.185301}%
  \BibitemOpen
  \bibfield  {author} {\bibinfo {author} {\bibfnamefont {M.}~\bibnamefont
  {Aidelsburger}}, \bibinfo {author} {\bibfnamefont {M.}~\bibnamefont {Atala}},
  \bibinfo {author} {\bibfnamefont {M.}~\bibnamefont {Lohse}}, \bibinfo
  {author} {\bibfnamefont {J.~T.}\ \bibnamefont {Barreiro}}, \bibinfo {author}
  {\bibfnamefont {B.}~\bibnamefont {Paredes}},\ and\ \bibinfo {author}
  {\bibfnamefont {I.}~\bibnamefont {Bloch}},\ }\bibfield  {title} {\bibinfo
  {title} {Realization of the hofstadter hamiltonian with ultracold atoms in
  optical lattices},\ }\href {https://doi.org/10.1103/PhysRevLett.111.185301}
  {\bibfield  {journal} {\bibinfo  {journal} {Phys. Rev. Lett.}\ }\textbf
  {\bibinfo {volume} {111}},\ \bibinfo {pages} {185301} (\bibinfo {year}
  {2013})}\BibitemShut {NoStop}%
\bibitem [{\citenamefont {Fläschner}\ \emph {et~al.}(2016)\citenamefont
  {Fläschner}, \citenamefont {Rem}, \citenamefont {Tarnowski}, \citenamefont
  {Vogel}, \citenamefont {Lühmann}, \citenamefont {Sengstock},\ and\
  \citenamefont {Weitenberg}}]{doi:10.1126/science.aad4568}%
  \BibitemOpen
  \bibfield  {author} {\bibinfo {author} {\bibfnamefont {N.}~\bibnamefont
  {Fläschner}}, \bibinfo {author} {\bibfnamefont {B.~S.}\ \bibnamefont {Rem}},
  \bibinfo {author} {\bibfnamefont {M.}~\bibnamefont {Tarnowski}}, \bibinfo
  {author} {\bibfnamefont {D.}~\bibnamefont {Vogel}}, \bibinfo {author}
  {\bibfnamefont {D.-S.}\ \bibnamefont {Lühmann}}, \bibinfo {author}
  {\bibfnamefont {K.}~\bibnamefont {Sengstock}},\ and\ \bibinfo {author}
  {\bibfnamefont {C.}~\bibnamefont {Weitenberg}},\ }\bibfield  {title}
  {\bibinfo {title} {Experimental reconstruction of the berry curvature in a
  floquet bloch band},\ }\href {https://doi.org/10.1126/science.aad4568}
  {\bibfield  {journal} {\bibinfo  {journal} {Science}\ }\textbf {\bibinfo
  {volume} {352}},\ \bibinfo {pages} {1091} (\bibinfo {year} {2016})},\ \Eprint
  {https://arxiv.org/abs/https://www.science.org/doi/pdf/10.1126/science.aad4568}
  {https://www.science.org/doi/pdf/10.1126/science.aad4568} \BibitemShut
  {NoStop}%
\bibitem [{\citenamefont {Meinert}\ \emph {et~al.}(2016)\citenamefont
  {Meinert}, \citenamefont {Mark}, \citenamefont {Lauber}, \citenamefont
  {Daley},\ and\ \citenamefont {N\"agerl}}]{PhysRevLett.116.205301}%
  \BibitemOpen
  \bibfield  {author} {\bibinfo {author} {\bibfnamefont {F.}~\bibnamefont
  {Meinert}}, \bibinfo {author} {\bibfnamefont {M.~J.}\ \bibnamefont {Mark}},
  \bibinfo {author} {\bibfnamefont {K.}~\bibnamefont {Lauber}}, \bibinfo
  {author} {\bibfnamefont {A.~J.}\ \bibnamefont {Daley}},\ and\ \bibinfo
  {author} {\bibfnamefont {H.-C.}\ \bibnamefont {N\"agerl}},\ }\bibfield
  {title} {\bibinfo {title} {Floquet engineering of correlated tunneling in the
  bose-hubbard model with ultracold atoms},\ }\href
  {https://doi.org/10.1103/PhysRevLett.116.205301} {\bibfield  {journal}
  {\bibinfo  {journal} {Phys. Rev. Lett.}\ }\textbf {\bibinfo {volume} {116}},\
  \bibinfo {pages} {205301} (\bibinfo {year} {2016})}\BibitemShut {NoStop}%
\bibitem [{\citenamefont {Schweizer}\ \emph {et~al.}(2019)\citenamefont
  {Schweizer}, \citenamefont {Grusdt}, \citenamefont {Berngruber},
  \citenamefont {Barbiero}, \citenamefont {Demler}, \citenamefont {Goldman},
  \citenamefont {Bloch},\ and\ \citenamefont
  {Aidelsburger}}]{schweizer2019floquet}%
  \BibitemOpen
  \bibfield  {author} {\bibinfo {author} {\bibfnamefont {C.}~\bibnamefont
  {Schweizer}}, \bibinfo {author} {\bibfnamefont {F.}~\bibnamefont {Grusdt}},
  \bibinfo {author} {\bibfnamefont {M.}~\bibnamefont {Berngruber}}, \bibinfo
  {author} {\bibfnamefont {L.}~\bibnamefont {Barbiero}}, \bibinfo {author}
  {\bibfnamefont {E.}~\bibnamefont {Demler}}, \bibinfo {author} {\bibfnamefont
  {N.}~\bibnamefont {Goldman}}, \bibinfo {author} {\bibfnamefont
  {I.}~\bibnamefont {Bloch}},\ and\ \bibinfo {author} {\bibfnamefont
  {M.}~\bibnamefont {Aidelsburger}},\ }\bibfield  {title} {\bibinfo {title}
  {Floquet approach to z2 lattice gauge theories with ultracold atoms in
  optical lattices},\ }\href@noop {} {\bibfield  {journal} {\bibinfo  {journal}
  {Nature Physics}\ }\textbf {\bibinfo {volume} {15}},\ \bibinfo {pages} {1168}
  (\bibinfo {year} {2019})}\BibitemShut {NoStop}%
\bibitem [{\citenamefont {Eckardt}(2017)}]{RevModPhys.89.011004}%
  \BibitemOpen
  \bibfield  {author} {\bibinfo {author} {\bibfnamefont {A.}~\bibnamefont
  {Eckardt}},\ }\bibfield  {title} {\bibinfo {title} {Colloquium: Atomic
  quantum gases in periodically driven optical lattices},\ }\href
  {https://doi.org/10.1103/RevModPhys.89.011004} {\bibfield  {journal}
  {\bibinfo  {journal} {Rev. Mod. Phys.}\ }\textbf {\bibinfo {volume} {89}},\
  \bibinfo {pages} {011004} (\bibinfo {year} {2017})}\BibitemShut {NoStop}%
\bibitem [{\citenamefont {Wintersperger}\ \emph {et~al.}(2020)\citenamefont
  {Wintersperger}, \citenamefont {Braun}, \citenamefont {{\"U}nal},
  \citenamefont {Eckardt}, \citenamefont {Liberto}, \citenamefont {Goldman},
  \citenamefont {Bloch},\ and\ \citenamefont
  {Aidelsburger}}]{wintersperger2020realization}%
  \BibitemOpen
  \bibfield  {author} {\bibinfo {author} {\bibfnamefont {K.}~\bibnamefont
  {Wintersperger}}, \bibinfo {author} {\bibfnamefont {C.}~\bibnamefont
  {Braun}}, \bibinfo {author} {\bibfnamefont {F.~N.}\ \bibnamefont {{\"U}nal}},
  \bibinfo {author} {\bibfnamefont {A.}~\bibnamefont {Eckardt}}, \bibinfo
  {author} {\bibfnamefont {M.~D.}\ \bibnamefont {Liberto}}, \bibinfo {author}
  {\bibfnamefont {N.}~\bibnamefont {Goldman}}, \bibinfo {author} {\bibfnamefont
  {I.}~\bibnamefont {Bloch}},\ and\ \bibinfo {author} {\bibfnamefont
  {M.}~\bibnamefont {Aidelsburger}},\ }\bibfield  {title} {\bibinfo {title}
  {Realization of an anomalous floquet topological system with ultracold
  atoms},\ }\href@noop {} {\bibfield  {journal} {\bibinfo  {journal} {Nature
  Physics}\ }\textbf {\bibinfo {volume} {16}},\ \bibinfo {pages} {1058}
  (\bibinfo {year} {2020})}\BibitemShut {NoStop}%
\bibitem [{\citenamefont {Choi}\ \emph {et~al.}(2020)\citenamefont {Choi},
  \citenamefont {Zhou}, \citenamefont {Knowles}, \citenamefont {Landig},
  \citenamefont {Choi},\ and\ \citenamefont {Lukin}}]{PhysRevX.10.031002}%
  \BibitemOpen
  \bibfield  {author} {\bibinfo {author} {\bibfnamefont {J.}~\bibnamefont
  {Choi}}, \bibinfo {author} {\bibfnamefont {H.}~\bibnamefont {Zhou}}, \bibinfo
  {author} {\bibfnamefont {H.~S.}\ \bibnamefont {Knowles}}, \bibinfo {author}
  {\bibfnamefont {R.}~\bibnamefont {Landig}}, \bibinfo {author} {\bibfnamefont
  {S.}~\bibnamefont {Choi}},\ and\ \bibinfo {author} {\bibfnamefont {M.~D.}\
  \bibnamefont {Lukin}},\ }\bibfield  {title} {\bibinfo {title} {Robust dynamic
  hamiltonian engineering of many-body spin systems},\ }\href
  {https://doi.org/10.1103/PhysRevX.10.031002} {\bibfield  {journal} {\bibinfo
  {journal} {Phys. Rev. X}\ }\textbf {\bibinfo {volume} {10}},\ \bibinfo
  {pages} {031002} (\bibinfo {year} {2020})}\BibitemShut {NoStop}%
\bibitem [{\citenamefont {Qiao}\ \emph {et~al.}(2021)\citenamefont {Qiao},
  \citenamefont {Kandel}, \citenamefont {Dyke}, \citenamefont {Fallahi},
  \citenamefont {Gardner}, \citenamefont {Manfra}, \citenamefont {Barnes},\
  and\ \citenamefont {Nichol}}]{qiao2021floquet}%
  \BibitemOpen
  \bibfield  {author} {\bibinfo {author} {\bibfnamefont {H.}~\bibnamefont
  {Qiao}}, \bibinfo {author} {\bibfnamefont {Y.~P.}\ \bibnamefont {Kandel}},
  \bibinfo {author} {\bibfnamefont {J.~S.~V.}\ \bibnamefont {Dyke}}, \bibinfo
  {author} {\bibfnamefont {S.}~\bibnamefont {Fallahi}}, \bibinfo {author}
  {\bibfnamefont {G.~C.}\ \bibnamefont {Gardner}}, \bibinfo {author}
  {\bibfnamefont {M.~J.}\ \bibnamefont {Manfra}}, \bibinfo {author}
  {\bibfnamefont {E.}~\bibnamefont {Barnes}},\ and\ \bibinfo {author}
  {\bibfnamefont {J.~M.}\ \bibnamefont {Nichol}},\ }\bibfield  {title}
  {\bibinfo {title} {Floquet-enhanced spin swaps},\ }\href@noop {} {\bibfield
  {journal} {\bibinfo  {journal} {Nature communications}\ }\textbf {\bibinfo
  {volume} {12}},\ \bibinfo {pages} {1} (\bibinfo {year} {2021})}\BibitemShut
  {NoStop}%
\bibitem [{\citenamefont {Sarkar}\ and\ \citenamefont
  {Buca}(2022)}]{https://doi.org/10.48550/arxiv.2204.13354}%
  \BibitemOpen
  \bibfield  {author} {\bibinfo {author} {\bibfnamefont {S.}~\bibnamefont
  {Sarkar}}\ and\ \bibinfo {author} {\bibfnamefont {B.}~\bibnamefont {Buca}},\
  }\href {https://doi.org/10.48550/ARXIV.2204.13354} {\bibinfo {title}
  {Protecting coherence from environment via stark many-body localization in a
  quantum-dot simulator}} (\bibinfo {year} {2022})\BibitemShut {NoStop}%
\bibitem [{\citenamefont {Abanin}\ \emph {et~al.}(2017)\citenamefont {Abanin},
  \citenamefont {De~Roeck}, \citenamefont {Ho},\ and\ \citenamefont
  {Huveneers}}]{abanin2017rigorous}%
  \BibitemOpen
  \bibfield  {author} {\bibinfo {author} {\bibfnamefont {D.}~\bibnamefont
  {Abanin}}, \bibinfo {author} {\bibfnamefont {W.}~\bibnamefont {De~Roeck}},
  \bibinfo {author} {\bibfnamefont {W.~W.}\ \bibnamefont {Ho}},\ and\ \bibinfo
  {author} {\bibfnamefont {F.}~\bibnamefont {Huveneers}},\ }\bibfield  {title}
  {\bibinfo {title} {A rigorous theory of many-body prethermalization for
  periodically driven and closed quantum systems},\ }\href@noop {} {\bibfield
  {journal} {\bibinfo  {journal} {Communications in Mathematical Physics}\
  }\textbf {\bibinfo {volume} {354}},\ \bibinfo {pages} {809} (\bibinfo {year}
  {2017})}\BibitemShut {NoStop}%
\bibitem [{\citenamefont {Else}\ \emph {et~al.}(2017)\citenamefont {Else},
  \citenamefont {Bauer},\ and\ \citenamefont {Nayak}}]{PhysRevX.7.011026}%
  \BibitemOpen
  \bibfield  {author} {\bibinfo {author} {\bibfnamefont {D.~V.}\ \bibnamefont
  {Else}}, \bibinfo {author} {\bibfnamefont {B.}~\bibnamefont {Bauer}},\ and\
  \bibinfo {author} {\bibfnamefont {C.}~\bibnamefont {Nayak}},\ }\bibfield
  {title} {\bibinfo {title} {Prethermal phases of matter protected by
  time-translation symmetry},\ }\href
  {https://doi.org/10.1103/PhysRevX.7.011026} {\bibfield  {journal} {\bibinfo
  {journal} {Phys. Rev. X}\ }\textbf {\bibinfo {volume} {7}},\ \bibinfo {pages}
  {011026} (\bibinfo {year} {2017})}\BibitemShut {NoStop}%
\bibitem [{\citenamefont {Mizuta}\ \emph {et~al.}(2019)\citenamefont {Mizuta},
  \citenamefont {Takasan},\ and\ \citenamefont
  {Kawakami}}]{PhysRevB.100.020301}%
  \BibitemOpen
  \bibfield  {author} {\bibinfo {author} {\bibfnamefont {K.}~\bibnamefont
  {Mizuta}}, \bibinfo {author} {\bibfnamefont {K.}~\bibnamefont {Takasan}},\
  and\ \bibinfo {author} {\bibfnamefont {N.}~\bibnamefont {Kawakami}},\
  }\bibfield  {title} {\bibinfo {title} {High-frequency expansion for floquet
  prethermal phases with emergent symmetries: Application to time crystals and
  floquet engineering},\ }\href {https://doi.org/10.1103/PhysRevB.100.020301}
  {\bibfield  {journal} {\bibinfo  {journal} {Phys. Rev. B}\ }\textbf {\bibinfo
  {volume} {100}},\ \bibinfo {pages} {020301} (\bibinfo {year}
  {2019})}\BibitemShut {NoStop}%
\bibitem [{\citenamefont
  {Claassen}(2021)}]{https://doi.org/10.48550/arxiv.2103.07485}%
  \BibitemOpen
  \bibfield  {author} {\bibinfo {author} {\bibfnamefont {M.}~\bibnamefont
  {Claassen}},\ }\href {https://doi.org/10.48550/ARXIV.2103.07485} {\bibinfo
  {title} {Flow renormalization and emergent prethermal regimes of
  periodically-driven quantum systems}} (\bibinfo {year} {2021})\BibitemShut
  {NoStop}%
\bibitem [{\citenamefont {Agarwal}\ and\ \citenamefont
  {Martin}(2020)}]{PhysRevLett.125.080602}%
  \BibitemOpen
  \bibfield  {author} {\bibinfo {author} {\bibfnamefont {K.}~\bibnamefont
  {Agarwal}}\ and\ \bibinfo {author} {\bibfnamefont {I.}~\bibnamefont
  {Martin}},\ }\bibfield  {title} {\bibinfo {title} {Dynamical enhancement of
  symmetries in many-body systems},\ }\href
  {https://doi.org/10.1103/PhysRevLett.125.080602} {\bibfield  {journal}
  {\bibinfo  {journal} {Phys. Rev. Lett.}\ }\textbf {\bibinfo {volume} {125}},\
  \bibinfo {pages} {080602} (\bibinfo {year} {2020})}\BibitemShut {NoStop}%
\bibitem [{\citenamefont {Halimeh}\ \emph
  {et~al.}(2021{\natexlab{a}})\citenamefont {Halimeh}, \citenamefont {Lang},
  \citenamefont {Mildenberger}, \citenamefont {Jiang},\ and\ \citenamefont
  {Hauke}}]{PRXQuantum.2.040311}%
  \BibitemOpen
  \bibfield  {author} {\bibinfo {author} {\bibfnamefont {J.~C.}\ \bibnamefont
  {Halimeh}}, \bibinfo {author} {\bibfnamefont {H.}~\bibnamefont {Lang}},
  \bibinfo {author} {\bibfnamefont {J.}~\bibnamefont {Mildenberger}}, \bibinfo
  {author} {\bibfnamefont {Z.}~\bibnamefont {Jiang}},\ and\ \bibinfo {author}
  {\bibfnamefont {P.}~\bibnamefont {Hauke}},\ }\bibfield  {title} {\bibinfo
  {title} {Gauge-symmetry protection using single-body terms},\ }\href
  {https://doi.org/10.1103/PRXQuantum.2.040311} {\bibfield  {journal} {\bibinfo
   {journal} {PRX Quantum}\ }\textbf {\bibinfo {volume} {2}},\ \bibinfo {pages}
  {040311} (\bibinfo {year} {2021}{\natexlab{a}})}\BibitemShut {NoStop}%
\bibitem [{\citenamefont {Halimeh}\ \emph {et~al.}(2022)\citenamefont
  {Halimeh}, \citenamefont {Lang},\ and\ \citenamefont
  {Hauke}}]{halimeh2022gauge}%
  \BibitemOpen
  \bibfield  {author} {\bibinfo {author} {\bibfnamefont {J.~C.}\ \bibnamefont
  {Halimeh}}, \bibinfo {author} {\bibfnamefont {H.}~\bibnamefont {Lang}},\ and\
  \bibinfo {author} {\bibfnamefont {P.}~\bibnamefont {Hauke}},\ }\bibfield
  {title} {\bibinfo {title} {Gauge protection in non-abelian lattice gauge
  theories},\ }\href@noop {} {\bibfield  {journal} {\bibinfo  {journal} {New
  Journal of Physics}\ }\textbf {\bibinfo {volume} {24}},\ \bibinfo {pages}
  {033015} (\bibinfo {year} {2022})}\BibitemShut {NoStop}%
\bibitem [{\citenamefont {Halimeh}\ \emph
  {et~al.}(2021{\natexlab{b}})\citenamefont {Halimeh}, \citenamefont {Homeier},
  \citenamefont {Schweizer}, \citenamefont {Aidelsburger}, \citenamefont
  {Hauke},\ and\ \citenamefont
  {Grusdt}}]{https://doi.org/10.48550/arxiv.2108.02203}%
  \BibitemOpen
  \bibfield  {author} {\bibinfo {author} {\bibfnamefont {J.~C.}\ \bibnamefont
  {Halimeh}}, \bibinfo {author} {\bibfnamefont {L.}~\bibnamefont {Homeier}},
  \bibinfo {author} {\bibfnamefont {C.}~\bibnamefont {Schweizer}}, \bibinfo
  {author} {\bibfnamefont {M.}~\bibnamefont {Aidelsburger}}, \bibinfo {author}
  {\bibfnamefont {P.}~\bibnamefont {Hauke}},\ and\ \bibinfo {author}
  {\bibfnamefont {F.}~\bibnamefont {Grusdt}},\ }\href
  {https://doi.org/10.48550/ARXIV.2108.02203} {\bibinfo {title} {Stabilizing
  lattice gauge theories through simplified local pseudo generators}} (\bibinfo
  {year} {2021}{\natexlab{b}})\BibitemShut {NoStop}%
\bibitem [{\citenamefont {Viola}\ \emph {et~al.}(1999)\citenamefont {Viola},
  \citenamefont {Knill},\ and\ \citenamefont {Lloyd}}]{PhysRevLett.82.2417}%
  \BibitemOpen
  \bibfield  {author} {\bibinfo {author} {\bibfnamefont {L.}~\bibnamefont
  {Viola}}, \bibinfo {author} {\bibfnamefont {E.}~\bibnamefont {Knill}},\ and\
  \bibinfo {author} {\bibfnamefont {S.}~\bibnamefont {Lloyd}},\ }\bibfield
  {title} {\bibinfo {title} {Dynamical decoupling of open quantum systems},\
  }\href {https://doi.org/10.1103/PhysRevLett.82.2417} {\bibfield  {journal}
  {\bibinfo  {journal} {Phys. Rev. Lett.}\ }\textbf {\bibinfo {volume} {82}},\
  \bibinfo {pages} {2417} (\bibinfo {year} {1999})}\BibitemShut {NoStop}%
\bibitem [{\citenamefont {Kasper}\ \emph {et~al.}(2020)\citenamefont {Kasper},
  \citenamefont {Zache}, \citenamefont {Jendrzejewski}, \citenamefont
  {Lewenstein},\ and\ \citenamefont
  {Zohar}}]{https://doi.org/10.48550/arxiv.2012.08620}%
  \BibitemOpen
  \bibfield  {author} {\bibinfo {author} {\bibfnamefont {V.}~\bibnamefont
  {Kasper}}, \bibinfo {author} {\bibfnamefont {T.~V.}\ \bibnamefont {Zache}},
  \bibinfo {author} {\bibfnamefont {F.}~\bibnamefont {Jendrzejewski}}, \bibinfo
  {author} {\bibfnamefont {M.}~\bibnamefont {Lewenstein}},\ and\ \bibinfo
  {author} {\bibfnamefont {E.}~\bibnamefont {Zohar}},\ }\href
  {https://doi.org/10.48550/ARXIV.2012.08620} {\bibinfo {title} {Non-abelian
  gauge invariance from dynamical decoupling}} (\bibinfo {year}
  {2020})\BibitemShut {NoStop}%
\bibitem [{\citenamefont {Fishman}\ \emph {et~al.}(2020)\citenamefont
  {Fishman}, \citenamefont {White},\ and\ \citenamefont
  {Stoudenmire}}]{itensor}%
  \BibitemOpen
  \bibfield  {author} {\bibinfo {author} {\bibfnamefont {M.}~\bibnamefont
  {Fishman}}, \bibinfo {author} {\bibfnamefont {S.~R.}\ \bibnamefont {White}},\
  and\ \bibinfo {author} {\bibfnamefont {E.~M.}\ \bibnamefont {Stoudenmire}},\
  }\href@noop {} {\bibinfo {title} {The \mbox{ITensor} software library for
  tensor network calculations}} (\bibinfo {year} {2020}),\ \Eprint
  {https://arxiv.org/abs/2007.14822} {arXiv:2007.14822} \BibitemShut {NoStop}%
\bibitem [{\citenamefont {White}(1992)}]{PhysRevLett.69.2863}%
  \BibitemOpen
  \bibfield  {author} {\bibinfo {author} {\bibfnamefont {S.~R.}\ \bibnamefont
  {White}},\ }\bibfield  {title} {\bibinfo {title} {Density matrix formulation
  for quantum renormalization groups},\ }\href
  {https://doi.org/10.1103/PhysRevLett.69.2863} {\bibfield  {journal} {\bibinfo
   {journal} {Phys. Rev. Lett.}\ }\textbf {\bibinfo {volume} {69}},\ \bibinfo
  {pages} {2863} (\bibinfo {year} {1992})}\BibitemShut {NoStop}%
\bibitem [{\citenamefont {White}(1993)}]{PhysRevB.48.10345}%
  \BibitemOpen
  \bibfield  {author} {\bibinfo {author} {\bibfnamefont {S.~R.}\ \bibnamefont
  {White}},\ }\bibfield  {title} {\bibinfo {title} {Density-matrix algorithms
  for quantum renormalization groups},\ }\href
  {https://doi.org/10.1103/PhysRevB.48.10345} {\bibfield  {journal} {\bibinfo
  {journal} {Phys. Rev. B}\ }\textbf {\bibinfo {volume} {48}},\ \bibinfo
  {pages} {10345} (\bibinfo {year} {1993})}\BibitemShut {NoStop}%
\bibitem [{\citenamefont {Haegeman}\ \emph {et~al.}(2011)\citenamefont
  {Haegeman}, \citenamefont {Cirac}, \citenamefont {Osborne}, \citenamefont
  {Pi\ifmmode~\check{z}\else \v{z}\fi{}orn}, \citenamefont {Verschelde},\ and\
  \citenamefont {Verstraete}}]{PhysRevLett.107.070601}%
  \BibitemOpen
  \bibfield  {author} {\bibinfo {author} {\bibfnamefont {J.}~\bibnamefont
  {Haegeman}}, \bibinfo {author} {\bibfnamefont {J.~I.}\ \bibnamefont {Cirac}},
  \bibinfo {author} {\bibfnamefont {T.~J.}\ \bibnamefont {Osborne}}, \bibinfo
  {author} {\bibfnamefont {I.}~\bibnamefont {Pi\ifmmode~\check{z}\else
  \v{z}\fi{}orn}}, \bibinfo {author} {\bibfnamefont {H.}~\bibnamefont
  {Verschelde}},\ and\ \bibinfo {author} {\bibfnamefont {F.}~\bibnamefont
  {Verstraete}},\ }\bibfield  {title} {\bibinfo {title} {Time-dependent
  variational principle for quantum lattices},\ }\href
  {https://doi.org/10.1103/PhysRevLett.107.070601} {\bibfield  {journal}
  {\bibinfo  {journal} {Phys. Rev. Lett.}\ }\textbf {\bibinfo {volume} {107}},\
  \bibinfo {pages} {070601} (\bibinfo {year} {2011})}\BibitemShut {NoStop}%
\bibitem [{\citenamefont {Haegeman}\ \emph {et~al.}(2016)\citenamefont
  {Haegeman}, \citenamefont {Lubich}, \citenamefont {Oseledets}, \citenamefont
  {Vandereycken},\ and\ \citenamefont {Verstraete}}]{PhysRevB.94.165116}%
  \BibitemOpen
  \bibfield  {author} {\bibinfo {author} {\bibfnamefont {J.}~\bibnamefont
  {Haegeman}}, \bibinfo {author} {\bibfnamefont {C.}~\bibnamefont {Lubich}},
  \bibinfo {author} {\bibfnamefont {I.}~\bibnamefont {Oseledets}}, \bibinfo
  {author} {\bibfnamefont {B.}~\bibnamefont {Vandereycken}},\ and\ \bibinfo
  {author} {\bibfnamefont {F.}~\bibnamefont {Verstraete}},\ }\bibfield  {title}
  {\bibinfo {title} {Unifying time evolution and optimization with matrix
  product states},\ }\href {https://doi.org/10.1103/PhysRevB.94.165116}
  {\bibfield  {journal} {\bibinfo  {journal} {Phys. Rev. B}\ }\textbf {\bibinfo
  {volume} {94}},\ \bibinfo {pages} {165116} (\bibinfo {year}
  {2016})}\BibitemShut {NoStop}%
\bibitem [{\citenamefont {Yang}\ and\ \citenamefont
  {White}(2020)}]{PhysRevB.102.094315}%
  \BibitemOpen
  \bibfield  {author} {\bibinfo {author} {\bibfnamefont {M.}~\bibnamefont
  {Yang}}\ and\ \bibinfo {author} {\bibfnamefont {S.~R.}\ \bibnamefont
  {White}},\ }\bibfield  {title} {\bibinfo {title} {Time-dependent variational
  principle with ancillary krylov subspace},\ }\href
  {https://doi.org/10.1103/PhysRevB.102.094315} {\bibfield  {journal} {\bibinfo
   {journal} {Phys. Rev. B}\ }\textbf {\bibinfo {volume} {102}},\ \bibinfo
  {pages} {094315} (\bibinfo {year} {2020})}\BibitemShut {NoStop}%
\bibitem [{\citenamefont {Vidal}(2003)}]{2003}%
  \BibitemOpen
  \bibfield  {author} {\bibinfo {author} {\bibfnamefont {G.}~\bibnamefont
  {Vidal}},\ }\bibfield  {title} {\bibinfo {title} {Efficient classical
  simulation of slightly entangled quantum computations},\ }\bibfield
  {journal} {\bibinfo  {journal} {Physical Review Letters}\ }\textbf {\bibinfo
  {volume} {91}},\ \href {https://doi.org/10.1103/physrevlett.91.147902}
  {10.1103/physrevlett.91.147902} (\bibinfo {year} {2003})\BibitemShut
  {NoStop}%
\bibitem [{\citenamefont {Vidal}(2004)}]{2004Eff}%
  \BibitemOpen
  \bibfield  {author} {\bibinfo {author} {\bibfnamefont {G.}~\bibnamefont
  {Vidal}},\ }\bibfield  {title} {\bibinfo {title} {Efficient simulation of
  one-dimensional quantum many-body systems},\ }\bibfield  {journal} {\bibinfo
  {journal} {Physical Review Letters}\ }\textbf {\bibinfo {volume} {93}},\
  \href {https://doi.org/10.1103/physrevlett.93.040502}
  {10.1103/physrevlett.93.040502} (\bibinfo {year} {2004})\BibitemShut
  {NoStop}%
\bibitem [{\citenamefont {Verstraete}\ \emph {et~al.}(2004)\citenamefont
  {Verstraete}, \citenamefont {García-Ripoll},\ and\ \citenamefont
  {Cirac}}]{2004MPS}%
  \BibitemOpen
  \bibfield  {author} {\bibinfo {author} {\bibfnamefont {F.}~\bibnamefont
  {Verstraete}}, \bibinfo {author} {\bibfnamefont {J.~J.}\ \bibnamefont
  {García-Ripoll}},\ and\ \bibinfo {author} {\bibfnamefont {J.~I.}\
  \bibnamefont {Cirac}},\ }\bibfield  {title} {\bibinfo {title} {Matrix product
  density operators: Simulation of finite-temperature and dissipative
  systems},\ }\bibfield  {journal} {\bibinfo  {journal} {Physical Review
  Letters}\ }\textbf {\bibinfo {volume} {93}},\ \href
  {https://doi.org/10.1103/physrevlett.93.207204}
  {10.1103/physrevlett.93.207204} (\bibinfo {year} {2004})\BibitemShut
  {NoStop}%
\bibitem [{\citenamefont {Heyl}(2018)}]{Heyl_2018}%
  \BibitemOpen
  \bibfield  {author} {\bibinfo {author} {\bibfnamefont {M.}~\bibnamefont
  {Heyl}},\ }\bibfield  {title} {\bibinfo {title} {Dynamical quantum phase
  transitions: a review},\ }\href {https://doi.org/10.1088/1361-6633/aaaf9a}
  {\bibfield  {journal} {\bibinfo  {journal} {Reports on Progress in Physics}\
  }\textbf {\bibinfo {volume} {81}},\ \bibinfo {pages} {054001} (\bibinfo
  {year} {2018})}\BibitemShut {NoStop}%
\bibitem [{\citenamefont {{\'A}lvarez}\ \emph {et~al.}(2006)\citenamefont
  {{\'A}lvarez}, \citenamefont {Danieli}, \citenamefont {Levstein},\ and\
  \citenamefont {Pastawski}}]{alvarez2006environmentally}%
  \BibitemOpen
  \bibfield  {author} {\bibinfo {author} {\bibfnamefont {G.~A.}\ \bibnamefont
  {{\'A}lvarez}}, \bibinfo {author} {\bibfnamefont {E.~P.}\ \bibnamefont
  {Danieli}}, \bibinfo {author} {\bibfnamefont {P.~R.}\ \bibnamefont
  {Levstein}},\ and\ \bibinfo {author} {\bibfnamefont {H.~M.}\ \bibnamefont
  {Pastawski}},\ }\bibfield  {title} {\bibinfo {title} {Environmentally induced
  quantum dynamical phase transition in the spin swapping operation},\
  }\href@noop {} {\bibfield  {journal} {\bibinfo  {journal} {The Journal of
  chemical physics}\ }\textbf {\bibinfo {volume} {124}},\ \bibinfo {pages}
  {194507} (\bibinfo {year} {2006})}\BibitemShut {NoStop}%
\bibitem [{\citenamefont {Lazarides}\ \emph {et~al.}(2014)\citenamefont
  {Lazarides}, \citenamefont {Das},\ and\ \citenamefont
  {Moessner}}]{PhysRevE.90.012110}%
  \BibitemOpen
  \bibfield  {author} {\bibinfo {author} {\bibfnamefont {A.}~\bibnamefont
  {Lazarides}}, \bibinfo {author} {\bibfnamefont {A.}~\bibnamefont {Das}},\
  and\ \bibinfo {author} {\bibfnamefont {R.}~\bibnamefont {Moessner}},\
  }\bibfield  {title} {\bibinfo {title} {Equilibrium states of generic quantum
  systems subject to periodic driving},\ }\href
  {https://doi.org/10.1103/PhysRevE.90.012110} {\bibfield  {journal} {\bibinfo
  {journal} {Phys. Rev. E}\ }\textbf {\bibinfo {volume} {90}},\ \bibinfo
  {pages} {012110} (\bibinfo {year} {2014})}\BibitemShut {NoStop}%
\bibitem [{\citenamefont {Abanin}\ \emph {et~al.}(2015)\citenamefont {Abanin},
  \citenamefont {Roeck},\ and\ \citenamefont {Huveneers}}]{Abanin_2015}%
  \BibitemOpen
  \bibfield  {author} {\bibinfo {author} {\bibfnamefont {D.~A.}\ \bibnamefont
  {Abanin}}, \bibinfo {author} {\bibfnamefont {W.~D.}\ \bibnamefont {Roeck}},\
  and\ \bibinfo {author} {\bibfnamefont {F.}~\bibnamefont {Huveneers}},\
  }\bibfield  {title} {\bibinfo {title} {Exponentially slow heating in
  periodically driven many-body systems},\ }\bibfield  {journal} {\bibinfo
  {journal} {Physical Review Letters}\ }\textbf {\bibinfo {volume} {115}},\
  \href {https://doi.org/10.1103/physrevlett.115.256803}
  {10.1103/physrevlett.115.256803} (\bibinfo {year} {2015})\BibitemShut
  {NoStop}%
\bibitem [{\citenamefont {Mori}\ \emph {et~al.}(2016)\citenamefont {Mori},
  \citenamefont {Kuwahara},\ and\ \citenamefont
  {Saito}}]{PhysRevLett.116.120401}%
  \BibitemOpen
  \bibfield  {author} {\bibinfo {author} {\bibfnamefont {T.}~\bibnamefont
  {Mori}}, \bibinfo {author} {\bibfnamefont {T.}~\bibnamefont {Kuwahara}},\
  and\ \bibinfo {author} {\bibfnamefont {K.}~\bibnamefont {Saito}},\ }\bibfield
   {title} {\bibinfo {title} {Rigorous bound on energy absorption and generic
  relaxation in periodically driven quantum systems},\ }\href
  {https://doi.org/10.1103/PhysRevLett.116.120401} {\bibfield  {journal}
  {\bibinfo  {journal} {Phys. Rev. Lett.}\ }\textbf {\bibinfo {volume} {116}},\
  \bibinfo {pages} {120401} (\bibinfo {year} {2016})}\BibitemShut {NoStop}%
\bibitem [{Note1()}]{Note1}%
  \BibitemOpen
  \bibinfo {note} {\protect \url
  {https://itconnect.uw.edu/research/hpc}}\BibitemShut {NoStop}%
\bibitem [{Note2()}]{Note2}%
  \BibitemOpen
  \bibinfo {note} {\protect \url {https://phys.washington.edu}}\BibitemShut
  {NoStop}%
\bibitem [{Note3()}]{Note3}%
  \BibitemOpen
  \bibinfo {note} {\protect \url
  {https://www.artsci.washington.edu}}\BibitemShut {NoStop}%
\end{thebibliography}%
